\documentclass[aps,prb,twocolumn,superscriptaddress,longbibliography]{revtex4-1}

\usepackage{graphicx,bm,amssymb,amsmath,xcolor,pifont,soul,multirow}
\usepackage[linktocpage=true,colorlinks=true,pdfborder={0 0 0},linkcolor=blue,citecolor=red,filecolor=yellow,urlcolor=blue,bookmarks,pdfauthor={},]{hyperref}
\usepackage[normalem]{ulem}
\usepackage{epstopdf}
\usepackage{epsfig}
\usepackage{blkarray}

\def\bk{{\bf k}}
\def\bkp{{\bf k'}}
\def\bq{{\bf q}}

\def\>{\rangle}
\def\<{\langle}
\usepackage{amsmath}
\usepackage{kbordermatrix}

\setstcolor{red}

\newcommand{\AK}[1]{\textcolor{black}{#1}}

\renewcommand{\kbldelim}{(}
\renewcommand{\kbrdelim}{)}

\begin{document}

\title{$Ab~initio$ study of Li-Mg-B superconductors}

\author{Gyanu P. Kafle}
\affiliation{Department of Physics, Applied Physics, and Astronomy, Binghamton University-SUNY, Binghamton, New York 13902, USA}
\author{Charlsey R. Tomassetti}
\affiliation{Department of Physics, Applied Physics, and Astronomy, Binghamton University-SUNY, Binghamton, New York 13902, USA}
\author{Igor I. Mazin}
\affiliation{Department of Physics and Astronomy and Quantum Science and Engineering Center, George Mason University, Fairfax, Virginia 22030, USA}
\author{Aleksey N. Kolmogorov}
\affiliation{Department of Physics, Applied Physics, and Astronomy, Binghamton University-SUNY, Binghamton, New York 13902, USA}
\author{Elena R. Margine}
\email{rmargine@binghamton.edu}
\affiliation{Department of Physics, Applied Physics, and Astronomy, Binghamton University-SUNY, Binghamton, New York 13902, USA}
\date{\today}

\begin{abstract}
LiB, a predicted layered compound analogous to the MgB$_2$ superconductor, has been recently synthesized via cold compression and quenched to ambient pressure, yet its superconducting properties have not been measured. According to prior isotropic superconductivity calculations, the critical temperature ($T_{\rm{c}}$) was expected to be only 10$-$15 K. Using the anisotropic Migdal-Eliashberg formalism, we show that the $T_{\rm{c}}$ may actually exceed 32~K. Our analysis of the contribution from different electronic states helps explain the detrimental effect of pressure and doping on the compound's superconducting properties. In the search for related superconductors, we screened Li-Mg-B binary and ternary layered materials and found metastable phases with $T_{\rm{c}}$ close to or even 10$-$20\% above the record 39 K value in MgB$_2$. Our reexamination of the Li-B binary phase stability reveals a possible route to synthesize the LiB superconductor at lower pressures readily achievable in multianvil cells.

\end{abstract}	

\maketitle

\section{Introduction}
\label{sec:introduction}

Materials with covalent honeycomb layers are a promising avenue in the search for thermodynamically stable high-temperature superconductors. Quintessential among these compounds is the conventional superconductor with the highest critical temperature ($T_{\rm{c}}$) at ambient pressure, MgB$_2$ ($T_{\rm{c}}$ = 39 K)~\cite{Nagamatsu2001}. Attempts to capitalize on the 2001 discovery have led to a large number of theoretical and experimental studies striving to find similar morphologies or appropriate doping and/or pressure conditions that would boost its record-breaking $T_{\rm{c}}$~\cite{Medvedeva2001, Tissen2001, Slusky2001, Zhao2001, Li2001,  Luo2002, Xiang2003, Cava2003,  Gasparov2004, Kazakov2005, ak09, Monni2006,  Bianconi2007,  Karpinski2008, Shein2008, Parisiades2009,  Daghero2009,  Barbero2017, Aydin2018, Pei2021, Yang2022}. This work has shown that metal diborides either lack stability, as in alkali or noble metal systems, or have relatively low $T_{\rm{c}}$ ($\sim$ 10 K) with non-MgB$_2$-type superconductivity, as in transition-metal binaries. The most significant $T_{\rm{c}}$ enhancement in this materials class has been recently observed in compressed MoB$_2$, with 32 K at $\sim 100$ GPa~\cite{Pei2021}. MgB$_2$ doping, in particular, somewhat surprisingly, with Li, has proven to be detrimental to superconductivity~\cite{Slusky2001,Zhao2001,Li2001,Luo2002,Xiang2003,Cava2003,Kazakov2005,Bianconi2007,Karpinski2008,Parisiades2009}. 

A simple phase in the binary Li-B system may offer an alternate path towards modulating MgB$_2$-type superconductivity. A ``metal-sandwich" (MS)-type structure, predicted to stabilize under pressure~\cite{ak08,ak09}, bears a resemblance to MgB$_2$, save for a second added triangular layer of metal atoms inserted in between the hexagonal boron sheets. The LiB phase has been experimentally synthesized and found to be metastable under ambient conditions \cite{Kolmogorov2015}. Its advantages as a superconductor have been enumerated in previous $ab~initio$ studies, mainly a higher degree of two dimensionality, a higher density of states (DOS) of key B-$\sigma$ states at the Fermi level ($E_{\rm F}$), and the possibility of additional electron-phonon coupling, reminiscent of that in CaC$_6$~\cite{Calandra2005}, from the Li-$s$ nearly-free-electron (NFE) states~\cite{Calandra2007, Liu2007,ak08}. Notwithstanding these promising features, LiB has been predicted to have a $T_{\rm{c}}$ of 10–15~K with the McMillan formula based on the standard Migdal-Eliashberg (ME) formalism~\cite{Calandra2007, Liu2007, Martinez2014}. Surprisingly, no anisotropic treatment has been considered for this material, despite the known significance of anisotropic effects in MgB$_2$.

In this study, we evaluate the \AK{thermodynamic} stability of \AK{LiB and} related Li-Mg-B phases and consider the superconductivity of the LiB, MgB, and intermediate ternary LiMgB$_2$ compounds using the fully anisotropic ME formalism. \AK{Based on prior findings revealing an unusual response of the LiB$_y$ and LiB compounds to compression and our present calculations of the Li-B phase stability at different ($T$,$P$) conditions, we conjecture that the desired layered LiB could be obtained in a metastable form at lower pressures. We show that alloying LiB with Mg does not lead to thermodynamically stable layered compounds but the study of their superconducting properties provides further insights into MgB$_2$-type superconductivity.} Our calculations predict a $T_{\rm{c}}$ for LiB under ambient pressure that is much higher than previously reported. To reduce the errors arising from the empirical $\mu$* parameter, we performed calculations with similar settings for LiB and MgB$_2$, and determined that the former should have a critical temperature near 80\% of the latter. At elevated pressures, we find that the $T_{\rm{c}}$ of LiB is strongly suppressed as a result of the concurring reduction of the density of states at the Fermi level and the hardening of the important low-frequency phonon modes. Similarly, with electron or hole doping, no improvements in critical temperature can be attained, suggesting that LiB, just as MgB$_2$~\cite{Slusky2001,Zhao2001,Xiang2003}, already represents an optimally doped material. For the isostructural metastable MgB compound, we show that the $T_{\rm{c}}$ may actually exceed that of MgB$_2$, while the pseudobinary LiMgB$_2$ nearly matches it. Finally, the decomposition of electron-phonon ($e$-ph) coupling by electronic character underlines the importance of B-$\sigma$ states to coupling in these materials, as well as the critical lack of B-$\pi$ states in LiB that is not compensated by the additional coupling from the NFE states. 

\section{Methods}
\label{sec:methods}
The density functional theory (DFT) calculations were performed with the nonlocal van der Waals (vdW) functional optB86b-vdW~\cite{optB86b, Thonhauser2007, Thonhauser2015, Berland2015, Langreth2009, Sabatini2012} to account for dispersive interactions important in the considered layered materials~\cite{Kolmogorov2015}. We used {\small VASP}~\cite{Kresse1996} with projector augmented wave potentials~\cite{Blochl1994} and a 500~eV plane-wave cutoff for the stability analysis of Li-Mg-B phases. The thermodynamic corrections due to vibrational entropy were evaluated within the finite displacement methods implemented in PHONOPY~\cite{Togo2015}. 

The Quantum \textsc{ESPRESSO} package~\cite{QE} with the relativistic Perdew-Burke-Ernzerhof parametrization~\cite{PBE} and norm-conserving  pseudopotentials from the Pseudo Dojo library~\cite{Dojo2018} were employed for calculations of ground-state properties related to superconductivity. A plane-wave cutoff value of 100~Ry, a Methfessel-Paxton smearing~\cite{Methfessel1989} value of 0.01~Ry, and a $\Gamma$-centered $18\times 18 \times 6$ Monkhorst-Pack~\cite{Monkhorst1976} \textbf{k}-mesh for LiB and MgB, $24\times 24 \times 24$ for MgB$_2$, and $18\times 4 \times 10$ for LiMgB$_2$ were used to describe the electronic structure. The dynamical matrices and the linear variation of the self-consistent potential were calculated within density-functional perturbation theory~\cite{Baroni2001} on the irreducible set of a regular $6 \times 6 \times 4$ \textbf{q}-mesh for LiB, MgB, and MgB$_2$, and $6 \times 2 \times 4$ for LiMgB$_2$. 

The EPW code~\cite{Giustino2007, EPW, Margine2013} was used to investigate the $e$-ph interactions and superconducting properties. The electronic wavefunctions required for the Wannier-Fourier interpolation~\cite{WANN1, WANN2} were calculated on a uniform $\Gamma$-centered $12 \times 12 \times 4$ \textbf{k}-grid for LiB and MgB, $6 \times 6 \times 4$ for MgB$_2$, and $12 \times 2 \times 8$ for LiMgB$_2$. Twenty orbitals were used to describe the electronic structure of LiB and MgB with maximally localized Wannier functions: three $sp^2$ and one $p_z$ orbital for each B atom, and one $s$ orbital for each Li or Mg atom. For MgB$_2$, we used one $p_z$ orbital for each B atom and three $s$ orbitals placed in the middle of the B-B bonds, while for LiMgB$_2$, we used three $sp^2$ orbitals for each B atom. Uniform $100 \times 100 \times 40$ \textbf{k}-point and  $50 \times 50 \times 20$ \textbf{q}-point grids for LiB and MgB, $100 \times 100 \times 80$ \textbf{k}-point and  $50 \times 50 \times 40$ \textbf{q}-point grids for MgB$_2$, and $80 \times 32 \times 48$ \textbf{k}-point and  $40 \times 16 \times 24$ \textbf{q}-point grids for LiMgB$_2$ were employed in the superconductivity calculations. For all systems, the Matsubara frequency cutoff was set to 1.0~eV and the Dirac deltas were replaced by Gaussians of width 50~meV (electrons) and 0.5~meV (phonons) when solving the Migdal-Eliashberg equations. 

\section{Results and discussions}
\subsection{Stability of LiB and \texorpdfstring{LiB$_y$}{Lg} compounds}

\begin{figure}[t]
	\centering
	\includegraphics[width=0.92\linewidth]{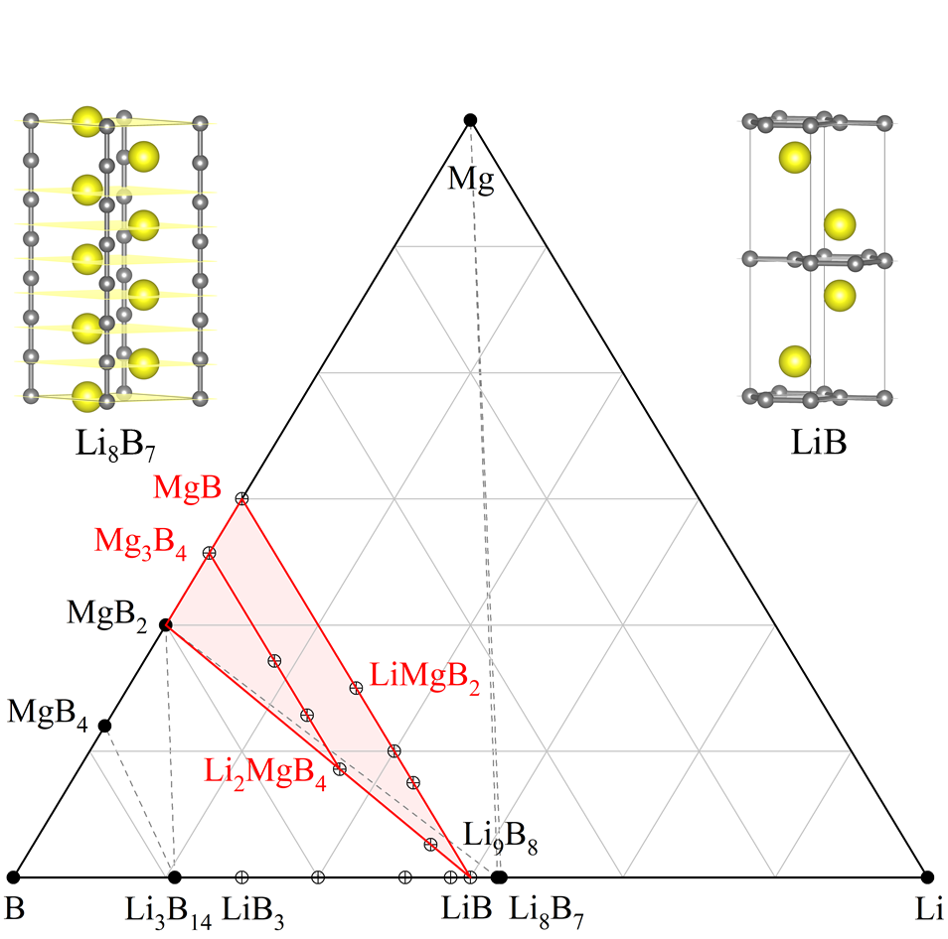}
	\caption{\label{fig01-LiMgB} Location of the known phases (in black) and some of the hypothetical phases (in red) on the Li-Mg-B ternary phase diagram. The open crossed symbols correspond to metastable phases at (0~K, 0~GPa). \AK{The gray dashed lines are the ternary convex hull facet boundaries projected onto the composition plane. The LiB$_{y\approx 0.9}$ compound with incommensurate one-dimensional (1D) B chains and the LiB compound with randomly stacked two-dimensional (2D) B layers are shown with short-period hP15-Li$_8$B$_7$ and hP8-LiB unit cells, respectively.}}
\end{figure}

\begin{figure*}[th!]
	\centering
	\includegraphics[width=0.75\linewidth]{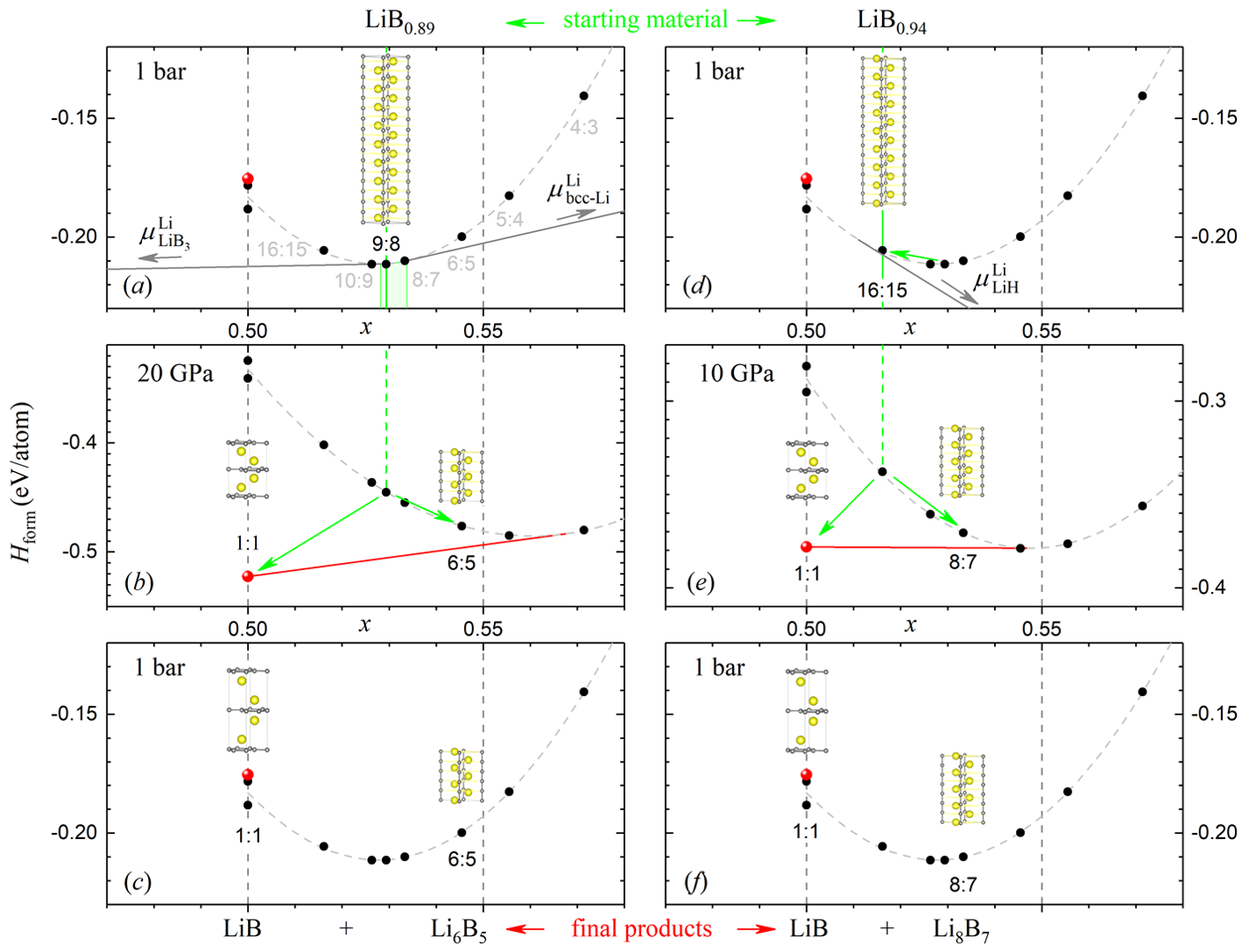}
	\caption{\label{fig02-LiBy} Formation enthalpies of Li$_x$B$_{1-x}$ compounds near  $x$=0.5 illustrating high-pressure LiB synthesis routes [(a)$-$(c)] inferred from a previous experiment~\cite{Kolmogorov2015} and [(d)$-$(f)] proposed in this study. The use of a delithiated LiB$_y$ starting material \AK{(Li$_{16}$B$_{15}\approx$ LiB$_{0.94}$ instead of Li$_{9}$B$_{8}\approx$ LiB$_{0.89}$)} is expected to lower the LiB formation pressure \AK{and lead to a higher fraction of LiB in the resulting multiphase sample (0.52 LiB + 0.48 L$_8$B$_7$ instead of 0.35 LiB + 0.65 Li$_6$B$_5$)}. The dashed curves are parabolic fits to the DFT formation energies of the LiB$_{y\approx0.9}$ compounds with linear boron chains. The straight lines are tangents to the parabolas defining the range of LiB$_y$ stability. \AK{$\mu^{\rm{Li}}_{\rm{LiB}_3}$, $\mu^{\rm{Li}}_{\rm{bcc-Li}}$, and $\mu^{\rm{Li}}_{\rm{LiH}}$ are chemical potentials of Li in LiB$_3$, bcc-Li, and LiH, respectively.}}
\end{figure*}

Structural characterization and thermodynamic stability analysis of Li-B \AK{binary compounds listed in Fig.~\ref{fig01-LiMgB}} have been the focus of numerous studies~\cite{ak08,ak09,Martinez2014,Hermann2012, Hermann2012-LiB, Van2014}. The ambient-pressure Li$_3$B$_{14}$ and LiB$_3$ phases were experimentally determined to have fractional occupancies of Li sites~\cite{Borgstedt2003, Kolmogorov2012-CaB6, Bialon2011} and represented well with ordered tP136 and tI16 unit cells in DFT calculations of the binary convex hull~\cite{Van2014, Curtarolo2005}. The near-stoichiometric LiB$_{y\approx0.9}$ was found to have boron chains incommensurate with the lithium sublattice~\cite{Worle2000}. Modeled with Li$_m$B$_n$ unit cells~\cite{ak09} \AK{(e.g., hP15-Li$_8$B$_7$ displayed in  Fig.~\ref{fig01-LiMgB})}, this unique material was shown to be thermodynamically stable in a finite range of compositions (0.874$\le$y$\le$0.90) at $zero$ temperature because its formation energies follow a near-perfect parabolic dependence (see Fig.~\ref{fig02-LiBy}). The layered LiB \AK{illustrated in Fig.~\ref{fig01-LiMgB}} was predicted to stabilize under pressure~\cite{ak08,ak09} and has been synthesized via cold compression above 20~GPa in diamond anvil cells (DACs)~\cite{Kolmogorov2015}. To describe our conjecture on how the desired LiB superconductor could be obtained at lower pressures, we overview and provide further interpretation of the relevant findings reported in previous studies~\cite{ak09,Kolmogorov2015,Kolmogorov2007}. 

It had been predicted that compressed LiB$_y$ would not  be fully transformed into LiB but would remain in the sample in a more Li-rich form~\cite{ak09}. It had also been suggested that the change in composition could be monitored by tracking certain x-ray diffraction (XRD) reflections from the Li sublattice~\cite{ak09, Hermann2012-LiB}. The collected XRD data confirmed the anticipated evolution of LiB$_y$ under compression~\cite{Kolmogorov2015}. As shown schematically in \AK{Figs.~\ref{fig02-LiBy}(a) and \ref{fig02-LiBy}(b)}, the starting composition of about \footnote{{{{T}he starting {LiB}$_y$ composition in {Ref}.~[\onlinecite{Kolmogorov2015}] was deduced to be around 10:9. {Given} the uncertainty in the determination of the absolute stoichiometry values, we picked the 9:8 starting ratio inside the stability range and illustrated its variation with pressure}}} 9:8 ($x$$\approx$0.529 or $y$$\approx$0.89)  at 1~bar shifted up to 6:5 at 20~GPa, while the appearance of the 002 peak signaled the formation of the layered LiB. Further pressure increase up to 33~GPa pushed the LiB$_y$ composition close to 4:3. Upon decompression down to 1~bar \AK{[Fig.~\ref{fig02-LiBy}(c)]}, the LiB$_y$ composition went partway back to about 6:5 and the two materials remained trapped in their now metastable configurations. A close relationship between the linear-chain and layered Li-B structures, along with the appearance of dynamical instability in LiB$_y$ phases in the 10$-$20~GPa range, helped rationalize why the transformation took place at room temperature~\cite{Kolmogorov2015}.

These observations reveal that LiB does not need to be thermodynamically stable with respect to LiB$_3$ and LiB$_y$ to form~\footnote{{T}he ordered {tI16}-{LiB$_3$} was found to be thermodynamically stable at 0 {K} in the previous {PBE} calculations~\cite{Van2014}. In the present {optB86b} treatment, it is metastable by 9 {meV}/atom at 0 {K} and by 5 {meV}/atom at 600 {K} when the vibrational entropy term is included. {The} phase may become stable at higher temperatures and/or upon inclusion of the configuration entropy contribution}. Because of the high strength of the boron chains in LiB$_y$ that can survive 1,200~K temperatures~\cite{Kolmogorov2007} and the direct transformation path between the two phases with the 1D and 2D boron frameworks, the LiB enthalpy needs to be just low enough to destabilize the starting LiB$_y$ compound. This happens when the line from LiB touches the LiB$_y$ formation enthalpy parabola above the starting composition [shown in red in Figs.~\ref{fig02-LiBy}(b) and \ref{fig02-LiBy}(e)]. According to the DFT results in Supplemental Material Fig.~S1~\cite{SM}, any LiB$_y$ stable at zero temperature and pressure should become unstable at 6~GPa.

\begin{figure*}[t]
	\centering
	\includegraphics[width=0.9\linewidth]{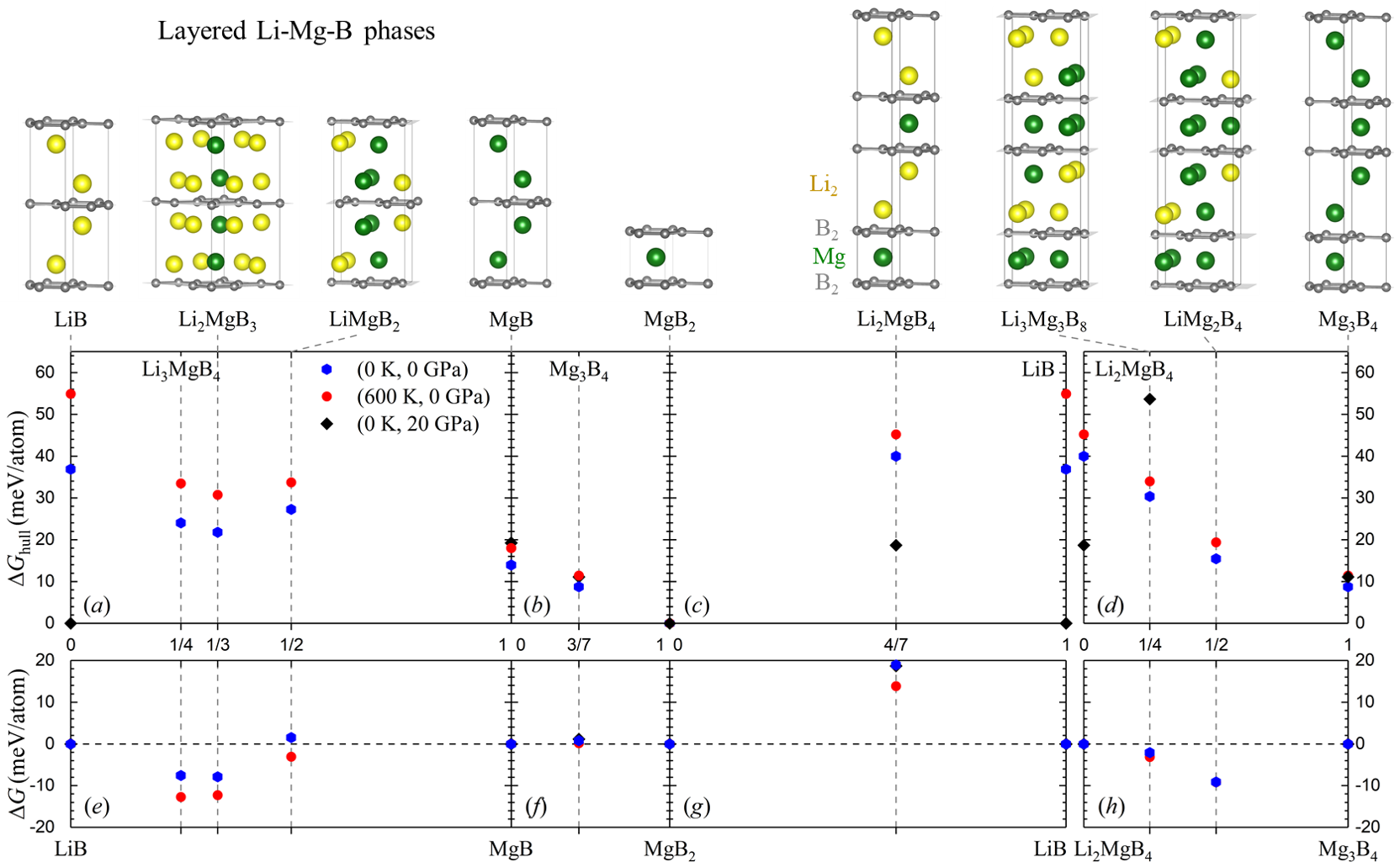}
	\caption{\label{fig03-LiMgB} \AK{Structure and stability of} layered Li-Mg-B phases obtained by combining the LiB and MgB$_2$ blocks. \AK{To illustrate the relation between the morphological and thermodynamic properties, the compounds are displayed along the four composition lines LiB-MgB-MgB$_2$-LiB and Li$_2$MgB$_4$-Mg$_3$B$_4$ highlighted in Fig.\ref{fig01-LiMgB},} with the intermediate phases marked as fractions of the end-point phases. \AK{[(a)$-$(d)]} The distance to the \AK{ternary} convex hull \AK{($\Delta G_{{\rm hull}}$)} at different ($T$,$P$) conditions. \AK{[(e)$-$(h)] The relative Gibbs free energies ($\Delta G$) of composite structures with respect to the end-point phases ({e.g.}, $\Delta G_{{\rm LiMgB}_2} = G_{{\rm LiMgB}_2} - 0.5 G_{{\rm LiB}} - 0.5 G_{{\rm MgB}}$). All end-point and select intermediate compounds are depicted above the panels, with the corresponding compositions marked by the dashed gray lines.}
    }
\end{figure*}

We propose that delithiation of LiB$_y$ prior to compression could promote the formation of LiB. The possibility of pushing LiB$_y$ outside its natural stability range to prime the compound for the synthesis of metastable LiBH and LiBH$_2$ derivatives was discussed in Ref.~[\onlinecite{Kolmogorov2007}]. The presented analysis~\cite{Kolmogorov2007} of a previous LiB$_y$ hydrogenation attempt~\cite{Reinosothesis} showed that the small amount of absorbed hydrogen and a minor change in the composition could be explained by the formation of LiH that drew Li from LiB$_y$ until the chemical potentials in the two materials equilibrated. As argued in Ref.~[\onlinecite{Kolmogorov2007}] and demonstrated in Fig.~\ref{fig02-LiBy}(d), the $\mu_{\rm{LiH}}^{\rm{Li}}$ value estimated at $-$0.83 eV at zero temperature is sufficient to move the lower stability boundary down to $x\approx$ 0.510 ($y\approx$ 0.96). A similar effect could be achieved by heat-treating the material at high temperatures, which has been successfully used to extract about half of Li from the LiBC compound~\cite{Fogg2006,Kalkan2019}.

In the scenario proposed in this work, a 16:15 sample would be destabilized at a noticeably lower pressure, around 3~GPa, and might ultimately yield a larger fraction of the desired LiB phase. The actual pressure needed to induce the transformation is difficult to evaluate from first principles because one would need to model the kinetics of the rebonding process. Performing the experiment at elevated temperatures would help the system overcome barriers but, at the same time, would make LiB less stable due to an unfavorable  vibrational entropy contribution and raise the transformation pressure by 1$-$2~GPa at 300~K (see Supplemental Fig.~S1~\cite{SM}). Considering that the LiB$_y$ compound's composition already started to change at 10~GPa but LiB reflections were not yet detectable at that pressure~\cite{Kolmogorov2015}, the LiB synthesis starting with the delithiated LiB$_y$ might not require much higher pressures and could be achievable in a multianvil setup. Magnetic susceptibility measurements could help identify the first appearance of the superconducting LiB phase.

\subsection{Stability of Li-Mg-B compounds}

Mixing different chemical elements is another way of stabilizing particular morphologies. A systematic screening of layered motifs in dozens of M-B and Li-M-B chemical systems uncovered only a few possibly synthesizable compounds, such as Li$_2$AlB$_4$, Li$_2$VB$_4$, and Li$_2$TaB$_4$~\cite{Kolmogorov2008}. The stabilization of the composite layered structures comprised of AlB$_2$ and LiB units (see Li$_2$MgB$_4$ in Fig.~\ref{fig03-LiMgB})  with electron-rich metals occurs because of the availability of unfilled B-$\sigma$ states in LiB. Unfortunately, transition metals have been shown to completely flood these electronic states and make the compounds poor superconducting candidates~\cite{Kolmogorov2008}. 

In the present study, we focus on exploring the part of the Li-Mg-B ternary phase space shaded in red in Fig.~\ref{fig01-LiMgB} that could host (near)-stable layered configurations. The choice of the monovalent Li and divalent Mg ensures that the resulting structures retain hole-doped B-$\sigma$ states. Compared to previous investigations~\cite{ak09,Kolmogorov2008}, we have now sampled configurations with a mixed population of sites within the metal layers and examined the thermodynamic stability at elevated temperatures and pressures.

To probe the propensity of the two metals to mix, we selected the 8-atom LiB and the 14-atom Li$_2$MgB$_4$ base structures and sampled possible decorations of the metal sites in laterally expanded supercells with up to 48 and 28 atoms, respectively. We found segregation of Li and Mg into homonuclear layers to be unfavorable. The lowest-energy structures shown in Fig.~\ref{fig03-LiMgB} have relatively small hexagonal or orthorhombic unit cells with mixed Li-Mg closed-packed layers. At the LiMgB$_2$ composition, for example, the B$_2$-(LiMg)$_{0.5}$-(MgLi)$_{0.5}$-B$_2$-(MgLi)$_{0.5}$-(LiMg)$_{0.5}$ sequence (oP16) is 13 meV/atom lower in energy than the B$_2$-Li-Li-B$_2$-Mg-Mg stacking (hP8) and 29 meV/atom lower than B$_2$-Li-Mg-B$_2$-Li-Mg (hP8). 

\AK{Figures~\ref{fig03-LiMgB}(a)$-$\ref{fig03-LiMgB}(d)} show the distance to the convex hull \AK{under different ($T$,$P$) conditions} for the best identified candidates \AK{[the set of planes defining the (0 K, 0 GPa) ternary convex hull in Fig. 1 is constructed using the thermodynamically stable phases shown with solid points]. The feasibility of obtaining metastable phases depends not only on how far they are from the convex hull but also on how they compare against morphologically related phases. For example,} Mg$_3$B$_4$ appears to be closest to  stability, being only 9~meV/atom at (0~K, 0~GPa). However, \AK{constructed as a stacking} of the known MgB$_2$~\cite{Nagamatsu2001} and the previously considered MgB~\cite{ak09} blocks,
\AK{it is actually slightly less favorable by 0.8 meV/atom than the mixture of these two phases. Given the usefulness of the relative stability analysis that has helped us identify the favorable LiB and Li-M-B layered motifs in our previous studies~\cite{ak09,Kolmogorov2008}, we have plotted energy differences in Figs.~\ref{fig03-LiMgB}(e)$-$\ref{fig03-LiMgB}(h) along select composition lines shown in Fig.~\ref{fig01-LiMgB}}. The results illustrate that most of the intermediate phases, especially LiMgB$_2$, Mg$_3$B$_4$, and Li$_2$MgB$_4$, have energies close to the linear interpolation of the end-point energies. The behavior indicates that the building blocks are fairly independent and cannot be assembled into particularly beneficial sequences. The largest gains of about 10~meV/atom for Li$_3$MgB$_4$, Li$_2$MgB$_3$, and LiMg$_2$B$_4$ are insufficient to bring the formation energies down below the boundaries of the convex hull at zero temperature and pressure.

Finally, we included the enthalpic and entropic contributions to examine the stability of the select Li-Mg-B compounds under elevated pressures and temperatures. According to the relative enthalpies at 20~GPa and relative free energies at 600~K shown in Fig.~\ref{fig03-LiMgB}, these external conditions do not favor any new binary or ternary phases. The substantial stabilization of LiB under pressure caused by the collapse of the $c$ axis discussed in Refs.~[\onlinecite{Calandra2007,ak08}] further increases the distance of the pseudobinary Li$_{1-x}$Mg$_x$B compounds from the convex hull. The notable destabilization of LiB under elevated temperatures, by 16~meV/atom at 600~K, lowers the free energy of the LiMgB$_2$ intermediate compound below the linear interpolation line. However, the vibrational entropy contribution has an overall destabilizing effect on the considered layered Li-Mg-B phases, as their relative free energies shown in red in Fig.~\ref{fig03-LiMgB} become more positive. \AK{The configurational entropy is not expected to stabilize the Li$_{1-x}$Mg$_x$B pseudobinary either: even if structures with different random populations of the metal sites by Li and Mg had the same energy, which we actually found to be dispersed by at least 30 meV/atom, the free energy would be lowered by at most $-18$~meV/atom at $x=0.5$ and 600 K.}

We conclude that the proposed layered Li-Mg-B morphologies are not thermodynamically stable to be obtained with standard bulk synthesis methods. It may be worth investigating the possibility of synthesizing these metastable phases with thin-film deposition methods. The following analysis of these hypothetical materials’ properties is aimed at establishing beneficial features for the MgB$_2$-type superconductivity.

\subsection{Superconductivity of LiB}

The presence of an additional metal layer in the MS structure sets it apart from the common AlB$_2$-type prototype~\cite{ak08,ak09}. The larger separation between the honeycomb sheets makes the B-dominated electronic features even more 2D and reduces the energy difference between various stackings. In fact, LiB was anticipated~\cite{ak09} and determined~\cite{Kolmogorov2015} to form random stacking sequences. Computationally, the phase has been represented with short-period MS1 (hR12, \AK{not shown}) and MS2 (hP8, \AK{shown in Fig.~\ref{fig01-LiMgB}}) structures \AK{with the ABC and AB stackings, respectively,} for the analysis of superconducting properties~\cite{Calandra2007, Liu2007}. In agreement with previous calculations~\cite{ak08, Liu2007}, we find MS2 to be slightly lower in energy by 1.8~meV/atom and use the 8-atom unit cell representation in our present superconductivity analysis. The fully optimized parameters of the $P6_3/mmc$ (No.~194) structure are $a$=3.047~$\text{\r{A}}$ and $c$=11.012~$\text{\r{A}}$ ($c$/$a$=3.614).

\begin{figure*}[t]
	\centering
	\includegraphics[width=0.98\linewidth]{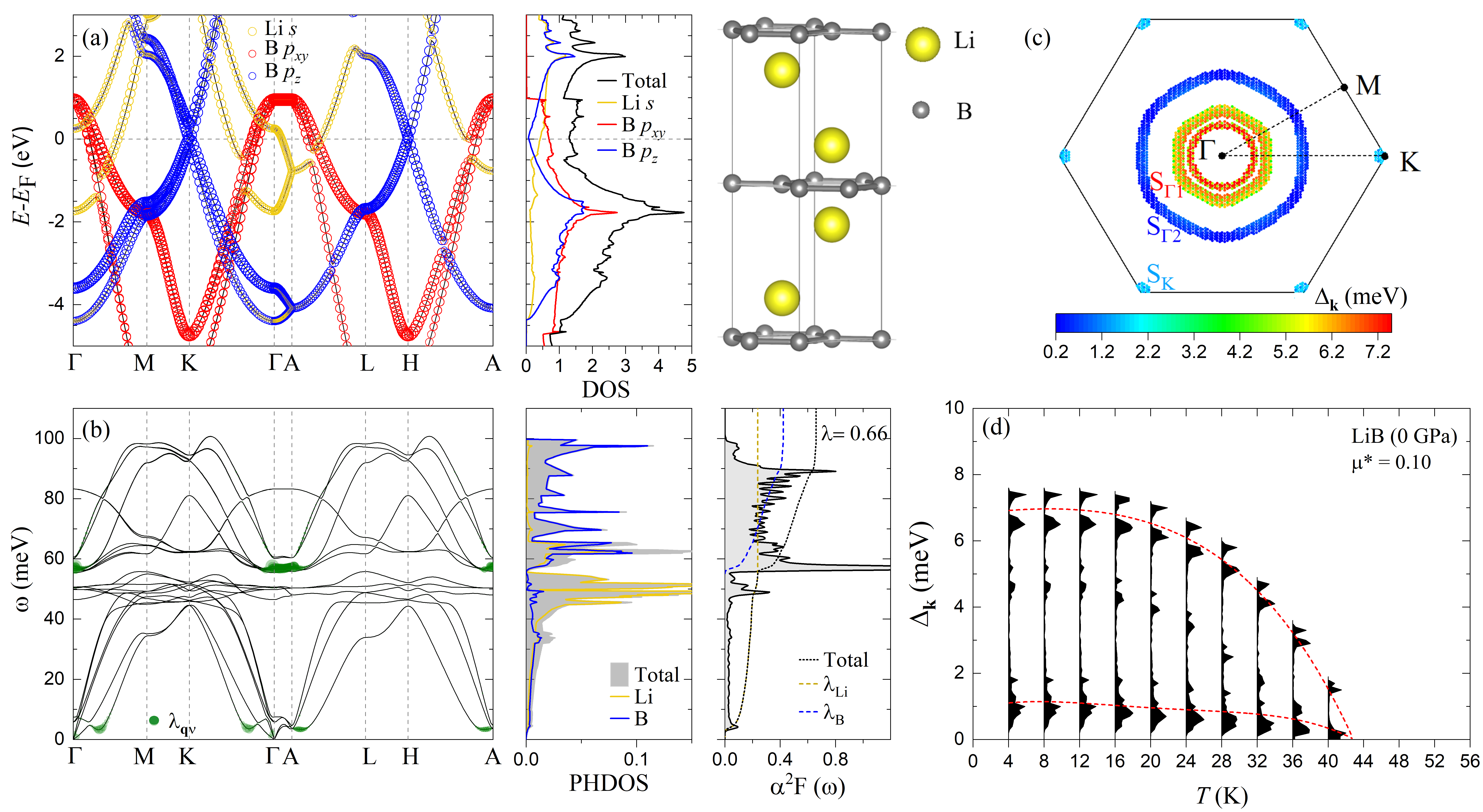}
	\caption{\label{fig04-LiB} Summary of LiB properties at 0 GPa. (a) Calculated band structure and DOS [states/(eV f.u.)]. The size of the symbols is proportional to the contribution of each orbital character. (b) Calculated phonon dispersion, phonon DOS, and Eliashberg spectral function $\alpha^2F(\omega)$. The phonon branches are broadened by the $e$-ph coupling strength $\lambda_{\bq\nu}$, showcasing the modes and high-symmetry directions that have the strongest $e$-ph coupling. (c) The distribution of superconducting gap at 4~K on the cross-section Fermi surface plane through $z$=0. (d) Calculated anisotropic superconducting gaps $\Delta_\bk$ as a function of temperature (red dashed curves are a guide to the eye).}
\end{figure*}

The electronic and vibrational properties of the MS phases have been discussed in detail in Refs.~[\onlinecite{ak08, Calandra2007, Liu2007}], and here we overview the key features. As shown in Fig.~\ref{fig04-LiB}(a), the main contribution to the DOS at the Fermi level ($N_{\rm F}$) comes from the B-$\sigma$ and Li-$s$ states (labeled as $\zeta$), while there is a notable lack of the B-$\pi$ states as the vertices of the Dirac cones at the $K$ and $H$ points lie nearly exactly at $E_{\rm F}$. In contrast, MgB$_2$ has no states of the intercalant atom at the Fermi level, and both B-$\sigma$ and B-$\pi$ bands are hole doped. Furthermore, in MgB$_2$, the two Dirac cones are asymmetrically shifted by $\sim$2~eV above and below $E_{\rm F}$ at the $K$ and $H$ points, and the $\sigma$ bands have a noticeable dispersion along the $\Gamma$-$A$ direction above $E_{\rm F}$ (Supplemental Fig.~S2(a)~\cite{SM}). Lastly, we observe that the B-$\sigma$ bands of LiB at the $\Gamma$ point are 0.7~eV higher in energy than in MgB$_2$, doubling the contribution of these states to the total $N_{\rm F}$. 

The phonon dispersion and the phonon density of states of LiB are shown in Fig.~\ref{fig04-LiB}(b). At ambient pressure, the optical region of the phonon spectrum is clearly separated by a small gap into two parts. As displayed by the characters in the phonon DOS, the modes in the  40-55~meV frequency range correspond to vibrations of Li atoms, whereas the modes above 55~meV are primarily related to B vibrations. As pointed out in Ref.~[\onlinecite{Calandra2007}], the Li-derived phonon modes are nearly dispersionless, behaving similar to Einstein modes. In MgB$_2$, there is a similar divide between the B and intercalant atom vibrations, although there is no frequency gap, and the Mg vibrations are not nearly as nondispersive, as illustrated in Supplemental Fig.~S2(b)~\cite{SM}.

The superconducting critical temperature of LiB under ambient conditions has been previously estimated using the McMillan expression for $T_{\rm{c}}$, which relies on the isotropic $e$-ph coupling strength~\cite{Calandra2007, Liu2007}. In this study, we go a step further and explore the role of anisotropy on the superconducting properties of LiB. To begin, in Fig.~\ref{fig04-LiB}(b), we show the isotropic Eliashberg spectral function $\alpha^2F(\omega)$ and the cumulative $e$-ph coupling strength $\lambda(\omega)$. There are three significant peaks in the spectral function, the largest corresponding to the E$_{2g}$ Raman mode at 58~meV at $\Gamma$ associated with the in-plane stretching of the B-B bonds [see Fig.~\ref{fig04-LiB}(b)]. The other two peaks at 6 and 90~meV are related to the soft out-of-plane vibrations of B and Li atoms, and the in-plane vibrations of B atoms, respectively. By examining $\lambda(\omega)$, one can infer that the coupling of the aforementioned E$_{2g}$ mode with the B-$\sigma$ bands comprises about 25\% of the total value ($\lambda$ = 0.66). In MgB$_2$, the E$_{2g}$ mode has a slightly higher energy of 72~meV at the zone center and makes up 45\% of $\lambda=0.66$. This difference can be understood by noting the dispersive distribution of the E$_{2g}$ branch along the $\Gamma$-$A$ direction in MgB$_2$ in contrast to the flat character in LiB.  In Fig.~\ref{fig04-LiB}(b) and Supplemental Fig.~S2(b), the strength of the $e$-ph coupling of this branch is illustrated by the size of the green dots on the phonon spectra. Solving the isotropic ME equations, we obtain $T_{\rm c}$=8$-$15~K for $\mu^*$=0.20$-$0.10, in line with previous estimates with the McMillan formula~\cite{Calandra2007, Liu2007}. 

Shifting to the anisotropic properties, we plot in Figs.~\ref{fig04-LiB}(c) and \ref{fig04-LiB}(d) the superconducting gap distribution on the Fermi surface (FS) and the superconducting gap as a function of temperature. The superconducting spectrum displays a multigap structure with separated, highly anisotropic domes, which can be associated with different FS sheets: the higher energy gap (above 4~meV) lies on the cylindrical FS centered along the $\Gamma$-$A$ direction originating from the B-$\sigma$ states, while the lower energy gap (below 2~meV) belongs to the ellipsoidal FS at $\Gamma$ arising from the intercalant NFE $\zeta$ states (see FS plots in Fig.~\ref{fig04-LiB} and Supplemental Fig.~S3~\cite{SM}). The $T_{\rm{c}}$ obtained from the fully ME anisotropic solutions is in the 34$-$42~K range for $\mu^*$=0.20$-$0.10 (Fig.~\ref{fig04-LiB}(d) and Supplemental Fig.~S4(a)~\cite{SM}). This value is three times larger than the value estimated with the isotropic ME formalism, and approximately 80\% that of MgB$_2$. 






As discussed in Ref.~[\onlinecite{Eilat}], one can introduce several different matrices to describe the anisotropy of the $e$-ph (or any other, for that matter), pairing interactions. The most fundamental is the matrix $\Lambda$ defined as
\begin{equation}
\Lambda_{ij} = \frac{2}{N_{\rm F}} \sum_{\bk \in i} \sum_{\bkp \in j}  \sum_{\nu} \omega^{-1}_{\bq \nu} |g_{\bk \bkp \nu}|^2\delta(\epsilon_{\bk}) \delta(\epsilon_{\bkp}), \end{equation}
where $i$ and $j$ are states on the $i{\rm th}$ and $j{\rm th}$ portions of the FS, and $g_{\bk \bkp \nu}$ is the $e$-ph matrix element for the scattering between the electronic states $\bk$ and $\bkp$ through  a phonon with wave vector $\bq = \bk-\bkp$, frequency $\omega_{\bq \nu}$, and branch index $\nu$. The total DOS is $N_F=\sum_{i}N_i$, with $N_i$ the DOS per spin at the Fermi level from the $i{\rm th}$ FS region.

Then, one can define the pairing interaction matrix,
\begin{equation}
V_{ij}=\Lambda_{ij} N_F/N_i N_j,
\end{equation}
as well as the asymmetric $e$-ph coupling matrix, 
\begin{equation}
\lambda_{ij}=V_{ij}N_j=\Lambda_{ij} N_F/N_i.
\end{equation}
Within the Bardeen-Cooper-Schrieffer (BCS) theory, the critical temperature is defined by the BCS equations with the largest eigenvalue of this matrix, $\lambda_{eff}=\max[ \mathrm{eigenvalue}\{\lambda_{ij}\}]$, and the isotropic  
coupling constant (which is always smaller) 
$\lambda_{\rm isotropic} = \sum_{ij} N_i\lambda_{ij}/N_F=\sum_{i, j} \Lambda_{ij}$, which also defines 
the $e$-ph specific heat renormalization. The anisotropic mass renormalization, measured, for instance, in quantum oscillations experiments, is given by $\lambda_i = \sum_{j}\lambda_{ij}= N_F\sum_j \Lambda_{ij}/N_j$. 

At 0~GPa, we separated the Fermi surface into three parts: two concentric cylinders along the $\Gamma$-$A$ direction of different radii and one along the $K$-$H$ direction, as denoted by the labeled FS regions in Fig.~\ref{fig04-LiB}(c). With this choice, the three regions include: (i) the B-$\sigma$ states plus the Li-$\zeta$ states closest to the $\Gamma$ point (S$_{\Gamma 1}$), (ii) the Li-$\zeta$ states further away from the $\Gamma$ point (S$_{\Gamma 2}$), and (iii) the B-$\pi$ states around the $K$ point (S$_K$). This decomposition leads to the following $\Lambda_{ij}$ matrix for LiB: 



\begin{equation}
  \rm {LiB}~(0~\rm {GPa}) = \kbordermatrix{
    & \mbox{\scriptsize{{$\sigma$} + $\zeta$}} & \mbox{\scriptsize{$\zeta$}} & \mbox{\scriptsize{$\pi$}} \\
    \mbox{\scriptsize{$\sigma$ + $\zeta$}} & 0.506 & 0.059 & 0.006 \\
    \mbox{\scriptsize{$\zeta$}} & 0.059 & 0.019 & 0.002 \\
    \mbox{\scriptsize{$\pi$}}  & 0.006 & 0.002 & 0.000
  },
\end{equation}
with $N_{\sigma + \zeta} = 0.507,\ N_{\zeta} = 0.173,$ and $N_{\pi} = 0.020.$
Our analysis of this matrix shows that the main contribution to $\lambda$ comes from the B-$\sigma$ states entangled with the intercalant $\zeta$ states (region S$_{\Gamma 1}$), with minor coupling from the intercalant $\zeta$ states (region S$_{\Gamma 2}$), and nothing from the B-$\pi$ states (region S$_K$). 

As a point of comparison, we also solved the fully anisotropic ME equations for MgB$_2$, and found, as expected, its $\pi$ and $\sigma$ band two-gap structure, as shown in Supplemental Fig.~S2(d)~\cite{SM}. The estimated $T_{\rm{c}}$ is 40 and 50~K for $\mu^*$ equal to 0.20 and 0.10, respectively. By defining a two-band model for the two sets of $\sigma$ and $\pi$ FS sheets (S$_{\Gamma 1}$ and S$_{\Gamma 2}$ as shown in Supplemental Fig.~S2(c)~\cite{SM}), we obtained the $\Lambda_{ij}$ matrix for MgB$_2$ to be



\begin{equation}
  \rm {MgB_2} = \kbordermatrix{
    & \mbox{\scriptsize{{$\sigma$}}} & \mbox{\scriptsize{$\pi$}} \\
    \mbox{\scriptsize{$\sigma$}} & 0.339 & 0.084  \\
    \mbox{\scriptsize{$\pi$}} & 0.084 & 0.149  \\
  },
\end{equation}
with $N_{\sigma} = 0.146$ and $N_{\pi} = 0.194$,
in very good agreement with the former studies~\cite{Mazin2004, Floris2007}. 

As mentioned above, LiB lacks any contribution to $\lambda$ from the $\pi$ states, while they give rise to 23\% of the coupling in MgB$_2$. Furthermore, we find that the off-diagonal terms representing the interband transitions in LiB are much smaller than the main diagonal term corresponding to the intraband scattering on the S$_{\Gamma 1}$ FS sheet. On the other hand, in MgB$_2$, the off-diagonal terms are about 25\% and 50\% of the two diagonal terms, respectively. 


\begin{figure}[t]
	\centering
	\includegraphics[width=0.8\linewidth]{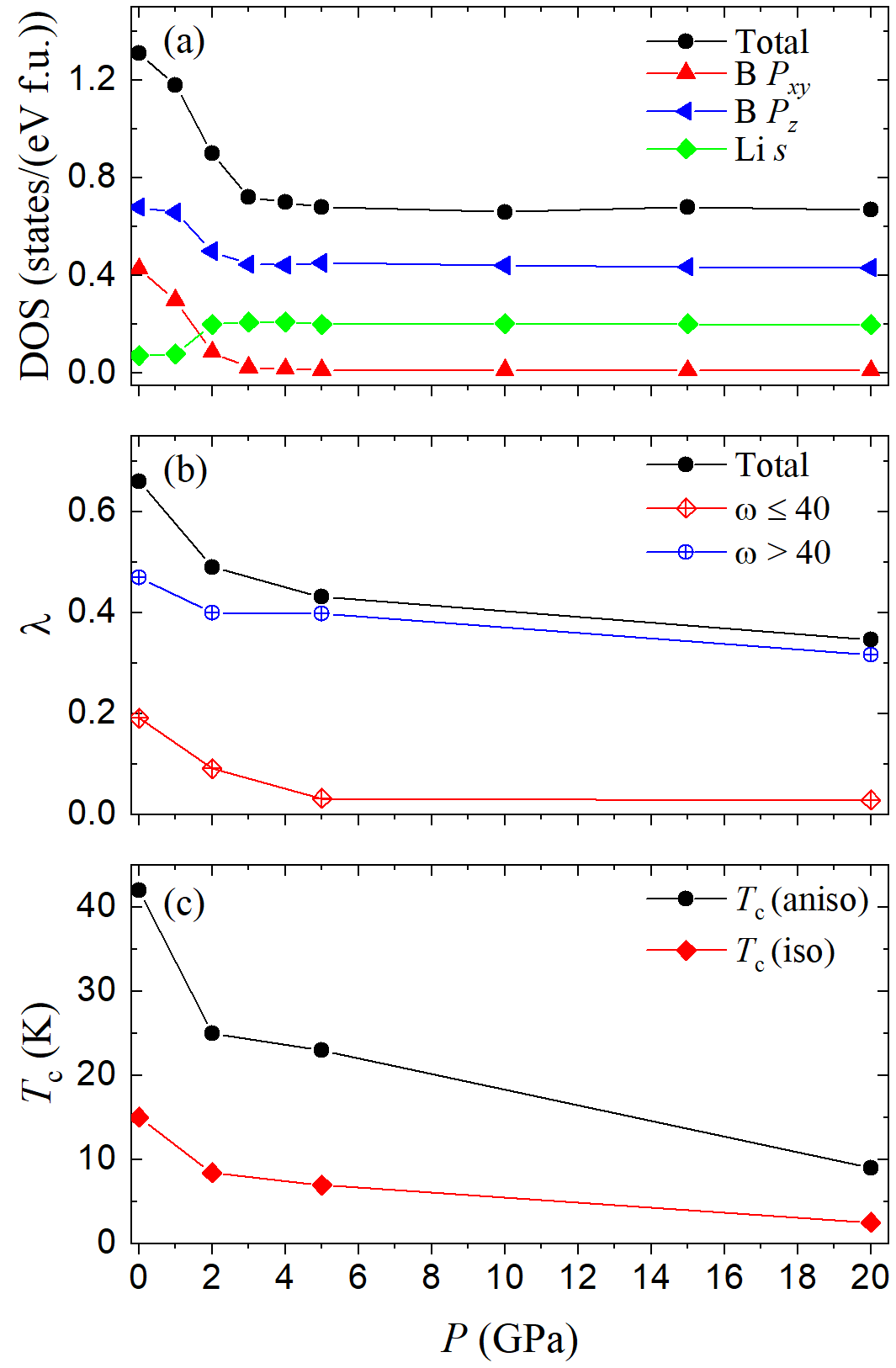}
	\caption{\label{fig05-dos-lam-tc} (a) Calculated total and projected DOS at the $E_{\rm F}$, (b) the total $\lambda$ and partition of the $\lambda$ according to the labeled frequency range, and (c) the anisotropic and isotropic $T_{\rm{c}}$ of LiB as a function of pressure.} 
\end{figure}

The following analysis reveals that compression has a dramatic effect on the compound's superconducting properties. As found previously~\cite{Calandra2007}, the lattice parameters vary nonlinearly under hydrostatic pressure. While $a$ expands by 3\% up to 3 GPa, before monotonically decreasing at higher pressures, $c$ decreases rapidly by about 30\% up to 3~GPa, before plateauing at higher pressures (see Supplemental Fig.~S5~\cite{SM}). Because the interaction between the layers is weak, the out-of-plane axis is readily compressible at low pressure until the intercalant atoms are close enough to start experiencing the hard-core repulsion. While the vdW corrections play a relatively small role in LiB~\cite{Kolmogorov2015}, it is still important to include them in order to accurately estimate the pressure point that separates the two distinctly different regimes of the LiB structural changes. Without vdW corrections, this pressure point is shifted from 3 to 5~GPa~\cite{Calandra2007}. 

The band structure of LiB also undergoes notable changes at low pressures up to 3~GPa (see Supplemental Fig.~S6~\cite{SM}).  As has been pointed out previously~\cite{Calandra2007}, this is caused by the charge redistribution from the Li to B states due to the rapid collapse of the interlayer distance. As a result, the $\sigma$ bands along the $\Gamma$-$A$ direction and the $\pi$ bands at the $K$- and $H$-points significantly lower in energy, while the Li-$\zeta$ bands are pushed up. A summary of these changes is shown in Fig.~\ref{fig05-dos-lam-tc}, exhibiting the aforementioned rapid decrease in projected DOS (PDOS) up to 3 GPa in all states except for B-$\pi$. While this increase in B-$\pi$ PDOS may seem promising, this contribution remains below the one observed in MgB$_2$, and cannot make up for the decrease in PDOS of the other states. Therefore, the net result is a reduction of the $N_F$ as a function of pressure. Between 3 and 20 GPa, there are few pertinent changes, the noteworthy points being that, even at the highest pressure in our study, the B-$\sigma$ and B-$\pi$ bands are both partially occupied, and the Li-$\zeta$ states have no more contribution to the $N_{\rm F}$ above 3~GPa. 

In Supplemental Fig.~S7~\cite{SM}, we report the evolution of the vibrational properties under pressure. As shown in the figure, the gap in the phonon spectrum has vanished as the optical phonon branches associated with the Li vibrations are no longer flat and localized to a specific frequency range, but have dispersed and mixed with the B modes within the 50$-$80~meV region. One thing to note is that the dispersion of the high-frequency modes along the $\Gamma$-$M$ and $K$-$\Gamma$ directions decreases with pressure beyond 5 GPa, after briefly increasing between 2 and 5~GPa. The overall effect of the increase in pressure is a shift of the phonon frequencies to higher values. For instance, between 0 and 20 GPa, the low-frequency optical mode at $\Gamma$ rises from 7 to 27~meV, while the peak of the highest frequency mode goes up from 100 to 118~meV.  

Upon increasing pressure, the $e$-ph coupling displays a monotonic decrease as both the DOS at the Fermi level and the contribution to $\lambda$ of the low-frequency phonons are strongly reduced. As shown in Fig.~\ref{fig05-dos-lam-tc}(b), the $\lambda$ value from the modes below 40 meV quickly drops and vanishes, while that of the other modes remains nearly constant over the full pressure range. In Supplemental Table~S1~\cite{SM}, we have also provided the band-resolved $\Lambda_{ij}$ matrices under pressure. While there is a considerable increase in the B-$\pi$ contribution, the diagonal component associated with the $\sigma$ and $\zeta$ states remains the dominant term overall, still accounting for 69\% of total $\lambda$=0.35 at 20~GPa.

The two-gap structure of LiB is retained under pressure although significant changes are observed in the FS topology. For example, as can be seen in Supplemental Fig.~S8~\cite{SM}, the FS sheet due to the NFE states is largely suppressed at 2~GPa and is completely absent at 5~GPa. At the same time, the $\Gamma$ pocket associated with the  B-$\sigma$ states shrinks, while the $K$ pocket linked to the B-$\pi$ states expands with pressure. The outcome is a strong reduction in the magnitude of the superconducting gap and the corresponding critical temperature.  For $\mu^*$=0.10, the  $T_{\rm{c}}$ obtained from the fully anisotropic solution of the ME is 25, 23, and 9~K at 2, 5, and 20~GPa, respectively, as shown in Supplemental Fig.~S9~\cite{SM}. 

Next, we examine if it is possible to favorably modulate the key electronic and vibrational features of LiB through electron ($e$) or hole (h) doping in order to further increase the $T_{\rm c}$. To simulate doping, we employed the jellium model, in which the extra charge added to the system is compensated by a uniform background of opposite charge. In our calculations, we have fixed the unit cell and allowed only the optimization of atomic positions. Under a doping level of 1/6 electrons or holes per unit cell at 0~GPa, the electronic structure of LiB undergoes only minor changes compared to the undoped case, as shown in Supplemental Fig.~S10~\cite{SM}. In particular, under h doping, the Li-$\zeta$ and B-$\pi$ bands rise slightly above the $E_{\rm F}$, while the opposite trend occurs for $e$ doping. The B-$\sigma$ states are not significantly affected either way. The result of these changes is a modest rise in the contribution from the B-$\pi$ states to the $N_{\rm F}$ for both types of doping, with the Li-$\zeta$ contribution increasing with $e$ doping, and decreasing with $h$-doping. Overall, the total DOS varies very little, from 1.32 in the undoped case to 1.34 and 1.35 states/(eV f.u.) for $e$ and h doping. 

As can be seen in Supplemental Fig.~S11~\cite{SM}, the most notable change in the phonon spectra arises from the hardening of the Raman active mode with E$_{2g}$ symmetry at 58~meV at $\Gamma$ by 15~meV with h doping and 5~meV with $e$ doping. As for the undoped system, this mode has the largest contribution to the $e$-ph coupling. We find that the total $\lambda$ decreases by 5\% with h doping ($\lambda$=0.63) and increases by 5\% with $e$ doping ($\lambda$=0.69), respectively. By dividing the $\lambda$ contributions from modes below and above 40 meV, we can further conclude that the increase in $\lambda$ for the $e$-doped case stems from the increase in coupling from modes below 40~meV that exceeds slightly the decrease in coupling from the higher frequency modes.  

The superconducting gap structures and $\Gamma$-plane cross-section Fermi surfaces for the two doping cases are shown in Supplemental Fig.~S12~\cite{SM}. Compared to the undoped system, the intensity of the two peaks in the upper superconducting gap is reversed, shifting from a maximum intensity at $\sim$ 7~meV to a maximum intensity at $\sim$ 4~meV. The superconducting gaps vanish at $T_{\rm{c}}$ = 41~K for $e$ doping, and $T_{\rm{c}}$ = 36~K for h doping, both lower than the value found for undoped LiB for the same $\mu$*=0.10. To understand the drop in $T_{\rm c}$, a summary of the band-resolved $e$-ph coupling between states for both doping cases is available in Supplemental Table~S2~\cite{SM}. This analysis highlights several pertinent points regarding the nature of the $e$-ph coupling in LiB, mainly that h doping the system may increase the coupling from the B-$\pi$ states, but only at the loss of coupling from the B-$\sigma$ and Li-$\zeta$ states. For $e$ doping, the only increase in $\lambda$ comes from the B-$\sigma$ and Li-$\zeta$ contribution, but is not sufficient to raise the critical temperature.

\subsection{Superconductivity of MgB}

\begin{figure*}[t]
	\centering
	\includegraphics[width=0.95\linewidth]{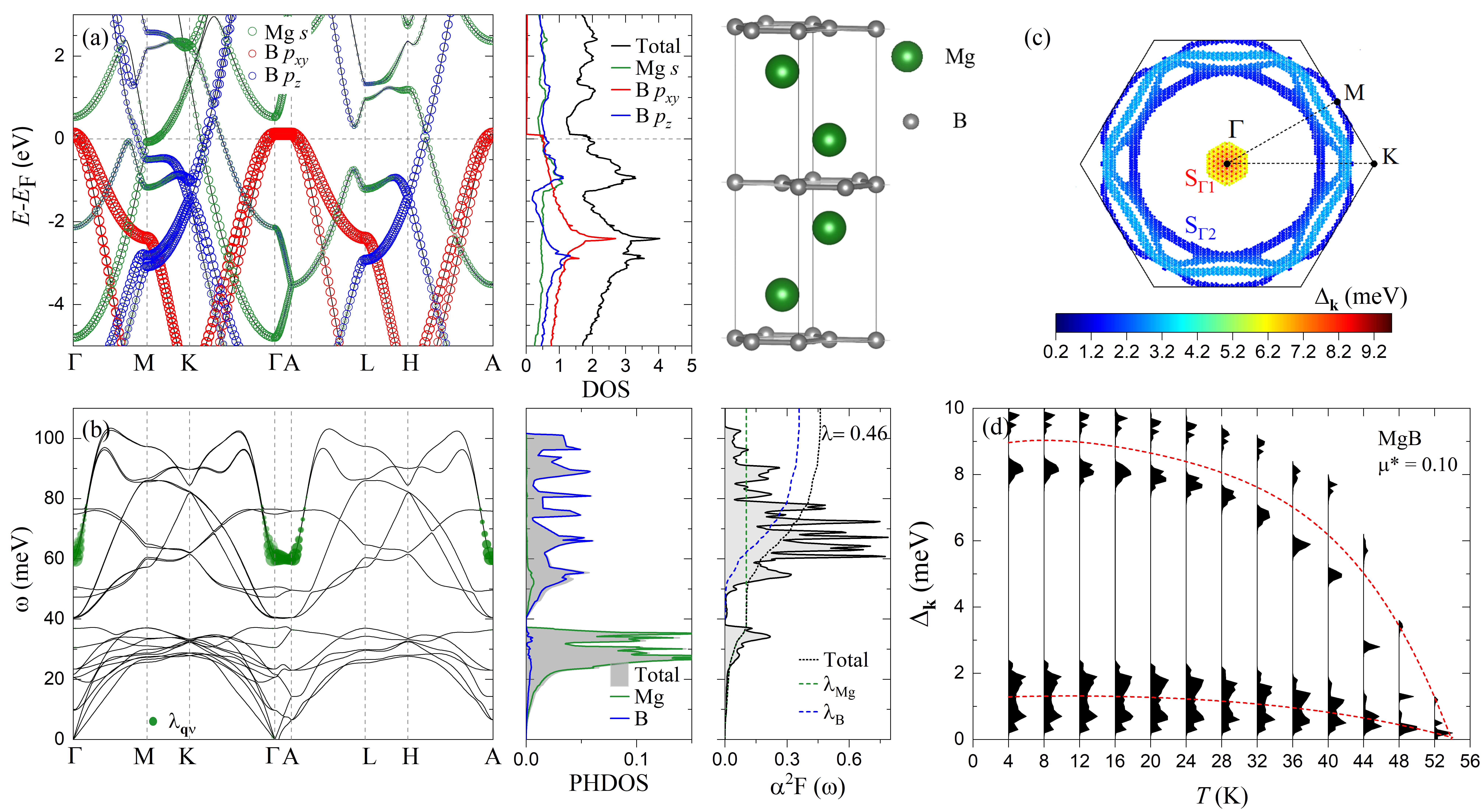}
	\caption{\label{fig06-MgB} Summary of MgB properties at 0~GPa. (a) Calculated band structure and DOS (states/(eV f.u.)). The size of the symbols is proportional to the contribution of each orbital character. (b) Calculated phonon dispersion, phonon DOS, and Eliashberg spectral function $\alpha^2F(\omega)$. The phonon branches are broadened by the $e$-ph coupling strength $\lambda_{\bq\nu}$, showcasing the modes and directions that have the strongest $e$-ph coupling. (c) The distribution of superconducting gap at 4~K on the cross-section Fermi surface plane through $z$=0. (d) Calculated anisotropic superconducting gaps $\Delta_\bk$ as a function of temperature (red dashed curves are a guide to the eye).}
\end{figure*}

In this section, we assess the electronic, vibrational, and superconducting properties of the hypothetical MS2-MgB phase, shown to be only metastable in previous \cite{ak09} and present studies, with the intent to further characterize helpful features for MgB$_2$-type superconductivity. Compared to LiB, both the in-plane and out-of-plane lattice parameters increase, reaching values of $a$=3.127~$\text{\r{A}}$ and $c$=11.943~$\text{\r{A}}$  ($c$/$a$=3.819), respectively. 

The electronic band structure and DOS of MgB are shown in Fig.~\ref{fig06-MgB}(a). Similarly to LiB and MgB$_2$, there are two hole-doped $\sigma$ bands derived from the in-plane B $p_{x, y}$ orbitals that cross the Fermi level. While  in LiB  the B $\pi$-$\pi$ Dirac cones lie exactly at $E_{\rm F}$, in MgB these states cross at approximately 1.5~eV below $E_{\rm F}$, leading to an increase in $N_{\rm F}$ by a factor of 1.5. The filling of the $\pi$ states is a key feature that could enhance the $e$-ph coupling as in MgB$_2$. It is also worth noting that the Dirac cones at $K$ and $H$ in MgB are symmetrically aligned with respect to the $E_{\rm F}$ as in LiB, rather than asymmetrically shifted as in MgB$_2$. Finally, compared to LiB, the interlayer $\zeta$ bands are significantly lower in energy relative to the $\sigma$ bands and hybridize more strongly with the B $p_z$ orbitals. This downward energy shift is related to the increase in the interlayer distance, thus providing more space for the electrons to occupy the $\zeta$ states which are located in the interlayer region where there is a strong attractive potential from the Mg ions. As a consequence, the Mg-$\zeta$ states have a  notably higher contribution to the $N_{\rm F}$ than the Li-$\zeta$ states of LiB. 

The phonon dispersion, isotropic Eliashberg spectral function $\alpha^2F(\omega)$, and cumulative $e$-ph coupling of MgB at ambient pressure are presented in Fig.~\ref{fig06-MgB}(b). Like in LiB, there is a small gap that divides the phonon spectrum in two parts associated primarily with Mg and B vibrations. However, there is a considerable change in the width of the two regions, the low- and high-frequency portions having shrunken and extended by approximately 15 and 20 meV, respectively.  As for the other phases investigated here, the $e$-ph coupling in MgB is mainly due to the E$_{2g}$ in-plane B stretching mode in the 60$-$70~meV frequency range. The value of $\lambda$=0.46 is 30\% smaller than that of LiB, a consequence of MgB's lack of a low frequency coupling mode. However, as shown in Fig.~\ref{fig05-dos-lam-tc}(b), excluding the contribution of the phonon modes under 40~meV in LiB also results in $\lambda$=0.46. We estimate, by solving the isotropic ME equations, a critical temperature in the range 5$-$8~K for $\mu^*$=0.15-0.10, considerably lower than that of LiB in the same isotropic limit.  

The superconducting gap on a cross-section FS plane through $\Gamma$ and the anisotropic gap $\Delta_\bk$ as a function of temperature are illustrated in Figs.~\ref{fig06-MgB}(c) and \ref{fig06-MgB}(d) (see Supplemental Fig.~S13~\cite{SM} for a three-dimensional (3D) FS plot). For $\mu^*$=0.10, the $T_{\rm c}$ obtained from the fully anisotropic solution ($T_{\rm c}$=54~K) is about seven times larger than the isotropic value. This is due to a highly anisotropic multigap structure, with minimum and maximum values at 0.2 and 10~meV in the low-temperature limit. In particular, the gap in the 0$-$2~meV energy range corresponds to the entangled B-$\pi$ and Mg-$\zeta$ states (region S$_{\Gamma 2}$), while the one in the 8$-$10~meV range is attributed to the B-$\sigma$ states (region S$_{\Gamma 1}$). 

Dividing the FS into two subsets, one for the B-$\sigma$ sheets and one for the B-$\pi$ and Mg-$\zeta$ sheets, we calculate the band-resolved symmetrized matrix $\Lambda_{ij}$: 


\begin{equation}
  \rm {MgB} = \kbordermatrix{
    & \mbox{\scriptsize{{$\sigma$}}} & \mbox{\scriptsize{$\pi$+$\zeta$}} \\
    \mbox{\scriptsize{$\sigma$}} & 0.218 & 0.050  \\
    \mbox{\scriptsize{$\pi$+$\zeta$}} & 0.050 & 0.142  \\
  },
\end{equation}
with $N_{\sigma}$ = 0.254 and $N_{\pi+\zeta}$ = 0.676. Like for MgB$_2$ and LiB, the dominant contribution to $\lambda$ comes from the coupling of the B-$\sigma$ states with the in-plane $E_{2g}$ phonons. Furthermore, MgB captures desired key features of the former, having both B-$\pi$ and intercalant electrons present at the Fermi level, that play important roles in indirectly strengthening the coupling of the $\sigma$ electrons and enhancing the $T_{\rm{c}}$. 

\subsection{Superconductivity of \texorpdfstring{LiMgB$_2$}{Lg}}

\begin{figure*}[t]
	\centering
	\includegraphics[width=0.95\linewidth]{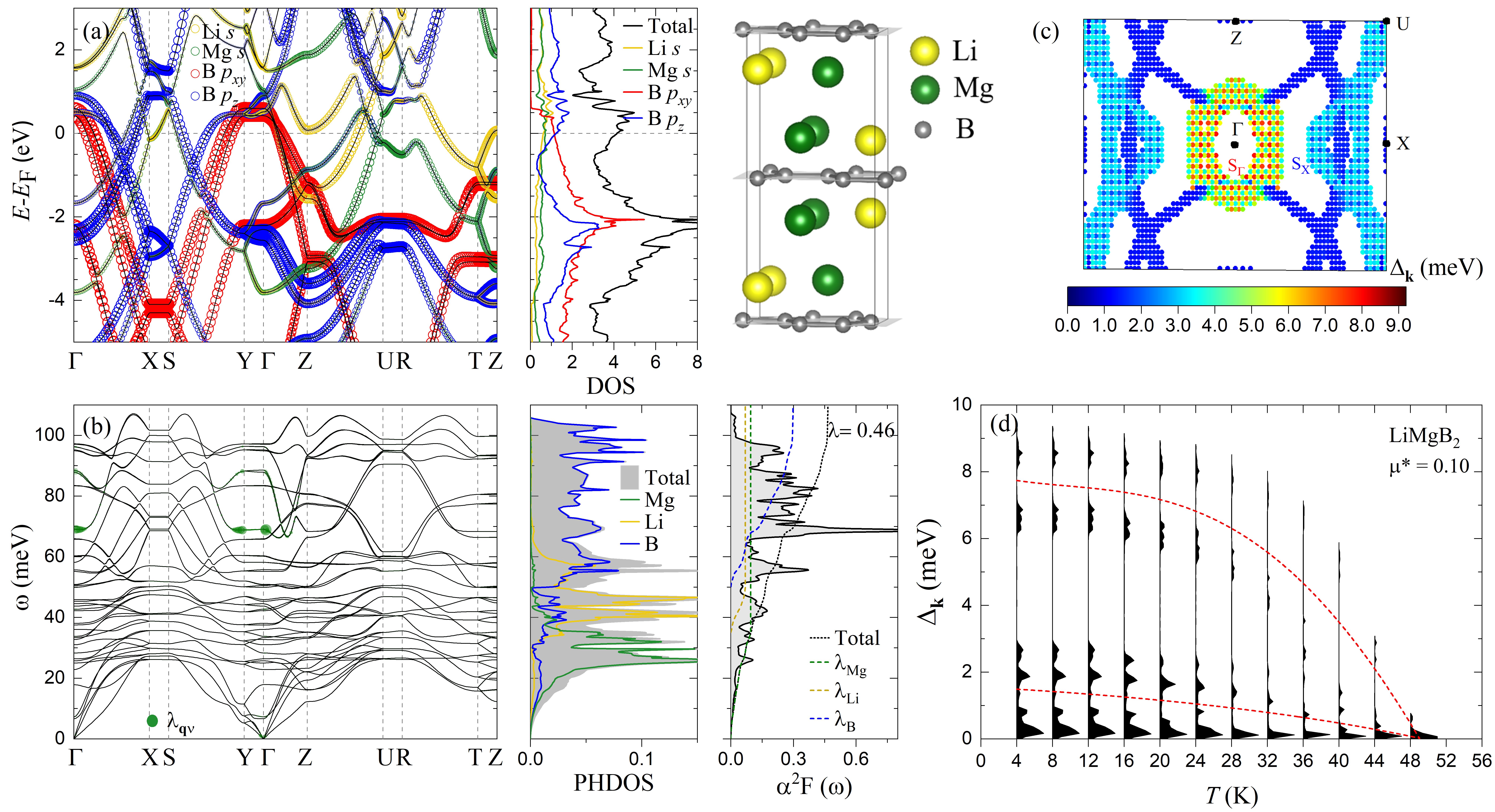}
	\caption{\label{fig07-LiMgB2} Summary of LiMgB$_2$ properties. (a) Calculated band structure and DOS (states/(eV f.u.)). The size of the symbols is proportional to the contribution of each orbital character. (b) Calculated phonon dispersion, phonon DOS, and Eliashberg spectral function $\alpha^2F(\omega)$. The phonon branches are broadened by the $e$-ph coupling strength $\lambda_{\bq\nu}$, showcasing the modes and directions that have the strongest $e$-ph coupling. (c) The distribution of superconducting gap at 4~K on the cross-section Fermi surface plane through $z$=0. (d) Calculated anisotropic superconducting gaps $\Delta_\bk$ as a function of temperature (red dashed curves are a guide to the eye).}
\end{figure*}

 In this section, we explore the electronic, vibrational, and superconducting properties of a predicted near-stable compound in the Li-Mg-B phase space: LiMgB$_2$, an orthorhombic structure with $Pmmn$ symmetry, in the B$_2$-(LiMg)-B$_2$-(LiMg) stacking sequence. This phase was chosen for further study because its composition bridges the divide between LiB and MgB, while having one of the smaller unit cells of the phases discussed in this work, with lattice parameters $a$=3.073~$\text{\r{A}}$, $b$=12.045~$\text{\r{A}}$, and $c$=5.291~$\text{\r{A}}$.
 
 Figure~\ref{fig07-LiMgB2}(a) shows the electronic structure and DOS of LiMgB$_2$, highlighting the four main contributions to the $N_{\rm F}$: Mg-$\zeta$, Li-$\zeta$, B-$\sigma$, and B-$\pi$. Compared to LiB, the Li-$\zeta$ bands have moved down by 2~eV such that a second band is now filled. In contrast, the Mg-$\zeta$ bands have emptied compared to MgB. With respect to the B-$\sigma$ bands, LiMgB$_2$ is the perfect hybrid phase, as its $\sigma$ bands have a degree of hole doping that is right in the middle of LiB and MgB. Another key feature of LiMgB$_2$ is that the B-$\pi$ Dirac cones do not lie at the Fermi level as in LiB, instead dropping approximately 1~eV below $E_{\rm F}$. Finally, we remark that the Mg-$\zeta$ and Li-$\zeta$ states have nearly equal contributions to the $N_{\rm F}$, while the B-$\pi$ contribution exceeds that of the B-$\sigma$ states, as is the case for MgB and MgB$_2$.
 
 
 The phonon dispersion of LiMgB$_2$ along with the phonon DOS and the isotropic Eliashberg spectral function are displayed in Fig.~\ref{fig07-LiMgB2}(b). As for LiB and MgB, the phonon spectrum is split into two regions associated with the metal and B vibrations by a narrow gap at around 50~meV. A closer look shows that in the low-frequency region the Mg modes lie below 40~meV, as in MgB, while the Li modes are confined in the 40$-$65~meV range, as in LiB. This can be linked to the fact that the mass of Mg is 3.5 times larger than that of Li, leading to frequencies which are on average 1.9 times softer. Examining the integrated spectral function, we find that these two sets of modes contribute together $\sim$0.16 (35\% of the total $\lambda$=0.46). The Raman active B$_{3g}$ mode at 69~meV corresponding to the in-plane stretching of B-B bonds is responsible for approximately 15\% of the $e$-ph coupling. For comparison, in the binary phases LiB, MgB and MgB$_2$, the in-plane B-B stretching mode accounts for 25\%, 35\%, and 45\% of their respective $e$-ph coupling. Lastly, we note that an equal $\lambda$ value of 0.46 was determined for LiMgB$_2$ and MgB, as they both miss the gain coming from the soft phonon branch present in LiB below 10~meV. For this $\lambda$ value, the isotropic $T_{\rm c}$ is estimated to be 4$-$7~K for $\mu^*$=0.15-0.10. 

Given the similarity of the key electronic and vibrational features that drive the $e$-ph coupling in MgB and LiMgB$_2$, it is not surprising to find that the superconducting multigap structure illustrated in Fig.~\ref{fig07-LiMgB2}(d) closely resembles that of MgB. As further shown by the gap structure on the cross-section plane (Fig.~\ref{fig07-LiMgB2}(c); a 3D view of the FS is given in Supplemental Fig.~S14~\cite{SM}), we can attribute the two high-energy gaps above 6~meV in the low-temperature limit to the B-$\sigma$ states on the S$_{\Gamma}$ pocket, and the lower-energy gaps to the complex S$_{X}$ sheets of entangled B-$\pi$, Li-$\zeta$, and Mg-$\zeta$ states. The estimated critical temperature for $\mu^*=0.10$ is $T_{\rm}=49$~K, which falls right in between the values obtained for LiB and MgB with similar settings. In particular, this value is 14\% higher than LiB, and 9\% lower than MgB.

Lastly, to quantify the contribution of different electronic states to the $e$-ph coupling in this compound, we can make a similar FS division to that of MgB: one for the B-$\sigma$ sheets and one for the B-$\pi$ and $\zeta$ surfaces [see Fig.~\ref{fig07-LiMgB2}(c)]. The calculated symmetric $\Lambda_{ij}$ matrix for LiMgB$_2$ is as follows:


\begin{equation}
  \rm {LiMgB_2} = \kbordermatrix{
    & \mbox{\scriptsize{{$\sigma$}}} & \mbox{\scriptsize{$\pi$+$\zeta$}} \\
    \mbox{\scriptsize{$\sigma$}} & 0.203 & 0.063  \\
    \mbox{\scriptsize{$\pi$+$\zeta$}} & 0.063 & 0.134  \\
  },
\end{equation}
with $N_{\sigma} = 0.714$ and $N_{\pi+\zeta} = 1.385$. Though MgB and LiMgB$_2$ have the same total $\lambda$, a smaller percentage in LiMgB$_2$ can be attributed to the $\sigma$ states, 44\% as opposed to 47\%. Interestingly, nearly the same contribution in these two compounds comes from the diagonal term corresponding to entangled $\pi$ and $\zeta$ states, $\sim$ 30\%, though LiMgB$_2$ has a larger degree of $e$-ph coupling arising from interband transitions. Therefore, the drop in $T_{\rm c}$, compared to MgB, stems mostly from the minute decrease in the B-$\sigma$ coupling triggered by the addition of Li. 

\begin{figure}[t]
	\centering
	\includegraphics[width=0.99\linewidth]{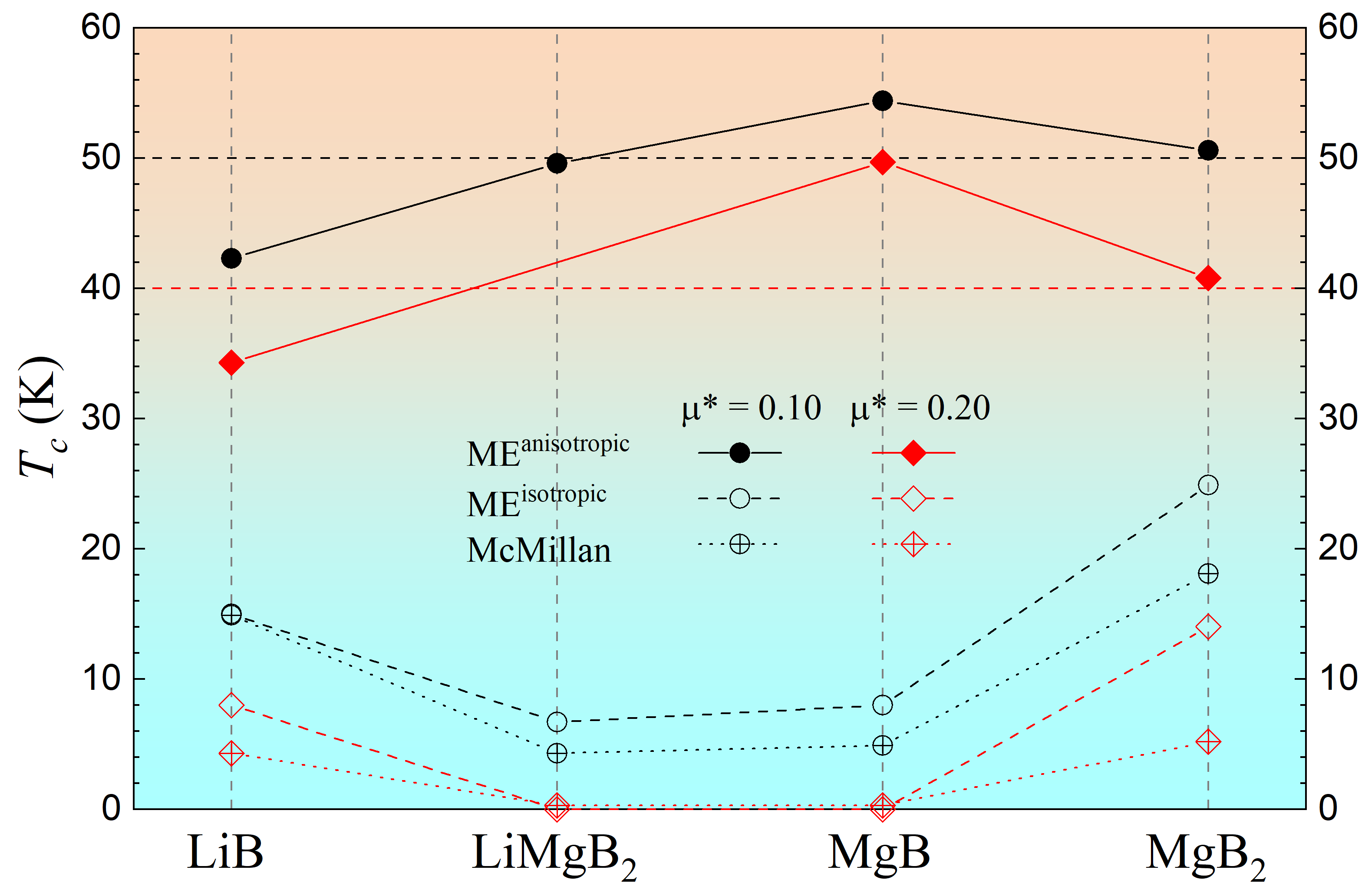}
	\caption{\label{fig08-Tc} Comparison of $T_{\rm{c}}$ estimated for observed (LiB and
MgB$_2$) and metastable (LiMgB$_2$ and MgB) layered borides at different levels of theory and different $\mu^*$ values.}
\end{figure}

\section{Conclusions}
\label{sec:conclusions}

In this work, we argue that the delithiation of the known starting LiB$_y$ material prior to compression would facilitate the synthesis of the LiB superconductor, leading to its formation at lower pressures readily achievable in multianvil cells.
Our examination of the LiB superconducting properties using the ME theory indicates that an anisotropic treatment is as essential in this compound as in MgB$_2$. We predict a critical temperature for LiB in the 32$-$42~K range, a factor of three times higher than estimates based on an isotropic description of the $e$-ph interaction. We further demonstrate that neither pressure nor doping can enhance superconductivity, with the $T_{\rm{c}}$ expected to drop below 10~K at 20~GPa. Our analysis of the Li-Mg-B phase space identifies several metastable layered materials with excellent superconducting properties. In particular, we find the critical temperature in MgB and LiMgB$_2$ to exceed or approach that of MgB$_2$, as summarized in Fig.~\ref{fig08-Tc}. These \AK{metastable} materials are \AK{not} likely to be synthesizable with standard bulk methods but \AK{might have a greater chance of forming in thin-film deposition experiments. Synthesizing these layered compounds may} challenge the supremacy of the MgB$_2$ superconductor.

\begin{acknowledgments} 
The authors acknowledge support from the National Science Foundation (NSF) (Awards No. DMR-2132586 and No. DMR-2132589). This work used the Expanse system at the San Diego Supercomputer Center via allocation TG-DMR180071 and the Frontera supercomputer at the Texas Advanced Computing Center via the Leadership Resource Allocation (LRAC) grant No. 2103991 (allocation DMR22004). Expanse is supported by the Extreme Science and Engineering Discovery Environment (XSEDE) program~\cite{XSEDE} through NSF Award No. ACI-1548562, and Frontera is supported by NSF Award No. OAC-1818253~\cite{Frontera}. 

\end{acknowledgments}

\bibliography{refs} 

\begin{thebibliography}{69}%
\makeatletter
\providecommand \@ifxundefined [1]{%
 \@ifx{#1\undefined}
}%
\providecommand \@ifnum [1]{%
 \ifnum #1\expandafter \@firstoftwo
 \else \expandafter \@secondoftwo
 \fi
}%
\providecommand \@ifx [1]{%
 \ifx #1\expandafter \@firstoftwo
 \else \expandafter \@secondoftwo
 \fi
}%
\providecommand \natexlab [1]{#1}%
\providecommand \enquote  [1]{``#1''}%
\providecommand \bibnamefont  [1]{#1}%
\providecommand \bibfnamefont [1]{#1}%
\providecommand \citenamefont [1]{#1}%
\providecommand \href@noop [0]{\@secondoftwo}%
\providecommand \href [0]{\begingroup \@sanitize@url \@href}%
\providecommand \@href[1]{\@@startlink{#1}\@@href}%
\providecommand \@@href[1]{\endgroup#1\@@endlink}%
\providecommand \@sanitize@url [0]{\catcode `\\12\catcode `\$12\catcode
  `\&12\catcode `\#12\catcode `\^12\catcode `\_12\catcode `\%12\relax}%
\providecommand \@@startlink[1]{}%
\providecommand \@@endlink[0]{}%
\providecommand \url  [0]{\begingroup\@sanitize@url \@url }%
\providecommand \@url [1]{\endgroup\@href {#1}{\urlprefix }}%
\providecommand \urlprefix  [0]{URL }%
\providecommand \Eprint [0]{\href }%
\providecommand \doibase [0]{http://dx.doi.org/}%
\providecommand \selectlanguage [0]{\@gobble}%
\providecommand \bibinfo  [0]{\@secondoftwo}%
\providecommand \bibfield  [0]{\@secondoftwo}%
\providecommand \translation [1]{[#1]}%
\providecommand \BibitemOpen [0]{}%
\providecommand \bibitemStop [0]{}%
\providecommand \bibitemNoStop [0]{.\EOS\space}%
\providecommand \EOS [0]{\spacefactor3000\relax}%
\providecommand \BibitemShut  [1]{\csname bibitem#1\endcsname}%
\let\auto@bib@innerbib\@empty
\bibitem [{\citenamefont {Nagamatsu}\ \emph {et~al.}(2001)\citenamefont
  {Nagamatsu}, \citenamefont {Nakagawa}, \citenamefont {Muranaka},
  \citenamefont {Zenitani},\ and\ \citenamefont {Akimitsu}}]{Nagamatsu2001}%
  \BibitemOpen
  \bibfield  {author} {\bibinfo {author} {\bibfnamefont {J.}~\bibnamefont
  {Nagamatsu}}, \bibinfo {author} {\bibfnamefont {N.}~\bibnamefont {Nakagawa}},
  \bibinfo {author} {\bibfnamefont {T.}~\bibnamefont {Muranaka}}, \bibinfo
  {author} {\bibfnamefont {Y.}~\bibnamefont {Zenitani}}, \ and\ \bibinfo
  {author} {\bibfnamefont {J.}~\bibnamefont {Akimitsu}},\ }\bibfield  {title}
  {\enquote {\bibinfo {title} {Superconductivity at 39 {K} in magnesium
  diboride},}\ }\href {http://www.nature.com/articles/35065039} {\bibfield
  {journal} {\bibinfo  {journal} {Nature}\ }\textbf {\bibinfo {volume} {410}},\
  \bibinfo {pages} {63} (\bibinfo {year} {2001})}\BibitemShut {NoStop}%
\bibitem [{\citenamefont {Medvedeva}\ \emph {et~al.}(2001)\citenamefont
  {Medvedeva}, \citenamefont {Ivanovskii}, \citenamefont {Medvedeva},\ and\
  \citenamefont {Freeman}}]{Medvedeva2001}%
  \BibitemOpen
  \bibfield  {author} {\bibinfo {author} {\bibfnamefont {N.~I.}\ \bibnamefont
  {Medvedeva}}, \bibinfo {author} {\bibfnamefont {A.~L.}\ \bibnamefont
  {Ivanovskii}}, \bibinfo {author} {\bibfnamefont {J.~E.}\ \bibnamefont
  {Medvedeva}}, \ and\ \bibinfo {author} {\bibfnamefont {A.~J.}\ \bibnamefont
  {Freeman}},\ }\bibfield  {title} {\enquote {\bibinfo {title} {Electronic
  structure of superconducting {MgB$_2$} and related binary and ternary
  borides},}\ }\href {https://link.aps.org/doi/10.1103/PhysRevB.64.020502}
  {\bibfield  {journal} {\bibinfo  {journal} {Phys. Rev. B}\ }\textbf {\bibinfo
  {volume} {64}},\ \bibinfo {pages} {020502} (\bibinfo {year}
  {2001})}\BibitemShut {NoStop}%
\bibitem [{\citenamefont {Tissen}\ \emph {et~al.}(2001)\citenamefont {Tissen},
  \citenamefont {Nefedova}, \citenamefont {Kolesnikov},\ and\ \citenamefont
  {Kulakov}}]{Tissen2001}%
  \BibitemOpen
  \bibfield  {author} {\bibinfo {author} {\bibfnamefont {V.~G.}\ \bibnamefont
  {Tissen}}, \bibinfo {author} {\bibfnamefont {M.~V.}\ \bibnamefont
  {Nefedova}}, \bibinfo {author} {\bibfnamefont {N.~N.}\ \bibnamefont
  {Kolesnikov}}, \ and\ \bibinfo {author} {\bibfnamefont {M.~P.}\ \bibnamefont
  {Kulakov}},\ }\bibfield  {title} {\enquote {\bibinfo {title} {Effect of
  pressure on the superconducting {T$_ c$} of {MgB$_2$}},}\ }\href
  {https://linkinghub.elsevier.com/retrieve/pii/S0921453401009182} {\bibfield
  {journal} {\bibinfo  {journal} {Physica C: Supercond.}\ }\textbf {\bibinfo
  {volume} {363}},\ \bibinfo {pages} {194} (\bibinfo {year}
  {2001})}\BibitemShut {NoStop}%
\bibitem [{\citenamefont {Slusky}\ \emph {et~al.}(2001)\citenamefont {Slusky},
  \citenamefont {Rogado}, \citenamefont {Regan}, \citenamefont {Hayward},
  \citenamefont {Khalifah}, \citenamefont {He}, \citenamefont {Inumaru},
  \citenamefont {Loureiro}, \citenamefont {Haas}, \citenamefont {Zandbergen},\
  and\ \citenamefont {Cava}}]{Slusky2001}%
  \BibitemOpen
  \bibfield  {author} {\bibinfo {author} {\bibfnamefont {J.~S.}\ \bibnamefont
  {Slusky}}, \bibinfo {author} {\bibfnamefont {N.}~\bibnamefont {Rogado}},
  \bibinfo {author} {\bibfnamefont {K.~A.}\ \bibnamefont {Regan}}, \bibinfo
  {author} {\bibfnamefont {M.~A.}\ \bibnamefont {Hayward}}, \bibinfo {author}
  {\bibfnamefont {P.}~\bibnamefont {Khalifah}}, \bibinfo {author}
  {\bibfnamefont {T.}~\bibnamefont {He}}, \bibinfo {author} {\bibfnamefont
  {K.}~\bibnamefont {Inumaru}}, \bibinfo {author} {\bibfnamefont {S.~M.}\
  \bibnamefont {Loureiro}}, \bibinfo {author} {\bibfnamefont {M.~K.}\
  \bibnamefont {Haas}}, \bibinfo {author} {\bibfnamefont {H.~W.}\ \bibnamefont
  {Zandbergen}}, \ and\ \bibinfo {author} {\bibfnamefont {R.~J.}\ \bibnamefont
  {Cava}},\ }\bibfield  {title} {\enquote {\bibinfo {title} {{Loss of
  superconductivity with the addition of Al to MgB$_2$ and a structural
  transition in Mg$_{1-x}$Al$_x$B$_2$}},}\ }\href {\doibase 10.1038/35066528}
  {\bibfield  {journal} {\bibinfo  {journal} {Nature}\ }\textbf {\bibinfo
  {volume} {410}},\ \bibinfo {pages} {343--345} (\bibinfo {year}
  {2001})}\BibitemShut {NoStop}%
\bibitem [{\citenamefont {Zhao}\ \emph {et~al.}(2001)\citenamefont {Zhao},
  \citenamefont {Zhang}, \citenamefont {Qiao}, \citenamefont {Zhang},
  \citenamefont {Jia}, \citenamefont {Cao}, \citenamefont {Zhu}, \citenamefont
  {Han}, \citenamefont {Wang},\ and\ \citenamefont {Gu}}]{Zhao2001}%
  \BibitemOpen
  \bibfield  {author} {\bibinfo {author} {\bibfnamefont {Y.~G.}\ \bibnamefont
  {Zhao}}, \bibinfo {author} {\bibfnamefont {X.~P.}\ \bibnamefont {Zhang}},
  \bibinfo {author} {\bibfnamefont {P.~T.}\ \bibnamefont {Qiao}}, \bibinfo
  {author} {\bibfnamefont {H.~T.}\ \bibnamefont {Zhang}}, \bibinfo {author}
  {\bibfnamefont {S.~L.}\ \bibnamefont {Jia}}, \bibinfo {author} {\bibfnamefont
  {B.~S.}\ \bibnamefont {Cao}}, \bibinfo {author} {\bibfnamefont {M.~H.}\
  \bibnamefont {Zhu}}, \bibinfo {author} {\bibfnamefont {Z.~H.}\ \bibnamefont
  {Han}}, \bibinfo {author} {\bibfnamefont {X.~L.}\ \bibnamefont {Wang}}, \
  and\ \bibinfo {author} {\bibfnamefont {B.~L.}\ \bibnamefont {Gu}},\
  }\bibfield  {title} {\enquote {\bibinfo {title} {{Effect of Li doping on
  structure and superconducting transition temperature of
  Mg$_{1-x}$Li$_x$B$_2$}},}\ }\href {\doibase 10.1016/S0921-4534(01)00907-8}
  {\bibfield  {journal} {\bibinfo  {journal} {Physica C: Supercond.}\ }\textbf
  {\bibinfo {volume} {361}},\ \bibinfo {pages} {91--94} (\bibinfo {year}
  {2001})}\BibitemShut {NoStop}%
\bibitem [{\citenamefont {Li}\ \emph {et~al.}(2001)\citenamefont {Li},
  \citenamefont {Xiong}, \citenamefont {Mo}, \citenamefont {Fan}, \citenamefont
  {Wang}, \citenamefont {Luo}, \citenamefont {Sun}, \citenamefont {Zhang},
  \citenamefont {Li}, \citenamefont {Cao},\ and\ \citenamefont
  {Chen}}]{Li2001}%
  \BibitemOpen
  \bibfield  {author} {\bibinfo {author} {\bibfnamefont {S.~Y.}\ \bibnamefont
  {Li}}, \bibinfo {author} {\bibfnamefont {Y.~M.}\ \bibnamefont {Xiong}},
  \bibinfo {author} {\bibfnamefont {W.~Q.}\ \bibnamefont {Mo}}, \bibinfo
  {author} {\bibfnamefont {R.}~\bibnamefont {Fan}}, \bibinfo {author}
  {\bibfnamefont {C.~H.}\ \bibnamefont {Wang}}, \bibinfo {author}
  {\bibfnamefont {X.~G.}\ \bibnamefont {Luo}}, \bibinfo {author} {\bibfnamefont
  {Z.}~\bibnamefont {Sun}}, \bibinfo {author} {\bibfnamefont {H.~T.}\
  \bibnamefont {Zhang}}, \bibinfo {author} {\bibfnamefont {L.}~\bibnamefont
  {Li}}, \bibinfo {author} {\bibfnamefont {L.~Z.}\ \bibnamefont {Cao}}, \ and\
  \bibinfo {author} {\bibfnamefont {X.~H.}\ \bibnamefont {Chen}},\ }\bibfield
  {title} {\enquote {\bibinfo {title} {{Alkali metal substitution effects in
  Mg$_{1-x}$A$_x$B$_2$ (A=Li and Na) }},}\ }\href {\doibase
  10.1016/S0921-4534(01)01071-1} {\bibfield  {journal} {\bibinfo  {journal}
  {Physica C: Supercond.}\ }\textbf {\bibinfo {volume} {363}},\ \bibinfo
  {pages} {219--223} (\bibinfo {year} {2001})}\BibitemShut {NoStop}%
\bibitem [{\citenamefont {Luo}\ \emph {et~al.}(2002)\citenamefont {Luo},
  \citenamefont {Li}, \citenamefont {Luo},\ and\ \citenamefont
  {Ding}}]{Luo2002}%
  \BibitemOpen
  \bibfield  {author} {\bibinfo {author} {\bibfnamefont {H.}~\bibnamefont
  {Luo}}, \bibinfo {author} {\bibfnamefont {C.~M.}\ \bibnamefont {Li}},
  \bibinfo {author} {\bibfnamefont {H.~M.}\ \bibnamefont {Luo}}, \ and\
  \bibinfo {author} {\bibfnamefont {S.~Y.}\ \bibnamefont {Ding}},\ }\bibfield
  {title} {\enquote {\bibinfo {title} {{Study of Al doping effect on
  superconductivity of Mg$_{1-x}$Al$_x$B$_2$}},}\ }\href {\doibase
  10.1063/1.1456421} {\bibfield  {journal} {\bibinfo  {journal} {J. Appl.
  Phys.}\ }\textbf {\bibinfo {volume} {91}},\ \bibinfo {pages} {7122} (\bibinfo
  {year} {2002})}\BibitemShut {NoStop}%
\bibitem [{\citenamefont {Xiang}\ \emph {et~al.}(2003)\citenamefont {Xiang},
  \citenamefont {Zheng}, \citenamefont {Li}, \citenamefont {Li}, \citenamefont
  {Wen},\ and\ \citenamefont {Zhao}}]{Xiang2003}%
  \BibitemOpen
  \bibfield  {author} {\bibinfo {author} {\bibfnamefont {J.~Y.}\ \bibnamefont
  {Xiang}}, \bibinfo {author} {\bibfnamefont {D.~N.}\ \bibnamefont {Zheng}},
  \bibinfo {author} {\bibfnamefont {J.~Q.}\ \bibnamefont {Li}}, \bibinfo
  {author} {\bibfnamefont {S.~L.}\ \bibnamefont {Li}}, \bibinfo {author}
  {\bibfnamefont {H.~H.}\ \bibnamefont {Wen}}, \ and\ \bibinfo {author}
  {\bibfnamefont {Z.~X.}\ \bibnamefont {Zhao}},\ }\bibfield  {title} {\enquote
  {\bibinfo {title} {{Effects of Al doping on the superconducting and
  structural properties of MgB$_2$}},}\ }\href {\doibase
  10.1016/S0921-4534(02)02175-5} {\bibfield  {journal} {\bibinfo  {journal}
  {Physica C: Supercond.}\ }\textbf {\bibinfo {volume} {386}},\ \bibinfo
  {pages} {611--615} (\bibinfo {year} {2003})}\BibitemShut {NoStop}%
\bibitem [{\citenamefont {Cava}\ \emph {et~al.}(2003)\citenamefont {Cava},
  \citenamefont {Zandbergen},\ and\ \citenamefont {Inumaru}}]{Cava2003}%
  \BibitemOpen
  \bibfield  {author} {\bibinfo {author} {\bibfnamefont {R.~J.}\ \bibnamefont
  {Cava}}, \bibinfo {author} {\bibfnamefont {H.~W.}\ \bibnamefont
  {Zandbergen}}, \ and\ \bibinfo {author} {\bibfnamefont {K.}~\bibnamefont
  {Inumaru}},\ }\bibfield  {title} {\enquote {\bibinfo {title} {{The
  substitutional chemistry of MgB$_2$}},}\ }\href {\doibase
  10.1016/S0921-4534(02)02327-4} {\bibfield  {journal} {\bibinfo  {journal}
  {Physica C: Supercond.}\ }\textbf {\bibinfo {volume} {385}},\ \bibinfo
  {pages} {8--15} (\bibinfo {year} {2003})}\BibitemShut {NoStop}%
\bibitem [{\citenamefont {Gasparov}\ \emph {et~al.}(2004)\citenamefont
  {Gasparov}, \citenamefont {Kulakov}, \citenamefont {Sidorov}, \citenamefont
  {Zve\'{r}kova}, \citenamefont {Filipov}, \citenamefont {Lyashenko},\ and\
  \citenamefont {Paderno}}]{Gasparov2004}%
  \BibitemOpen
  \bibfield  {author} {\bibinfo {author} {\bibfnamefont {V.~A.}\ \bibnamefont
  {Gasparov}}, \bibinfo {author} {\bibfnamefont {M.~P.}\ \bibnamefont
  {Kulakov}}, \bibinfo {author} {\bibfnamefont {N.~S.}\ \bibnamefont
  {Sidorov}}, \bibinfo {author} {\bibfnamefont {I.~I.}\ \bibnamefont
  {Zve\'{r}kova}}, \bibinfo {author} {\bibfnamefont {V.~B.}\ \bibnamefont
  {Filipov}}, \bibinfo {author} {\bibfnamefont {A.~B.}\ \bibnamefont
  {Lyashenko}}, \ and\ \bibinfo {author} {\bibfnamefont {Y.~B.}\ \bibnamefont
  {Paderno}},\ }\bibfield  {title} {\enquote {\bibinfo {title} {On electron
  transport in {ZrB$_{12}$}, {ZrB$_2$}, and {MgB$_2$} in normal state},}\
  }\href {http://link.springer.com/10.1134/1.1825116} {\bibfield  {journal}
  {\bibinfo  {journal} {J. Exp. Theor. Phys. Lett.}\ }\textbf {\bibinfo
  {volume} {80}},\ \bibinfo {pages} {330} (\bibinfo {year} {2004})}\BibitemShut
  {NoStop}%
\bibitem [{\citenamefont {Kazakov}\ \emph {et~al.}(2005)\citenamefont
  {Kazakov}, \citenamefont {Puzniak}, \citenamefont {Rogacki}, \citenamefont
  {Mironov}, \citenamefont {Zhigadlo}, \citenamefont {Jun}, \citenamefont
  {Soltmann}, \citenamefont {Batlogg},\ and\ \citenamefont
  {Karpinski}}]{Kazakov2005}%
  \BibitemOpen
  \bibfield  {author} {\bibinfo {author} {\bibfnamefont {S.~M.}\ \bibnamefont
  {Kazakov}}, \bibinfo {author} {\bibfnamefont {R.}~\bibnamefont {Puzniak}},
  \bibinfo {author} {\bibfnamefont {K.}~\bibnamefont {Rogacki}}, \bibinfo
  {author} {\bibfnamefont {A.~V.}\ \bibnamefont {Mironov}}, \bibinfo {author}
  {\bibfnamefont {N.~D.}\ \bibnamefont {Zhigadlo}}, \bibinfo {author}
  {\bibfnamefont {J.}~\bibnamefont {Jun}}, \bibinfo {author} {\bibfnamefont
  {C.}~\bibnamefont {Soltmann}}, \bibinfo {author} {\bibfnamefont
  {B.}~\bibnamefont {Batlogg}}, \ and\ \bibinfo {author} {\bibfnamefont
  {J.}~\bibnamefont {Karpinski}},\ }\bibfield  {title} {\enquote {\bibinfo
  {title} {{Carbon substitution in MgB$_2$ single crystals: Structural and
  superconducting properties}},}\ }\href {\doibase 10.1103/PhysRevB.71.024533}
  {\bibfield  {journal} {\bibinfo  {journal} {Phys. Rev. B}\ }\textbf {\bibinfo
  {volume} {71}},\ \bibinfo {pages} {024533} (\bibinfo {year}
  {2005})}\BibitemShut {NoStop}%
\bibitem [{\citenamefont {Kolmogorov}\ and\ \citenamefont
  {Curtarolo}(2006{\natexlab{a}})}]{ak09}%
  \BibitemOpen
  \bibfield  {author} {\bibinfo {author} {\bibfnamefont {A.~N.}\ \bibnamefont
  {Kolmogorov}}\ and\ \bibinfo {author} {\bibfnamefont {S.}~\bibnamefont
  {Curtarolo}},\ }\bibfield  {title} {\enquote {\bibinfo {title} {Theoretical
  study of metal borides stability},}\ }\href
  {https://link.aps.org/doi/10.1103/PhysRevB.74.224507} {\bibfield  {journal}
  {\bibinfo  {journal} {Phys. Rev. B}\ }\textbf {\bibinfo {volume} {74}},\
  \bibinfo {pages} {224507} (\bibinfo {year} {2006}{\natexlab{a}})}\BibitemShut
  {NoStop}%
\bibitem [{\citenamefont {Monni}\ \emph {et~al.}(2006)\citenamefont {Monni},
  \citenamefont {Ferdeghini}, \citenamefont {Putti}, \citenamefont
  {Manfrinetti}, \citenamefont {Palenzona}, \citenamefont {Affronte},
  \citenamefont {Postorino}, \citenamefont {Lavagnini}, \citenamefont
  {Sacchetti}, \citenamefont {{Di Castro}}, \citenamefont {Sacchetti},
  \citenamefont {Petrillo},\ and\ \citenamefont {Orecchini}}]{Monni2006}%
  \BibitemOpen
  \bibfield  {author} {\bibinfo {author} {\bibfnamefont {M.}~\bibnamefont
  {Monni}}, \bibinfo {author} {\bibfnamefont {C.}~\bibnamefont {Ferdeghini}},
  \bibinfo {author} {\bibfnamefont {M.}~\bibnamefont {Putti}}, \bibinfo
  {author} {\bibfnamefont {P.}~\bibnamefont {Manfrinetti}}, \bibinfo {author}
  {\bibfnamefont {A.}~\bibnamefont {Palenzona}}, \bibinfo {author}
  {\bibfnamefont {M.}~\bibnamefont {Affronte}}, \bibinfo {author}
  {\bibfnamefont {P.}~\bibnamefont {Postorino}}, \bibinfo {author}
  {\bibfnamefont {M.}~\bibnamefont {Lavagnini}}, \bibinfo {author}
  {\bibfnamefont {A.}~\bibnamefont {Sacchetti}}, \bibinfo {author}
  {\bibfnamefont {D.}~\bibnamefont {{Di Castro}}}, \bibinfo {author}
  {\bibfnamefont {F.}~\bibnamefont {Sacchetti}}, \bibinfo {author}
  {\bibfnamefont {C.}~\bibnamefont {Petrillo}}, \ and\ \bibinfo {author}
  {\bibfnamefont {A.}~\bibnamefont {Orecchini}},\ }\bibfield  {title} {\enquote
  {\bibinfo {title} {{Role of charge doping and lattice distortions in codoped
  Mg$_{1-x}$(Al, Li)$_x$ B$_2$}},}\ }\href {\doibase
  10.1103/PhysRevB.73.214508} {\bibfield  {journal} {\bibinfo  {journal} {Phys.
  Rev. B}\ }\textbf {\bibinfo {volume} {73}},\ \bibinfo {pages} {214508}
  (\bibinfo {year} {2006})}\BibitemShut {NoStop}%
\bibitem [{\citenamefont {Bianconi}\ \emph {et~al.}(2007)\citenamefont
  {Bianconi}, \citenamefont {Busby}, \citenamefont {Fratini}, \citenamefont
  {Palmisano}, \citenamefont {Simonelli}, \citenamefont {Filippi},
  \citenamefont {Sanna}, \citenamefont {Congiu}, \citenamefont {Saccone},
  \citenamefont {Giovannini},\ and\ \citenamefont {{De Negri}}}]{Bianconi2007}%
  \BibitemOpen
  \bibfield  {author} {\bibinfo {author} {\bibfnamefont {A.}~\bibnamefont
  {Bianconi}}, \bibinfo {author} {\bibfnamefont {Y.}~\bibnamefont {Busby}},
  \bibinfo {author} {\bibfnamefont {M.}~\bibnamefont {Fratini}}, \bibinfo
  {author} {\bibfnamefont {V.}~\bibnamefont {Palmisano}}, \bibinfo {author}
  {\bibfnamefont {L.}~\bibnamefont {Simonelli}}, \bibinfo {author}
  {\bibfnamefont {M.}~\bibnamefont {Filippi}}, \bibinfo {author} {\bibfnamefont
  {S.}~\bibnamefont {Sanna}}, \bibinfo {author} {\bibfnamefont
  {F.}~\bibnamefont {Congiu}}, \bibinfo {author} {\bibfnamefont
  {A.}~\bibnamefont {Saccone}}, \bibinfo {author} {\bibfnamefont
  {M.}~\bibnamefont {Giovannini}}, \ and\ \bibinfo {author} {\bibfnamefont
  {S.}~\bibnamefont {{De Negri}}},\ }\bibfield  {title} {\enquote {\bibinfo
  {title} {{Controlling the Critical Temperature in Mg$_{1-x}$Al$_x$B$_2$}},}\
  }\href {http://link.springer.com/10.1007/s10948-007-0279-7} {\bibfield
  {journal} {\bibinfo  {journal} {J. Supercond. Nov. Magn.}\ }\textbf {\bibinfo
  {volume} {20}},\ \bibinfo {pages} {495--501} (\bibinfo {year}
  {2007})}\BibitemShut {NoStop}%
\bibitem [{\citenamefont {Karpinski}\ \emph {et~al.}(2008)\citenamefont
  {Karpinski}, \citenamefont {Zhigadlo}, \citenamefont {Katrych}, \citenamefont
  {Rogacki}, \citenamefont {Batlogg}, \citenamefont {Tortello},\ and\
  \citenamefont {Puzniak}}]{Karpinski2008}%
  \BibitemOpen
  \bibfield  {author} {\bibinfo {author} {\bibfnamefont {J.}~\bibnamefont
  {Karpinski}}, \bibinfo {author} {\bibfnamefont {N.~D.}\ \bibnamefont
  {Zhigadlo}}, \bibinfo {author} {\bibfnamefont {S.}~\bibnamefont {Katrych}},
  \bibinfo {author} {\bibfnamefont {K.}~\bibnamefont {Rogacki}}, \bibinfo
  {author} {\bibfnamefont {B.}~\bibnamefont {Batlogg}}, \bibinfo {author}
  {\bibfnamefont {M.}~\bibnamefont {Tortello}}, \ and\ \bibinfo {author}
  {\bibfnamefont {R.}~\bibnamefont {Puzniak}},\ }\bibfield  {title} {\enquote
  {\bibinfo {title} {{MgB$_2$ single crystals substituted with Li and with
  Li-C: Structural and superconducting properties}},}\ }\href {\doibase
  10.1103/PhysRevB.77.214507} {\bibfield  {journal} {\bibinfo  {journal} {Phys.
  Rev. B}\ }\textbf {\bibinfo {volume} {77}},\ \bibinfo {pages} {214507}
  (\bibinfo {year} {2008})}\BibitemShut {NoStop}%
\bibitem [{\citenamefont {Shein}\ and\ \citenamefont
  {Ivanovskii}(2008)}]{Shein2008}%
  \BibitemOpen
  \bibfield  {author} {\bibinfo {author} {\bibfnamefont {I.~R.}\ \bibnamefont
  {Shein}}\ and\ \bibinfo {author} {\bibfnamefont {A.~L.}\ \bibnamefont
  {Ivanovskii}},\ }\bibfield  {title} {\enquote {\bibinfo {title} {Elastic
  properties of mono-and polycrystalline hexagonal {AlB$_2$}-like diborides of
  s, p and d metals from first-principles calculations},}\ }\href
  {https://iopscience.iop.org/article/10.1088/0953-8984/20/41/415218}
  {\bibfield  {journal} {\bibinfo  {journal} {J. Phys: Condens. Matter}\
  }\textbf {\bibinfo {volume} {20}},\ \bibinfo {pages} {415218} (\bibinfo
  {year} {2008})}\BibitemShut {NoStop}%
\bibitem [{\citenamefont {Parisiades}\ \emph {et~al.}(2009)\citenamefont
  {Parisiades}, \citenamefont {Liarokapis}, \citenamefont {Zhigadlo},
  \citenamefont {Katrych},\ and\ \citenamefont {Karpinski}}]{Parisiades2009}%
  \BibitemOpen
  \bibfield  {author} {\bibinfo {author} {\bibfnamefont {P.}~\bibnamefont
  {Parisiades}}, \bibinfo {author} {\bibfnamefont {E.}~\bibnamefont
  {Liarokapis}}, \bibinfo {author} {\bibfnamefont {N.~D.}\ \bibnamefont
  {Zhigadlo}}, \bibinfo {author} {\bibfnamefont {S.}~\bibnamefont {Katrych}}, \
  and\ \bibinfo {author} {\bibfnamefont {J.}~\bibnamefont {Karpinski}},\
  }\bibfield  {title} {\enquote {\bibinfo {title} {{Raman Investigations of C-,
  Li- and Mn-Doped MgB$_2$}},}\ }\href {\doibase 10.1007/s10948-008-0402-4}
  {\bibfield  {journal} {\bibinfo  {journal} {J. Supercond. Nov. Magn.}\
  }\textbf {\bibinfo {volume} {22}},\ \bibinfo {pages} {169--172} (\bibinfo
  {year} {2009})}\BibitemShut {NoStop}%
\bibitem [{\citenamefont {Daghero}\ \emph {et~al.}(2009)\citenamefont
  {Daghero}, \citenamefont {Ummarino}, \citenamefont {Tortello}, \citenamefont
  {Delaude}, \citenamefont {Gonnelli}, \citenamefont {Stepanov}, \citenamefont
  {Monni},\ and\ \citenamefont {Palenzona}}]{Daghero2009}%
  \BibitemOpen
  \bibfield  {author} {\bibinfo {author} {\bibfnamefont {D.}~\bibnamefont
  {Daghero}}, \bibinfo {author} {\bibfnamefont {G.~A.}\ \bibnamefont
  {Ummarino}}, \bibinfo {author} {\bibfnamefont {M.}~\bibnamefont {Tortello}},
  \bibinfo {author} {\bibfnamefont {D.}~\bibnamefont {Delaude}}, \bibinfo
  {author} {\bibfnamefont {R.~S.}\ \bibnamefont {Gonnelli}}, \bibinfo {author}
  {\bibfnamefont {V.~A.}\ \bibnamefont {Stepanov}}, \bibinfo {author}
  {\bibfnamefont {M.}~\bibnamefont {Monni}}, \ and\ \bibinfo {author}
  {\bibfnamefont {A.}~\bibnamefont {Palenzona}},\ }\bibfield  {title} {\enquote
  {\bibinfo {title} {{Effect of Li-Al co-doping on the energy gaps of
  MgB$_2$}},}\ }\href {\doibase 10.1088/0953-2048/22/2/025012} {\bibfield
  {journal} {\bibinfo  {journal} {Supercond. Sci. Technol.}\ }\textbf {\bibinfo
  {volume} {22}},\ \bibinfo {pages} {025012} (\bibinfo {year}
  {2009})}\BibitemShut {NoStop}%
\bibitem [{\citenamefont {Barbero}\ \emph {et~al.}(2017)\citenamefont
  {Barbero}, \citenamefont {Shiroka}, \citenamefont {Delley}, \citenamefont
  {Grant}, \citenamefont {Machado}, \citenamefont {Fisk}, \citenamefont {Ott},\
  and\ \citenamefont {Mesot}}]{Barbero2017}%
  \BibitemOpen
  \bibfield  {author} {\bibinfo {author} {\bibfnamefont {N.}~\bibnamefont
  {Barbero}}, \bibinfo {author} {\bibfnamefont {T.}~\bibnamefont {Shiroka}},
  \bibinfo {author} {\bibfnamefont {B.}~\bibnamefont {Delley}}, \bibinfo
  {author} {\bibfnamefont {T.}~\bibnamefont {Grant}}, \bibinfo {author}
  {\bibfnamefont {A.~J.~S.}\ \bibnamefont {Machado}}, \bibinfo {author}
  {\bibfnamefont {Z.}~\bibnamefont {Fisk}}, \bibinfo {author} {\bibfnamefont
  {H.-R.}\ \bibnamefont {Ott}}, \ and\ \bibinfo {author} {\bibfnamefont
  {J.}~\bibnamefont {Mesot}},\ }\bibfield  {title} {\enquote {\bibinfo {title}
  {{Doping-induced superconductivity of ZrB$_2$ and HfB$_2$}},}\ }\href
  {\doibase 10.1103/PhysRevB.95.094505} {\bibfield  {journal} {\bibinfo
  {journal} {Phys. Rev. B}\ }\textbf {\bibinfo {volume} {95}},\ \bibinfo
  {pages} {094505} (\bibinfo {year} {2017})}\BibitemShut {NoStop}%
\bibitem [{\citenamefont {Aydin}\ and\ \citenamefont
  {Şimşek}(2018)}]{Aydin2018}%
  \BibitemOpen
  \bibfield  {author} {\bibinfo {author} {\bibfnamefont {S.}~\bibnamefont
  {Aydin}}\ and\ \bibinfo {author} {\bibfnamefont {M.}~\bibnamefont
  {Şimşek}},\ }\bibfield  {title} {\enquote {\bibinfo {title} {{Stability and
  superconductivity properties of metal substituted aluminum diborides
  (M$_{0.5}$Al$_{0.5}$B$_2$)}},}\ }\href {\doibase
  10.1016/j.commatsci.2018.08.005} {\bibfield  {journal} {\bibinfo  {journal}
  {Comp. Mater. Sci.}\ }\textbf {\bibinfo {volume} {154}},\ \bibinfo {pages}
  {234--242} (\bibinfo {year} {2018})}\BibitemShut {NoStop}%
\bibitem [{\citenamefont {Pei}\ \emph {et~al.}(2021)\citenamefont {Pei},
  \citenamefont {Zhang}, \citenamefont {Wang}, \citenamefont {Zhao},
  \citenamefont {Gao}, \citenamefont {Gong}, \citenamefont {Tian},
  \citenamefont {Luo}, \citenamefont {Lu}, \citenamefont {Lei}, \citenamefont
  {Liu},\ and\ \citenamefont {Qi}}]{Pei2021}%
  \BibitemOpen
  \bibfield  {author} {\bibinfo {author} {\bibfnamefont {C.}~\bibnamefont
  {Pei}}, \bibinfo {author} {\bibfnamefont {J.}~\bibnamefont {Zhang}}, \bibinfo
  {author} {\bibfnamefont {Q.}~\bibnamefont {Wang}}, \bibinfo {author}
  {\bibfnamefont {Y.}~\bibnamefont {Zhao}}, \bibinfo {author} {\bibfnamefont
  {L.}~\bibnamefont {Gao}}, \bibinfo {author} {\bibfnamefont {C.}~\bibnamefont
  {Gong}}, \bibinfo {author} {\bibfnamefont {S.}~\bibnamefont {Tian}}, \bibinfo
  {author} {\bibfnamefont {R.}~\bibnamefont {Luo}}, \bibinfo {author}
  {\bibfnamefont {Z.-y}\ \bibnamefont {Lu}}, \bibinfo {author} {\bibfnamefont
  {H.}~\bibnamefont {Lei}}, \bibinfo {author} {\bibfnamefont {K.}~\bibnamefont
  {Liu}}, \ and\ \bibinfo {author} {\bibfnamefont {Y.}~\bibnamefont {Qi}},\
  }\bibfield  {title} {\enquote {\bibinfo {title} {{Pressure-induced
  Superconductivity at 32 K in MoB$_2$}},}\ }\href
  {http://arxiv.org/abs/2105.13250} {\  (\bibinfo {year} {2021})},\ \Eprint
  {http://arxiv.org/abs/2105.13250} {arXiv:2105.13250} \BibitemShut {NoStop}%
\bibitem [{\citenamefont {Yang}\ \emph {et~al.}(2022)\citenamefont {Yang},
  \citenamefont {Xiao}, \citenamefont {Zhu}, \citenamefont {Cui}, \citenamefont
  {Song}, \citenamefont {Cao},\ and\ \citenamefont {Ren}}]{Yang2022}%
  \BibitemOpen
  \bibfield  {author} {\bibinfo {author} {\bibfnamefont {W.}~\bibnamefont
  {Yang}}, \bibinfo {author} {\bibfnamefont {G.}~\bibnamefont {Xiao}}, \bibinfo
  {author} {\bibfnamefont {Q.}~\bibnamefont {Zhu}}, \bibinfo {author}
  {\bibfnamefont {Y.}~\bibnamefont {Cui}}, \bibinfo {author} {\bibfnamefont
  {S.}~\bibnamefont {Song}}, \bibinfo {author} {\bibfnamefont {G.-H.}\
  \bibnamefont {Cao}}, \ and\ \bibinfo {author} {\bibfnamefont
  {Z.}~\bibnamefont {Ren}},\ }\bibfield  {title} {\enquote {\bibinfo {title}
  {{Stabilization and superconductivity of AlB$_2$-type nonstoichiometric
  molybdenum diboride by Sc doping}},}\ }\href
  {https://doi.org/10.1016/j.ceramint.2022.03.272} {\bibfield  {journal}
  {\bibinfo  {journal} {Ceram. Int.}\ }\textbf {\bibinfo {volume} {48}},\
  \bibinfo {pages} {19971--19977} (\bibinfo {year} {2022})}\BibitemShut
  {NoStop}%
\bibitem [{\citenamefont {Kolmogorov}\ and\ \citenamefont
  {Curtarolo}(2006{\natexlab{b}})}]{ak08}%
  \BibitemOpen
  \bibfield  {author} {\bibinfo {author} {\bibfnamefont {A.~N.}\ \bibnamefont
  {Kolmogorov}}\ and\ \bibinfo {author} {\bibfnamefont {S.}~\bibnamefont
  {Curtarolo}},\ }\bibfield  {title} {\enquote {\bibinfo {title} {{Prediction
  of different crystal structure phases in metal borides: A lithium monoboride
  analog to MgB$_2$}},}\ }\href
  {https://link.aps.org/doi/10.1103/PhysRevB.73.180501} {\bibfield  {journal}
  {\bibinfo  {journal} {Phys. Rev. B}\ }\textbf {\bibinfo {volume} {73}},\
  \bibinfo {pages} {180501} (\bibinfo {year} {2006}{\natexlab{b}})}\BibitemShut
  {NoStop}%
\bibitem [{\citenamefont {Kolmogorov}\ \emph {et~al.}(2015)\citenamefont
  {Kolmogorov}, \citenamefont {Hajinazar}, \citenamefont {Angyal},
  \citenamefont {Kuznetsov},\ and\ \citenamefont {Jephcoat}}]{Kolmogorov2015}%
  \BibitemOpen
  \bibfield  {author} {\bibinfo {author} {\bibfnamefont {A.~N.}\ \bibnamefont
  {Kolmogorov}}, \bibinfo {author} {\bibfnamefont {S.}~\bibnamefont
  {Hajinazar}}, \bibinfo {author} {\bibfnamefont {C.}~\bibnamefont {Angyal}},
  \bibinfo {author} {\bibfnamefont {V.~L.}\ \bibnamefont {Kuznetsov}}, \ and\
  \bibinfo {author} {\bibfnamefont {A.~P.}\ \bibnamefont {Jephcoat}},\
  }\bibfield  {title} {\enquote {\bibinfo {title} {Synthesis of a predicted
  layered {LiB} via cold compression},}\ }\href
  {https://link.aps.org/doi/10.1103/PhysRevB.92.144110} {\bibfield  {journal}
  {\bibinfo  {journal} {Phys. Rev. B}\ }\textbf {\bibinfo {volume} {92}},\
  \bibinfo {pages} {144110} (\bibinfo {year} {2015})}\BibitemShut {NoStop}%
\bibitem [{\citenamefont {Calandra}\ and\ \citenamefont
  {Mauri}(2005)}]{Calandra2005}%
  \BibitemOpen
  \bibfield  {author} {\bibinfo {author} {\bibfnamefont {M.}~\bibnamefont
  {Calandra}}\ and\ \bibinfo {author} {\bibfnamefont {F.}~\bibnamefont
  {Mauri}},\ }\bibfield  {title} {\enquote {\bibinfo {title} {{Theoretical
  explanation of superconductivity in C$_6$Ca}},}\ }\href
  {https://doi.org/10.48550/arXiv.cond-mat/0506082} {\bibfield  {journal}
  {\bibinfo  {journal} {Phys. Rev. Lett.}\ }\textbf {\bibinfo {volume} {95}},\
  \bibinfo {pages} {237002} (\bibinfo {year} {2005})}\BibitemShut {NoStop}%
\bibitem [{\citenamefont {Calandra}\ \emph {et~al.}(2007)\citenamefont
  {Calandra}, \citenamefont {Kolmogorov},\ and\ \citenamefont
  {Curtarolo}}]{Calandra2007}%
  \BibitemOpen
  \bibfield  {author} {\bibinfo {author} {\bibfnamefont {M.}~\bibnamefont
  {Calandra}}, \bibinfo {author} {\bibfnamefont {A.~N.}\ \bibnamefont
  {Kolmogorov}}, \ and\ \bibinfo {author} {\bibfnamefont {S.}~\bibnamefont
  {Curtarolo}},\ }\bibfield  {title} {\enquote {\bibinfo {title} {Search for
  high {T$_ c$} in layered structures: {The} case of {LiB}},}\ }\href
  {https://link.aps.org/doi/10.1103/PhysRevB.75.144506} {\bibfield  {journal}
  {\bibinfo  {journal} {Phys. Rev. B}\ }\textbf {\bibinfo {volume} {75}},\
  \bibinfo {pages} {144506} (\bibinfo {year} {2007})}\BibitemShut {NoStop}%
\bibitem [{\citenamefont {Liu}\ and\ \citenamefont {Mazin}(2007)}]{Liu2007}%
  \BibitemOpen
  \bibfield  {author} {\bibinfo {author} {\bibfnamefont {A.~Y.}\ \bibnamefont
  {Liu}}\ and\ \bibinfo {author} {\bibfnamefont {I.~I.}\ \bibnamefont
  {Mazin}},\ }\bibfield  {title} {\enquote {\bibinfo {title} {Combining the
  advantages of superconducting {MgB$_2$} and {CaC$_6$} in one material:
  {Suggestions} from first-principles calculations},}\ }\href
  {https://link.aps.org/doi/10.1103/PhysRevB.75.064510} {\bibfield  {journal}
  {\bibinfo  {journal} {Phys. Rev. B}\ }\textbf {\bibinfo {volume} {75}},\
  \bibinfo {pages} {064510} (\bibinfo {year} {2007})}\BibitemShut {NoStop}%
\bibitem [{\citenamefont {Mart{\'\i}nez-Guerra}\ \emph
  {et~al.}(2014)\citenamefont {Mart{\'\i}nez-Guerra}, \citenamefont
  {Ort{\'\i}z-Chi}, \citenamefont {Curtarolo},\ and\ \citenamefont
  {de~Coss}}]{Martinez2014}%
  \BibitemOpen
  \bibfield  {author} {\bibinfo {author} {\bibfnamefont {E.}~\bibnamefont
  {Mart{\'\i}nez-Guerra}}, \bibinfo {author} {\bibfnamefont {F.}~\bibnamefont
  {Ort{\'\i}z-Chi}}, \bibinfo {author} {\bibfnamefont {S.}~\bibnamefont
  {Curtarolo}}, \ and\ \bibinfo {author} {\bibfnamefont {R.}~\bibnamefont
  {de~Coss}},\ }\bibfield  {title} {\enquote {\bibinfo {title} {Pressure
  effects on the electronic structure and superconducting critical temperature
  of {Li$_2$B$_2$}},}\ }\href
  {https://iopscience.iop.org/article/10.1088/0953-8984/26/11/115701}
  {\bibfield  {journal} {\bibinfo  {journal} {J. Phys: Condens. Matter}\
  }\textbf {\bibinfo {volume} {26}},\ \bibinfo {pages} {115701} (\bibinfo
  {year} {2014})}\BibitemShut {NoStop}%
\bibitem [{\citenamefont {Klime{\v{s}}}\ \emph {et~al.}(2011)\citenamefont
  {Klime{\v{s}}}, \citenamefont {Bowler},\ and\ \citenamefont
  {Michaelides}}]{optB86b}%
  \BibitemOpen
  \bibfield  {author} {\bibinfo {author} {\bibfnamefont {J.}~\bibnamefont
  {Klime{\v{s}}}}, \bibinfo {author} {\bibfnamefont {D.~R.}\ \bibnamefont
  {Bowler}}, \ and\ \bibinfo {author} {\bibfnamefont {A.}~\bibnamefont
  {Michaelides}},\ }\bibfield  {title} {\enquote {\bibinfo {title} {Van der
  {Waals} density functionals applied to solids},}\ }\href
  {https://journals.aps.org/prb/abstract/10.1103/PhysRevB.83.195131} {\bibfield
   {journal} {\bibinfo  {journal} {Phys. Rev. B}\ }\textbf {\bibinfo {volume}
  {83}},\ \bibinfo {pages} {195131} (\bibinfo {year} {2011})}\BibitemShut
  {NoStop}%
\bibitem [{\citenamefont {Thonhauser}\ \emph {et~al.}(2007)\citenamefont
  {Thonhauser}, \citenamefont {Cooper}, \citenamefont {Li}, \citenamefont
  {Puzder}, \citenamefont {Hyldgaard},\ and\ \citenamefont
  {Langreth}}]{Thonhauser2007}%
  \BibitemOpen
  \bibfield  {author} {\bibinfo {author} {\bibfnamefont {T.}~\bibnamefont
  {Thonhauser}}, \bibinfo {author} {\bibfnamefont {V.~R.}\ \bibnamefont
  {Cooper}}, \bibinfo {author} {\bibfnamefont {S.}~\bibnamefont {Li}}, \bibinfo
  {author} {\bibfnamefont {A.}~\bibnamefont {Puzder}}, \bibinfo {author}
  {\bibfnamefont {P.}~\bibnamefont {Hyldgaard}}, \ and\ \bibinfo {author}
  {\bibfnamefont {D.~C.}\ \bibnamefont {Langreth}},\ }\bibfield  {title}
  {\enquote {\bibinfo {title} {Van der {Waals} density functional:
  {Self}-consistent potential and the nature of the van der {Waals} bond},}\
  }\href {https://journals.aps.org/prb/abstract/10.1103/PhysRevB.76.125112}
  {\bibfield  {journal} {\bibinfo  {journal} {Phys. Rev. B}\ }\textbf {\bibinfo
  {volume} {76}},\ \bibinfo {pages} {125112} (\bibinfo {year}
  {2007})}\BibitemShut {NoStop}%
\bibitem [{\citenamefont {Thonhauser}\ \emph {et~al.}(2015)\citenamefont
  {Thonhauser}, \citenamefont {Zuluaga}, \citenamefont {Arter}, \citenamefont
  {Berland}, \citenamefont {Schr{\"o}der},\ and\ \citenamefont
  {Hyldgaard}}]{Thonhauser2015}%
  \BibitemOpen
  \bibfield  {author} {\bibinfo {author} {\bibfnamefont {T.}~\bibnamefont
  {Thonhauser}}, \bibinfo {author} {\bibfnamefont {S.}~\bibnamefont {Zuluaga}},
  \bibinfo {author} {\bibfnamefont {C.~A.}\ \bibnamefont {Arter}}, \bibinfo
  {author} {\bibfnamefont {K.}~\bibnamefont {Berland}}, \bibinfo {author}
  {\bibfnamefont {E.}~\bibnamefont {Schr{\"o}der}}, \ and\ \bibinfo {author}
  {\bibfnamefont {P.}~\bibnamefont {Hyldgaard}},\ }\bibfield  {title} {\enquote
  {\bibinfo {title} {Spin signature of nonlocal correlation binding in
  metal-organic frameworks},}\ }\href
  {https://journals.aps.org/prl/abstract/10.1103/PhysRevLett.115.136402}
  {\bibfield  {journal} {\bibinfo  {journal} {Phys. Rev. Lett.}\ }\textbf
  {\bibinfo {volume} {115}},\ \bibinfo {pages} {136402} (\bibinfo {year}
  {2015})}\BibitemShut {NoStop}%
\bibitem [{\citenamefont {Berland}\ \emph {et~al.}(2015)\citenamefont
  {Berland}, \citenamefont {Cooper}, \citenamefont {Lee}, \citenamefont
  {Schr{\"o}der}, \citenamefont {Thonhauser}, \citenamefont {Hyldgaard},\ and\
  \citenamefont {Lundqvist}}]{Berland2015}%
  \BibitemOpen
  \bibfield  {author} {\bibinfo {author} {\bibfnamefont {K.}~\bibnamefont
  {Berland}}, \bibinfo {author} {\bibfnamefont {V.~R.}\ \bibnamefont {Cooper}},
  \bibinfo {author} {\bibfnamefont {K.}~\bibnamefont {Lee}}, \bibinfo {author}
  {\bibfnamefont {E.}~\bibnamefont {Schr{\"o}der}}, \bibinfo {author}
  {\bibfnamefont {T.}~\bibnamefont {Thonhauser}}, \bibinfo {author}
  {\bibfnamefont {P.}~\bibnamefont {Hyldgaard}}, \ and\ \bibinfo {author}
  {\bibfnamefont {B.~I.}\ \bibnamefont {Lundqvist}},\ }\bibfield  {title}
  {\enquote {\bibinfo {title} {van der {Waals} forces in density functional
  theory: a review of the {vdW-DF} method},}\ }\href
  {https://iopscience.iop.org/article/10.1088/0034-4885/78/6/066501/meta}
  {\bibfield  {journal} {\bibinfo  {journal} {Rep. Prog. Phys.}\ }\textbf
  {\bibinfo {volume} {78}},\ \bibinfo {pages} {066501} (\bibinfo {year}
  {2015})}\BibitemShut {NoStop}%
\bibitem [{\citenamefont {Langreth}\ \emph {et~al.}(2009)\citenamefont
  {Langreth}, \citenamefont {Lundqvist}, \citenamefont {Chakarova-K{\"a}ck},
  \citenamefont {Cooper}, \citenamefont {Dion}, \citenamefont {Hyldgaard},
  \citenamefont {Kelkkanen}, \citenamefont {Kleis}, \citenamefont {Kong},\ and\
  \citenamefont {Li~$et. al$}}]{Langreth2009}%
  \BibitemOpen
  \bibfield  {author} {\bibinfo {author} {\bibfnamefont {D.~C.}\ \bibnamefont
  {Langreth}}, \bibinfo {author} {\bibfnamefont {B.~I.}\ \bibnamefont
  {Lundqvist}}, \bibinfo {author} {\bibfnamefont {S.~D.}\ \bibnamefont
  {Chakarova-K{\"a}ck}}, \bibinfo {author} {\bibfnamefont {V.~R.}\ \bibnamefont
  {Cooper}}, \bibinfo {author} {\bibfnamefont {M.}~\bibnamefont {Dion}},
  \bibinfo {author} {\bibfnamefont {P.}~\bibnamefont {Hyldgaard}}, \bibinfo
  {author} {\bibfnamefont {A.}~\bibnamefont {Kelkkanen}}, \bibinfo {author}
  {\bibfnamefont {J.}~\bibnamefont {Kleis}}, \bibinfo {author} {\bibfnamefont
  {L.}~\bibnamefont {Kong}}, \ and\ \bibinfo {author} {\bibfnamefont
  {S.}~\bibnamefont {Li~$et. al$}},\ }\bibfield  {title} {\enquote {\bibinfo
  {title} {A density functional for sparse matter},}\ }\href
  {https://iopscience.iop.org/article/10.1088/0953-8984/21/8/084203/meta}
  {\bibfield  {journal} {\bibinfo  {journal} {J. Phys: Condens. Matter}\
  }\textbf {\bibinfo {volume} {21}},\ \bibinfo {pages} {084203} (\bibinfo
  {year} {2009})}\BibitemShut {NoStop}%
\bibitem [{\citenamefont {Sabatini}\ \emph {et~al.}(2012)\citenamefont
  {Sabatini}, \citenamefont {K{\"u}{\c{c}}{\"u}kbenli}, \citenamefont {Kolb},
  \citenamefont {Thonhauser},\ and\ \citenamefont
  {De~Gironcoli}}]{Sabatini2012}%
  \BibitemOpen
  \bibfield  {author} {\bibinfo {author} {\bibfnamefont {R.}~\bibnamefont
  {Sabatini}}, \bibinfo {author} {\bibfnamefont {E.}~\bibnamefont
  {K{\"u}{\c{c}}{\"u}kbenli}}, \bibinfo {author} {\bibfnamefont
  {B.}~\bibnamefont {Kolb}}, \bibinfo {author} {\bibfnamefont {T.}~\bibnamefont
  {Thonhauser}}, \ and\ \bibinfo {author} {\bibfnamefont {S.}~\bibnamefont
  {De~Gironcoli}},\ }\bibfield  {title} {\enquote {\bibinfo {title} {Structural
  evolution of amino acid crystals under stress from a non-empirical density
  functional},}\ }\href
  {https://iopscience.iop.org/article/10.1088/0953-8984/24/42/424209/meta}
  {\bibfield  {journal} {\bibinfo  {journal} {J. Phys: Condens. Matter}\
  }\textbf {\bibinfo {volume} {24}},\ \bibinfo {pages} {424209} (\bibinfo
  {year} {2012})}\BibitemShut {NoStop}%
\bibitem [{\citenamefont {Kresse}\ and\ \citenamefont
  {Furthm{\"u}ller}(1996)}]{Kresse1996}%
  \BibitemOpen
  \bibfield  {author} {\bibinfo {author} {\bibfnamefont {G.}~\bibnamefont
  {Kresse}}\ and\ \bibinfo {author} {\bibfnamefont {J.}~\bibnamefont
  {Furthm{\"u}ller}},\ }\bibfield  {title} {\enquote {\bibinfo {title}
  {{Efficient iterative schemes for $ab~initio$ total-energy calculations using
  a plane-wave basis set}},}\ }\href
  {https://doi.org/10.1103/PhysRevB.54.11169} {\bibfield  {journal} {\bibinfo
  {journal} {Phys. Rev. B}\ }\textbf {\bibinfo {volume} {54}},\ \bibinfo
  {pages} {11169} (\bibinfo {year} {1996})}\BibitemShut {NoStop}%
\bibitem [{\citenamefont {Bl{\"o}chl}(1994)}]{Blochl1994}%
  \BibitemOpen
  \bibfield  {author} {\bibinfo {author} {\bibfnamefont {P.~E.}\ \bibnamefont
  {Bl{\"o}chl}},\ }\bibfield  {title} {\enquote {\bibinfo {title} {Projector
  augmented-wave method},}\ }\href {https://doi.org/10.1103/PhysRevB.50.17953}
  {\bibfield  {journal} {\bibinfo  {journal} {Phys. Rev. B}\ }\textbf {\bibinfo
  {volume} {50}},\ \bibinfo {pages} {17953} (\bibinfo {year}
  {1994})}\BibitemShut {NoStop}%
\bibitem [{\citenamefont {Togo}\ and\ \citenamefont {Tanaka}(2015)}]{Togo2015}%
  \BibitemOpen
  \bibfield  {author} {\bibinfo {author} {\bibfnamefont {A.}~\bibnamefont
  {Togo}}\ and\ \bibinfo {author} {\bibfnamefont {I.}~\bibnamefont {Tanaka}},\
  }\bibfield  {title} {\enquote {\bibinfo {title} {{First principles phonon
  calculations in materials science}},}\ }\href
  {https://doi.org/10.1016/j.scriptamat.2015.07.021} {\bibfield  {journal}
  {\bibinfo  {journal} {Scr. Mater.}\ }\textbf {\bibinfo {volume} {108}},\
  \bibinfo {pages} {1--5} (\bibinfo {year} {2015})}\BibitemShut {NoStop}%
\bibitem [{\citenamefont {Giannozzi}\ \emph {et~al.}(2017)\citenamefont
  {Giannozzi}, \citenamefont {Andreussi}, \citenamefont {Brumme}, \citenamefont
  {Bunau}, \citenamefont {Nardelli}, \citenamefont {Calandra}, \citenamefont
  {Car}, \citenamefont {Cavazzoni}, \citenamefont {Ceresoli},\ and\
  \citenamefont {Cococcioni~$et. al$}}]{QE}%
  \BibitemOpen
  \bibfield  {author} {\bibinfo {author} {\bibfnamefont {P.}~\bibnamefont
  {Giannozzi}}, \bibinfo {author} {\bibfnamefont {O.}~\bibnamefont
  {Andreussi}}, \bibinfo {author} {\bibfnamefont {T.}~\bibnamefont {Brumme}},
  \bibinfo {author} {\bibfnamefont {O.}~\bibnamefont {Bunau}}, \bibinfo
  {author} {\bibfnamefont {M.~B.}\ \bibnamefont {Nardelli}}, \bibinfo {author}
  {\bibfnamefont {M.}~\bibnamefont {Calandra}}, \bibinfo {author}
  {\bibfnamefont {R.}~\bibnamefont {Car}}, \bibinfo {author} {\bibfnamefont
  {C.}~\bibnamefont {Cavazzoni}}, \bibinfo {author} {\bibfnamefont
  {D.}~\bibnamefont {Ceresoli}}, \ and\ \bibinfo {author} {\bibfnamefont
  {M.}~\bibnamefont {Cococcioni~$et. al$}},\ }\bibfield  {title} {\enquote
  {\bibinfo {title} {Advanced capabilities for materials modelling with
  {Quantum} {ESPRESSO}},}\ }\href
  {https://iopscience.iop.org/article/10.1088/1361-648X/aa8f79/meta} {\bibfield
   {journal} {\bibinfo  {journal} {J. Phys: Condens. Matter}\ }\textbf
  {\bibinfo {volume} {29}},\ \bibinfo {pages} {465901} (\bibinfo {year}
  {2017})}\BibitemShut {NoStop}%
\bibitem [{\citenamefont {Perdew}\ \emph {et~al.}(1996)\citenamefont {Perdew},
  \citenamefont {Burke},\ and\ \citenamefont {Ernzerhof}}]{PBE}%
  \BibitemOpen
  \bibfield  {author} {\bibinfo {author} {\bibfnamefont {J.~P.}\ \bibnamefont
  {Perdew}}, \bibinfo {author} {\bibfnamefont {K.}~\bibnamefont {Burke}}, \
  and\ \bibinfo {author} {\bibfnamefont {M.}~\bibnamefont {Ernzerhof}},\
  }\bibfield  {title} {\enquote {\bibinfo {title} {Generalized gradient
  approximation made simple},}\ }\href
  {https://journals.aps.org/prl/abstract/10.1103/PhysRevLett.77.3865}
  {\bibfield  {journal} {\bibinfo  {journal} {Phys. Rev. Lett.}\ }\textbf
  {\bibinfo {volume} {77}},\ \bibinfo {pages} {3865} (\bibinfo {year}
  {1996})}\BibitemShut {NoStop}%
\bibitem [{\citenamefont {van Setten}\ \emph {et~al.}(2018)\citenamefont {van
  Setten}, \citenamefont {Giantomassi}, \citenamefont {Bousquet}, \citenamefont
  {Verstraete}, \citenamefont {Hamann}, \citenamefont {Gonze},\ and\
  \citenamefont {Rignanese}}]{Dojo2018}%
  \BibitemOpen
  \bibfield  {author} {\bibinfo {author} {\bibfnamefont {M.~J.}\ \bibnamefont
  {van Setten}}, \bibinfo {author} {\bibfnamefont {M.}~\bibnamefont
  {Giantomassi}}, \bibinfo {author} {\bibfnamefont {E.}~\bibnamefont
  {Bousquet}}, \bibinfo {author} {\bibfnamefont {M.~J.}\ \bibnamefont
  {Verstraete}}, \bibinfo {author} {\bibfnamefont {D.~R.}\ \bibnamefont
  {Hamann}}, \bibinfo {author} {\bibfnamefont {X.}~\bibnamefont {Gonze}}, \
  and\ \bibinfo {author} {\bibfnamefont {G.-M.}\ \bibnamefont {Rignanese}},\
  }\bibfield  {title} {\enquote {\bibinfo {title} {The {PseudoDojo}: {Training}
  and grading a 85 element optimized norm-conserving pseudopotential table},}\
  }\href {https://doi.org/10.1016/j.cpc.2018.01.012} {\bibfield  {journal}
  {\bibinfo  {journal} {Comput. Phys. Commun.}\ }\textbf {\bibinfo {volume}
  {226}},\ \bibinfo {pages} {39} (\bibinfo {year} {2018})}\BibitemShut
  {NoStop}%
\bibitem [{\citenamefont {Methfessel}\ and\ \citenamefont
  {Paxton}(1989)}]{Methfessel1989}%
  \BibitemOpen
  \bibfield  {author} {\bibinfo {author} {\bibfnamefont {M.}~\bibnamefont
  {Methfessel}}\ and\ \bibinfo {author} {\bibfnamefont {A.~T.}\ \bibnamefont
  {Paxton}},\ }\bibfield  {title} {\enquote {\bibinfo {title} {High-precision
  sampling for {Brillouin}-zone integration in metals},}\ }\href
  {https://journals.aps.org/prb/abstract/10.1103/PhysRevB.40.3616} {\bibfield
  {journal} {\bibinfo  {journal} {Phys. Rev. B}\ }\textbf {\bibinfo {volume}
  {40}},\ \bibinfo {pages} {3616} (\bibinfo {year} {1989})}\BibitemShut
  {NoStop}%
\bibitem [{\citenamefont {Monkhorst}\ and\ \citenamefont
  {Pack}(1976)}]{Monkhorst1976}%
  \BibitemOpen
  \bibfield  {author} {\bibinfo {author} {\bibfnamefont {H.~J.}\ \bibnamefont
  {Monkhorst}}\ and\ \bibinfo {author} {\bibfnamefont {J.~D.}\ \bibnamefont
  {Pack}},\ }\bibfield  {title} {\enquote {\bibinfo {title} {Special points for
  {Brillouin}-zone integrations},}\ }\href
  {https://journals.aps.org/prb/abstract/10.1103/PhysRevB.13.5188} {\bibfield
  {journal} {\bibinfo  {journal} {Phys. Rev. B}\ }\textbf {\bibinfo {volume}
  {13}},\ \bibinfo {pages} {5188} (\bibinfo {year} {1976})}\BibitemShut
  {NoStop}%
\bibitem [{\citenamefont {Baroni}\ \emph {et~al.}(2001)\citenamefont {Baroni},
  \citenamefont {De~Gironcoli}, \citenamefont {Dal~Corso},\ and\ \citenamefont
  {Giannozzi}}]{Baroni2001}%
  \BibitemOpen
  \bibfield  {author} {\bibinfo {author} {\bibfnamefont {S.}~\bibnamefont
  {Baroni}}, \bibinfo {author} {\bibfnamefont {S.}~\bibnamefont
  {De~Gironcoli}}, \bibinfo {author} {\bibfnamefont {A.}~\bibnamefont
  {Dal~Corso}}, \ and\ \bibinfo {author} {\bibfnamefont {P.}~\bibnamefont
  {Giannozzi}},\ }\bibfield  {title} {\enquote {\bibinfo {title} {Phonons and
  related crystal properties from density-functional perturbation theory},}\
  }\href {https://journals.aps.org/rmp/abstract/10.1103/RevModPhys.73.515}
  {\bibfield  {journal} {\bibinfo  {journal} {Rev. Mod. Phys.}\ }\textbf
  {\bibinfo {volume} {73}},\ \bibinfo {pages} {515} (\bibinfo {year}
  {2001})}\BibitemShut {NoStop}%
\bibitem [{\citenamefont {Giustino}\ \emph {et~al.}(2007)\citenamefont
  {Giustino}, \citenamefont {Cohen},\ and\ \citenamefont
  {Louie}}]{Giustino2007}%
  \BibitemOpen
  \bibfield  {author} {\bibinfo {author} {\bibfnamefont {F.}~\bibnamefont
  {Giustino}}, \bibinfo {author} {\bibfnamefont {M.~L.}\ \bibnamefont {Cohen}},
  \ and\ \bibinfo {author} {\bibfnamefont {S.~G.}\ \bibnamefont {Louie}},\
  }\bibfield  {title} {\enquote {\bibinfo {title} {Electron-phonon interaction
  using {Wannier} functions},}\ }\href
  {https://journals.aps.org/prb/abstract/10.1103/PhysRevB.76.165108} {\bibfield
   {journal} {\bibinfo  {journal} {Phys. Rev. B}\ }\textbf {\bibinfo {volume}
  {76}},\ \bibinfo {pages} {165108} (\bibinfo {year} {2007})}\BibitemShut
  {NoStop}%
\bibitem [{\citenamefont {Ponc{\'e}}\ \emph {et~al.}(2016)\citenamefont
  {Ponc{\'e}}, \citenamefont {Margine}, \citenamefont {Verdi},\ and\
  \citenamefont {Giustino}}]{EPW}%
  \BibitemOpen
  \bibfield  {author} {\bibinfo {author} {\bibfnamefont {S.}~\bibnamefont
  {Ponc{\'e}}}, \bibinfo {author} {\bibfnamefont {E.~R.}\ \bibnamefont
  {Margine}}, \bibinfo {author} {\bibfnamefont {C.}~\bibnamefont {Verdi}}, \
  and\ \bibinfo {author} {\bibfnamefont {F.}~\bibnamefont {Giustino}},\
  }\bibfield  {title} {\enquote {\bibinfo {title} {{EPW}: {Electron}-phonon
  coupling, transport and superconducting properties using maximally localized
  {Wannier} functions},}\ }\href
  {https://www.sciencedirect.com/science/article/pii/S0010465516302260}
  {\bibfield  {journal} {\bibinfo  {journal} {Comput. Phys. Commun.}\ }\textbf
  {\bibinfo {volume} {209}},\ \bibinfo {pages} {116} (\bibinfo {year}
  {2016})}\BibitemShut {NoStop}%
\bibitem [{\citenamefont {Margine}\ and\ \citenamefont
  {Giustino}(2013)}]{Margine2013}%
  \BibitemOpen
  \bibfield  {author} {\bibinfo {author} {\bibfnamefont {E.~R.}\ \bibnamefont
  {Margine}}\ and\ \bibinfo {author} {\bibfnamefont {F.}~\bibnamefont
  {Giustino}},\ }\bibfield  {title} {\enquote {\bibinfo {title} {{Anisotropic
  Migdal-Eliashberg theory using Wannier functions}},}\ }\href
  {https://journals.aps.org/prb/abstract/10.1103/PhysRevB.87.024505} {\bibfield
   {journal} {\bibinfo  {journal} {Phys. Rev. B}\ }\textbf {\bibinfo {volume}
  {87}},\ \bibinfo {pages} {024505} (\bibinfo {year} {2013})}\BibitemShut
  {NoStop}%
\bibitem [{\citenamefont {Marzari}\ \emph {et~al.}(2012)\citenamefont
  {Marzari}, \citenamefont {Mostofi}, \citenamefont {Yates}, \citenamefont
  {Souza},\ and\ \citenamefont {Vanderbilt}}]{WANN1}%
  \BibitemOpen
  \bibfield  {author} {\bibinfo {author} {\bibfnamefont {N.}~\bibnamefont
  {Marzari}}, \bibinfo {author} {\bibfnamefont {A.~A.}\ \bibnamefont
  {Mostofi}}, \bibinfo {author} {\bibfnamefont {J.~R.}\ \bibnamefont {Yates}},
  \bibinfo {author} {\bibfnamefont {I.}~\bibnamefont {Souza}}, \ and\ \bibinfo
  {author} {\bibfnamefont {D.}~\bibnamefont {Vanderbilt}},\ }\bibfield  {title}
  {\enquote {\bibinfo {title} {Maximally localized {Wannier} functions:
  {Theory} and applications},}\ }\href
  {https://journals.aps.org/rmp/abstract/10.1103/RevModPhys.84.1419} {\bibfield
   {journal} {\bibinfo  {journal} {Rev. Mod. Phys.}\ }\textbf {\bibinfo
  {volume} {84}},\ \bibinfo {pages} {1419} (\bibinfo {year}
  {2012})}\BibitemShut {NoStop}%
\bibitem [{\citenamefont {Pizzi}\ \emph {et~al.}(2020)\citenamefont {Pizzi},
  \citenamefont {Vitale}, \citenamefont {Arita}, \citenamefont {Bl{\"u}gel},
  \citenamefont {Freimuth}, \citenamefont {G{\'e}ranton}, \citenamefont
  {Gibertini}, \citenamefont {Gresch}, \citenamefont {Johnson},\ and\
  \citenamefont {Koretsune~$et. al$}}]{WANN2}%
  \BibitemOpen
  \bibfield  {author} {\bibinfo {author} {\bibfnamefont {G.}~\bibnamefont
  {Pizzi}}, \bibinfo {author} {\bibfnamefont {V.}~\bibnamefont {Vitale}},
  \bibinfo {author} {\bibfnamefont {R.}~\bibnamefont {Arita}}, \bibinfo
  {author} {\bibfnamefont {S.}~\bibnamefont {Bl{\"u}gel}}, \bibinfo {author}
  {\bibfnamefont {F.}~\bibnamefont {Freimuth}}, \bibinfo {author}
  {\bibfnamefont {G.}~\bibnamefont {G{\'e}ranton}}, \bibinfo {author}
  {\bibfnamefont {M.}~\bibnamefont {Gibertini}}, \bibinfo {author}
  {\bibfnamefont {D.}~\bibnamefont {Gresch}}, \bibinfo {author} {\bibfnamefont
  {C.}~\bibnamefont {Johnson}}, \ and\ \bibinfo {author} {\bibfnamefont
  {T.}~\bibnamefont {Koretsune~$et. al$}},\ }\bibfield  {title} {\enquote
  {\bibinfo {title} {Wannier90 as a community code: new features and
  applications},}\ }\href
  {https://iopscience.iop.org/article/10.1088/1361-648X/ab51ff?hootPostID=8865030f3411ebd77f127a8addfbbdce}
  {\bibfield  {journal} {\bibinfo  {journal} {J. Phys: Condens. Matter}\
  }\textbf {\bibinfo {volume} {32}},\ \bibinfo {pages} {165902} (\bibinfo
  {year} {2020})}\BibitemShut {NoStop}%
\bibitem [{\citenamefont {Hermann}\ \emph
  {et~al.}(2012{\natexlab{a}})\citenamefont {Hermann}, \citenamefont
  {{McSorley}}, \citenamefont {Ashcroft},\ and\ \citenamefont
  {Hoffmann}}]{Hermann2012}%
  \BibitemOpen
  \bibfield  {author} {\bibinfo {author} {\bibfnamefont {A.}~\bibnamefont
  {Hermann}}, \bibinfo {author} {\bibfnamefont {A.}~\bibnamefont {{McSorley}}},
  \bibinfo {author} {\bibfnamefont {N.~W.}\ \bibnamefont {Ashcroft}}, \ and\
  \bibinfo {author} {\bibfnamefont {R.}~\bibnamefont {Hoffmann}},\ }\bibfield
  {title} {\enquote {\bibinfo {title} {{From {Wade-Mingos to Zintl-Klemm} at
  100 GPa: binary compounds of boron and lithium}},}\ }\href
  {https://pubs.acs.org/doi/abs/10.1021/ja308492g} {\bibfield  {journal}
  {\bibinfo  {journal} {J. Am. Chem. Soc.}\ }\textbf {\bibinfo {volume}
  {134}},\ \bibinfo {pages} {18606} (\bibinfo {year}
  {2012}{\natexlab{a}})}\BibitemShut {NoStop}%
\bibitem [{\citenamefont {Hermann}\ \emph
  {et~al.}(2012{\natexlab{b}})\citenamefont {Hermann}, \citenamefont
  {Suarez-{Alcubilla}}, \citenamefont {Gurtubay}, \citenamefont {Yang},
  \citenamefont {Bergara}, \citenamefont {Ashcroft},\ and\ \citenamefont
  {Hoffmann}}]{Hermann2012-LiB}%
  \BibitemOpen
  \bibfield  {author} {\bibinfo {author} {\bibfnamefont {A.}~\bibnamefont
  {Hermann}}, \bibinfo {author} {\bibfnamefont {A.}~\bibnamefont
  {Suarez-{Alcubilla}}}, \bibinfo {author} {\bibfnamefont {I.~G.}\ \bibnamefont
  {Gurtubay}}, \bibinfo {author} {\bibfnamefont {L.-M.}\ \bibnamefont {Yang}},
  \bibinfo {author} {\bibfnamefont {A.}~\bibnamefont {Bergara}}, \bibinfo
  {author} {\bibfnamefont {N.~W.}\ \bibnamefont {Ashcroft}}, \ and\ \bibinfo
  {author} {\bibfnamefont {R.}~\bibnamefont {Hoffmann}},\ }\bibfield  {title}
  {\enquote {\bibinfo {title} {{LiB and its boron-deficient variants under
  pressure}},}\ }\href
  {https://journals.aps.org/prb/abstract/10.1103/PhysRevB.86.144110} {\bibfield
   {journal} {\bibinfo  {journal} {Phys. Rev. B}\ }\textbf {\bibinfo {volume}
  {86}},\ \bibinfo {pages} {144110} (\bibinfo {year}
  {2012}{\natexlab{b}})}\BibitemShut {NoStop}%
\bibitem [{\citenamefont {Van Der~Geest}\ and\ \citenamefont
  {Kolmogorov}(2014)}]{Van2014}%
  \BibitemOpen
  \bibfield  {author} {\bibinfo {author} {\bibfnamefont {A.~G.}\ \bibnamefont
  {Van Der~Geest}}\ and\ \bibinfo {author} {\bibfnamefont {A.~N.}\ \bibnamefont
  {Kolmogorov}},\ }\bibfield  {title} {\enquote {\bibinfo {title} {Stability of
  41 metal-boron systems at 0 {GPa} and 30 {GPa} from first principles},}\
  }\href {https://doi.org/10.1016/j.calphad.2014.03.005} {\bibfield  {journal}
  {\bibinfo  {journal} {Calphad}\ }\textbf {\bibinfo {volume} {46}},\ \bibinfo
  {pages} {184} (\bibinfo {year} {2014})}\BibitemShut {NoStop}%
\bibitem [{\citenamefont {Borgstedt}\ and\ \citenamefont
  {Guminski}(2003)}]{Borgstedt2003}%
  \BibitemOpen
  \bibfield  {author} {\bibinfo {author} {\bibfnamefont {H.~B.}\ \bibnamefont
  {Borgstedt}}\ and\ \bibinfo {author} {\bibfnamefont {C.}~\bibnamefont
  {Guminski}},\ }\bibfield  {title} {\enquote {\bibinfo {title} {The {B-Li}
  (boron-lithium) system},}\ }\href
  {https://doi.org/10.1361/105497103772084723} {\bibfield  {journal} {\bibinfo
  {journal} {J. Phase Equilibria}\ }\textbf {\bibinfo {volume} {24}},\ \bibinfo
  {pages} {572} (\bibinfo {year} {2003})}\BibitemShut {NoStop}%
\bibitem [{\citenamefont {Kolmogorov}\ \emph {et~al.}(2012)\citenamefont
  {Kolmogorov}, \citenamefont {Shah}, \citenamefont {Margine}, \citenamefont
  {Kleppe},\ and\ \citenamefont {Jephcoat}}]{Kolmogorov2012-CaB6}%
  \BibitemOpen
  \bibfield  {author} {\bibinfo {author} {\bibfnamefont {A.~N.}\ \bibnamefont
  {Kolmogorov}}, \bibinfo {author} {\bibfnamefont {S.}~\bibnamefont {Shah}},
  \bibinfo {author} {\bibfnamefont {E.~R.}\ \bibnamefont {Margine}}, \bibinfo
  {author} {\bibfnamefont {A.~K.}\ \bibnamefont {Kleppe}}, \ and\ \bibinfo
  {author} {\bibfnamefont {A.~P.}\ \bibnamefont {Jephcoat}},\ }\bibfield
  {title} {\enquote {\bibinfo {title} {Pressure-driven evolution of the
  covalent network in {CaB$_6$}},}\ }\href
  {https://journals.aps.org/prl/abstract/10.1103/PhysRevLett.109.075501}
  {\bibfield  {journal} {\bibinfo  {journal} {Phys. Rev. Lett.}\ }\textbf
  {\bibinfo {volume} {109}},\ \bibinfo {pages} {075501} (\bibinfo {year}
  {2012})}\BibitemShut {NoStop}%
\bibitem [{\citenamefont {Bialon}\ \emph {et~al.}(2011)\citenamefont {Bialon},
  \citenamefont {Hammerschmidt}, \citenamefont {Drautz}, \citenamefont {Shah},
  \citenamefont {Margine},\ and\ \citenamefont {Kolmogorov}}]{Bialon2011}%
  \BibitemOpen
  \bibfield  {author} {\bibinfo {author} {\bibfnamefont {A.~F.}\ \bibnamefont
  {Bialon}}, \bibinfo {author} {\bibfnamefont {T.}~\bibnamefont
  {Hammerschmidt}}, \bibinfo {author} {\bibfnamefont {R.}~\bibnamefont
  {Drautz}}, \bibinfo {author} {\bibfnamefont {S.}~\bibnamefont {Shah}},
  \bibinfo {author} {\bibfnamefont {E.~R.}\ \bibnamefont {Margine}}, \ and\
  \bibinfo {author} {\bibfnamefont {A.~N.}\ \bibnamefont {Kolmogorov}},\
  }\bibfield  {title} {\enquote {\bibinfo {title} {{Possible routes for
  synthesis of new boron-rich {Fe-B} and Fe$_{1- x}${Cr$_ x$}{B$_4$}
  compounds}},}\ }\href {https://doi.org/10.1063/1.3556564} {\bibfield
  {journal} {\bibinfo  {journal} {Appl. Phys. Lett.}\ }\textbf {\bibinfo
  {volume} {98}},\ \bibinfo {pages} {081901} (\bibinfo {year}
  {2011})}\BibitemShut {NoStop}%
\bibitem [{\citenamefont {Curtarolo}\ \emph {et~al.}(2005)\citenamefont
  {Curtarolo}, \citenamefont {Morgan},\ and\ \citenamefont
  {Ceder}}]{Curtarolo2005}%
  \BibitemOpen
  \bibfield  {author} {\bibinfo {author} {\bibfnamefont {S.}~\bibnamefont
  {Curtarolo}}, \bibinfo {author} {\bibfnamefont {D.}~\bibnamefont {Morgan}}, \
  and\ \bibinfo {author} {\bibfnamefont {G.}~\bibnamefont {Ceder}},\ }\bibfield
   {title} {\enquote {\bibinfo {title} {Accuracy of $ab~initio$ methods in
  predicting the crystal structures of metals: {A} review of 80 binary
  alloys},}\ }\href {https://doi.org/10.1016/j.calphad.2005.01.002} {\bibfield
  {journal} {\bibinfo  {journal} {Calphad}\ }\textbf {\bibinfo {volume} {29}},\
  \bibinfo {pages} {163} (\bibinfo {year} {2005})}\BibitemShut {NoStop}%
\bibitem [{\citenamefont {W{\"o}rle}\ and\ \citenamefont
  {Nesper}(2000)}]{Worle2000}%
  \BibitemOpen
  \bibfield  {author} {\bibinfo {author} {\bibfnamefont {M.}~\bibnamefont
  {W{\"o}rle}}\ and\ \bibinfo {author} {\bibfnamefont {R.}~\bibnamefont
  {Nesper}},\ }\bibfield  {title} {\enquote {\bibinfo {title} {{Infinite,
  linear, unbranched borynide chains in LiB$_x$—Isoelectronic to polyyne and
  polycumulene}},}\ }\href
  {https://doi.org/10.1002/1521-3757(20000703)112:13<2439::AID-ANGE2439>3.0.CO;2-Q}
  {\bibfield  {journal} {\bibinfo  {journal} {Angewandte Chemie}\ }\textbf
  {\bibinfo {volume} {112}},\ \bibinfo {pages} {2439} (\bibinfo {year}
  {2000})}\BibitemShut {NoStop}%
\bibitem [{\citenamefont {Kolmogorov}\ \emph {et~al.}(2007)\citenamefont
  {Kolmogorov}, \citenamefont {Drautz},\ and\ \citenamefont
  {Pettifor}}]{Kolmogorov2007}%
  \BibitemOpen
  \bibfield  {author} {\bibinfo {author} {\bibfnamefont {A.~N~.}\ \bibnamefont
  {Kolmogorov}}, \bibinfo {author} {\bibfnamefont {R.}~\bibnamefont {Drautz}},
  \ and\ \bibinfo {author} {\bibfnamefont {D.~G.}\ \bibnamefont {Pettifor}},\
  }\bibfield  {title} {\enquote {\bibinfo {title} {{\it Ab-initio} modeling of
  {Li-B-H} boron-chain alloys for hydrogen storage applications},}\ }\href
  {https://journals.aps.org/prb/abstract/10.1103/PhysRevB.76.184102} {\bibfield
   {journal} {\bibinfo  {journal} {Phys. Rev. B}\ }\textbf {\bibinfo {volume}
  {76}},\ \bibinfo {pages} {184102} (\bibinfo {year} {2007})}\BibitemShut
  {NoStop}%
\bibitem [{Note1()}]{Note1}%
  \BibitemOpen
  \bibinfo {note} {{{{T}he starting {LiB}$_y$ composition in {Ref}.~[\protect
  \rev@citealpnum {Kolmogorov2015}] was deduced to be around 10:9. {Given} the
  uncertainty in the determination of the absolute stoichiometry values, we
  picked the 9:8 starting ratio inside the stability range and illustrated its
  variation with pressure}}}\BibitemShut {NoStop}%
\bibitem [{Note2()}]{Note2}%
  \BibitemOpen
  \bibinfo {note} {{T}he ordered {tI16}-{LiB$_3$} was found to be
  thermodynamically stable at 0 {K} in the previous {PBE} calculations~\cite
  {Van2014}. In the present {optB86b} treatment, it is metastable by 9
  {meV}/atom at 0 {K} and by 5 {meV}/atom at 600 {K} when the vibrational
  entropy term is included. {The} phase may become stable at higher
  temperatures and/or upon inclusion of the configuration entropy
  contribution}\BibitemShut {NoStop}%
\bibitem [{SM()}]{SM}%
  \BibitemOpen
  \href@noop {} {\bibinfo  {journal} {{See Supplemental Material for
  Figs.~S1-S14 and Table~S1-S2}. A description to each figure is given in the
  body of the paper, wherever Ref. [60] is cited}\ }\BibitemShut {NoStop}%
\bibitem [{\citenamefont {Reinoso}(2005)}]{Reinosothesis}%
  \BibitemOpen
\bibfield  {journal} {  }\bibfield  {author} {\bibinfo {author} {\bibfnamefont
  {J.~M.}\ \bibnamefont {Reinoso}},\ }\emph {\bibinfo {title} {{Untersuchungen
  zu Wasserstoffspeicherung in ausgewählten anorganischen Materialien}}},\
  \href {https://doi.org/10.3929/ethz-a-004945659} {Ph.D. thesis},\ \bibinfo
  {school} {ETH Zürich}, \bibinfo {address} {Zürich} (\bibinfo {year}
  {2005})\BibitemShut {NoStop}%
\bibitem [{\citenamefont {Fogg}\ \emph {et~al.}(2006)\citenamefont {Fogg},
  \citenamefont {Meldrum}, \citenamefont {Darling}, \citenamefont {Claridge},\
  and\ \citenamefont {Rosseinsky}}]{Fogg2006}%
  \BibitemOpen
  \bibfield  {author} {\bibinfo {author} {\bibfnamefont {A.~M.}\ \bibnamefont
  {Fogg}}, \bibinfo {author} {\bibfnamefont {J.}~\bibnamefont {Meldrum}},
  \bibinfo {author} {\bibfnamefont {G.~R.}\ \bibnamefont {Darling}}, \bibinfo
  {author} {\bibfnamefont {J.~B.}\ \bibnamefont {Claridge}}, \ and\ \bibinfo
  {author} {\bibfnamefont {M.~J.}\ \bibnamefont {Rosseinsky}},\ }\bibfield
  {title} {\enquote {\bibinfo {title} {{Chemical control of electronic
  structure and superconductivity in layered borides and borocarbides:
  understanding the absence of superconductivity in {Li$_x$BC}}},}\ }\href
  {https://pubs.acs.org/doi/10.1021/ja0578449} {\bibfield  {journal} {\bibinfo
  {journal} {J. Am. Chem. Soc.}\ }\textbf {\bibinfo {volume} {128}},\ \bibinfo
  {pages} {10043} (\bibinfo {year} {2006})}\BibitemShut {NoStop}%
\bibitem [{\citenamefont {Kalkan}\ and\ \citenamefont
  {Ozdas}(2019)}]{Kalkan2019}%
  \BibitemOpen
  \bibfield  {author} {\bibinfo {author} {\bibfnamefont {B.}~\bibnamefont
  {Kalkan}}\ and\ \bibinfo {author} {\bibfnamefont {E.}~\bibnamefont {Ozdas}},\
  }\bibfield  {title} {\enquote {\bibinfo {title} {Staging {Phenomena} in
  {Lithium-Intercalated Boron--Carbon}},}\ }\href
  {https://doi.org/10.1021/acsami.8b19142} {\bibfield  {journal} {\bibinfo
  {journal} {ACS Appl. Mater. Interfaces}\ }\textbf {\bibinfo {volume} {11}},\
  \bibinfo {pages} {4111} (\bibinfo {year} {2019})}\BibitemShut {NoStop}%
\bibitem [{\citenamefont {Kolmogorov}\ \emph {et~al.}(2008)\citenamefont
  {Kolmogorov}, \citenamefont {Calandra},\ and\ \citenamefont
  {Curtarolo}}]{Kolmogorov2008}%
  \BibitemOpen
  \bibfield  {author} {\bibinfo {author} {\bibfnamefont {A.~N.}\ \bibnamefont
  {Kolmogorov}}, \bibinfo {author} {\bibfnamefont {M.}~\bibnamefont
  {Calandra}}, \ and\ \bibinfo {author} {\bibfnamefont {S.}~\bibnamefont
  {Curtarolo}},\ }\bibfield  {title} {\enquote {\bibinfo {title}
  {{Thermodynamic stabilities of ternary metal borides: An $ab~initio$ guide
  for synthesizing layered superconductors}},}\ }\href
  {https://link.aps.org/doi/10.1103/PhysRevB.78.094520} {\bibfield  {journal}
  {\bibinfo  {journal} {Phys. Rev. B}\ }\textbf {\bibinfo {volume} {78}},\
  \bibinfo {pages} {094520} (\bibinfo {year} {2008})}\BibitemShut {NoStop}%
\bibitem [{\citenamefont {Mazin}\ \emph {et~al.}(1993)\citenamefont {Mazin},
  \citenamefont {Liechtenstein}, \citenamefont {Rodriguez}, \citenamefont
  {Jepsen},\ and\ \citenamefont {Andersen}}]{Eilat}%
  \BibitemOpen
  \bibfield  {author} {\bibinfo {author} {\bibfnamefont {I.~I.}\ \bibnamefont
  {Mazin}}, \bibinfo {author} {\bibfnamefont {A.~I.}\ \bibnamefont
  {Liechtenstein}}, \bibinfo {author} {\bibfnamefont {C.~O.}\ \bibnamefont
  {Rodriguez}}, \bibinfo {author} {\bibfnamefont {O.}~\bibnamefont {Jepsen}}, \
  and\ \bibinfo {author} {\bibfnamefont {O.~K.}\ \bibnamefont {Andersen}},\
  }\bibfield  {title} {\enquote {\bibinfo {title} {{Superconducting and
  transport electron-phonon coupling constants in YBa$_2$Cu$_3$O$_7$: effect of
  the interband anisotropy}},}\ }\href {\doibase
  https://doi.org/10.1016/0921-4534(93)90886-U} {\bibfield  {journal} {\bibinfo
   {journal} {Physica C: Supercond.}\ }\textbf {\bibinfo {volume} {209}},\
  \bibinfo {pages} {125--128} (\bibinfo {year} {1993})}\BibitemShut {NoStop}%
\bibitem [{\citenamefont {Mazin}\ \emph {et~al.}(2004)\citenamefont {Mazin},
  \citenamefont {Andersen}, \citenamefont {Jepsen}, \citenamefont {Golubov},
  \citenamefont {Dolgov},\ and\ \citenamefont {Kortus}}]{Mazin2004}%
  \BibitemOpen
  \bibfield  {author} {\bibinfo {author} {\bibfnamefont {I.~I.}\ \bibnamefont
  {Mazin}}, \bibinfo {author} {\bibfnamefont {O.~K.}\ \bibnamefont {Andersen}},
  \bibinfo {author} {\bibfnamefont {O.}~\bibnamefont {Jepsen}}, \bibinfo
  {author} {\bibfnamefont {A.~A.}\ \bibnamefont {Golubov}}, \bibinfo {author}
  {\bibfnamefont {O.~V.}\ \bibnamefont {Dolgov}}, \ and\ \bibinfo {author}
  {\bibfnamefont {J.}~\bibnamefont {Kortus}},\ }\bibfield  {title} {\enquote
  {\bibinfo {title} {{Comment on `First-principles calculation of the
  superconducting transition in MgB$_2$ within the anisotropic Eliashberg
  formalism'}},}\ }\href
  {https://journals.aps.org/prb/abstract/10.1103/PhysRevB.69.056501} {\bibfield
   {journal} {\bibinfo  {journal} {Phys. Rev. B}\ }\textbf {\bibinfo {volume}
  {69}},\ \bibinfo {pages} {056501} (\bibinfo {year} {2004})}\BibitemShut
  {NoStop}%
\bibitem [{\citenamefont {Floris}\ \emph {et~al.}(2007)\citenamefont {Floris},
  \citenamefont {Sanna}, \citenamefont {L{\"u}ders}, \citenamefont {Profeta},
  \citenamefont {Lathiotakis}, \citenamefont {Marques}, \citenamefont
  {Franchini}, \citenamefont {Gross}, \citenamefont {Continenza},\ and\
  \citenamefont {Massidda}}]{Floris2007}%
  \BibitemOpen
  \bibfield  {author} {\bibinfo {author} {\bibfnamefont {A.}~\bibnamefont
  {Floris}}, \bibinfo {author} {\bibfnamefont {A.}~\bibnamefont {Sanna}},
  \bibinfo {author} {\bibfnamefont {M.}~\bibnamefont {L{\"u}ders}}, \bibinfo
  {author} {\bibfnamefont {G.}~\bibnamefont {Profeta}}, \bibinfo {author}
  {\bibfnamefont {N.~N.}\ \bibnamefont {Lathiotakis}}, \bibinfo {author}
  {\bibfnamefont {M.~A.~L.}\ \bibnamefont {Marques}}, \bibinfo {author}
  {\bibfnamefont {C.}~\bibnamefont {Franchini}}, \bibinfo {author}
  {\bibfnamefont {E.~K.~U.}\ \bibnamefont {Gross}}, \bibinfo {author}
  {\bibfnamefont {A.}~\bibnamefont {Continenza}}, \ and\ \bibinfo {author}
  {\bibfnamefont {S.}~\bibnamefont {Massidda}},\ }\bibfield  {title} {\enquote
  {\bibinfo {title} {{Superconducting properties of MgB$_2$ from first
  principles}},}\ }\href {https://doi.org/10.1016/j.physc.2007.01.026}
  {\bibfield  {journal} {\bibinfo  {journal} {Physica C: Supercond.}\ }\textbf
  {\bibinfo {volume} {456}},\ \bibinfo {pages} {45} (\bibinfo {year}
  {2007})}\BibitemShut {NoStop}%
\bibitem [{\citenamefont {Towns}\ \emph {et~al.}(2014)\citenamefont {Towns},
  \citenamefont {Cockerill}, \citenamefont {Dahan}, \citenamefont {Foster},
  \citenamefont {Gaither}, \citenamefont {Grimshaw}, \citenamefont {Hazlewood},
  \citenamefont {Lathrop}, \citenamefont {Lifka}, \citenamefont {Peterson},
  \citenamefont {Roskies}, \citenamefont {Scott},\ and\ \citenamefont
  {Wilkins-Diehr}}]{XSEDE}%
  \BibitemOpen
  \bibfield  {author} {\bibinfo {author} {\bibfnamefont {J.}~\bibnamefont
  {Towns}}, \bibinfo {author} {\bibfnamefont {T.}~\bibnamefont {Cockerill}},
  \bibinfo {author} {\bibfnamefont {M.}~\bibnamefont {Dahan}}, \bibinfo
  {author} {\bibfnamefont {I.}~\bibnamefont {Foster}}, \bibinfo {author}
  {\bibfnamefont {K.}~\bibnamefont {Gaither}}, \bibinfo {author} {\bibfnamefont
  {A.}~\bibnamefont {Grimshaw}}, \bibinfo {author} {\bibfnamefont
  {V.}~\bibnamefont {Hazlewood}}, \bibinfo {author} {\bibfnamefont
  {S.}~\bibnamefont {Lathrop}}, \bibinfo {author} {\bibfnamefont
  {D.}~\bibnamefont {Lifka}}, \bibinfo {author} {\bibfnamefont {G.~D.}\
  \bibnamefont {Peterson}}, \bibinfo {author} {\bibfnamefont {R.}~\bibnamefont
  {Roskies}}, \bibinfo {author} {\bibfnamefont {J.~R.}\ \bibnamefont {Scott}},
  \ and\ \bibinfo {author} {\bibfnamefont {N.}~\bibnamefont {Wilkins-Diehr}},\
  }\bibfield  {title} {\enquote {\bibinfo {title} {{XSEDE}: Accelerating
  scientific discovery},}\ }\href {\doibase 10.1109/MCSE.2014.80} {\bibfield
  {journal} {\bibinfo  {journal} {Comput. Sci. {\&} Eng.}\ }\textbf {\bibinfo
  {volume} {16}},\ \bibinfo {pages} {62--74} (\bibinfo {year}
  {2014})}\BibitemShut {NoStop}%
\bibitem [{\citenamefont {Stanzione}\ \emph {et~al.}(2020)\citenamefont
  {Stanzione}, \citenamefont {West}, \citenamefont {Evans}, \citenamefont
  {Minyard}, \citenamefont {Ghattas},\ and\ \citenamefont {Panda}}]{Frontera}%
  \BibitemOpen
  \bibfield  {author} {\bibinfo {author} {\bibfnamefont {Dan}\ \bibnamefont
  {Stanzione}}, \bibinfo {author} {\bibfnamefont {John}\ \bibnamefont {West}},
  \bibinfo {author} {\bibfnamefont {R.~Todd}\ \bibnamefont {Evans}}, \bibinfo
  {author} {\bibfnamefont {Tommy}\ \bibnamefont {Minyard}}, \bibinfo {author}
  {\bibfnamefont {Omar}\ \bibnamefont {Ghattas}}, \ and\ \bibinfo {author}
  {\bibfnamefont {Dhabaleswar~K.}\ \bibnamefont {Panda}},\ }\enquote {\bibinfo
  {title} {Frontera: The evolution of leadership computing at the national
  science foundation},}\ in\ \href {https://doi.org/10.1145/3311790.3396656}
  {\emph {\bibinfo {booktitle} {Practice and Experience in Advanced Research
  Computing}}}\ (\bibinfo  {publisher} {ACM},\ \bibinfo {address} {New York,
  NY, USA},\ \bibinfo {year} {2020})\ pp.\ \bibinfo {pages}
  {106--111}\BibitemShut {NoStop}%
\end{thebibliography}%

\clearpage
\onecolumngrid

\begin{center}
	\textbf{\large Supplemental Material \\$Ab~initio$ study of Li-Mg-B superconductors}\\[.5cm]
	
	Gyanu P. Kafle,$^{1}$ Charlsey R. Tomassetti,$^{1}$ Igor I. Mazin,$^{2}$ Aleksey N. Kolmogorov$^{1}$ and Elena R. Margine$^{1, \textcolor{blue}{*}}$\\[.1cm]
	{\itshape ${}^1$Department of Physics, Applied Physics, and Astronomy,\\ Binghamton University-SUNY, Binghamton, New York 13902, USA\\
		${}^2$Department of Physics and Astronomy and Quantum Science and Engineering Center,\\ George Mason University, Fairfax, Virginia 22030, USA\\}
	(Dated: \today)\\[1cm]
	
\end{center}

\setcounter{equation}{0}
\setcounter{figure}{0}
\setcounter{table}{0}
\setcounter{page}{1}
\renewcommand{\thefigure}{S\arabic{figure}}
\renewcommand{\thetable}{S\arabic{table}}
\renewcommand{\arraystretch}{1.3}
\renewcommand{\bibnumfmt}[1]{[S#1]}
\renewcommand{\citenumfont}[1]{S#1}

\begin{figure*}[h]
	\centering
	\includegraphics[width=0.5\linewidth]{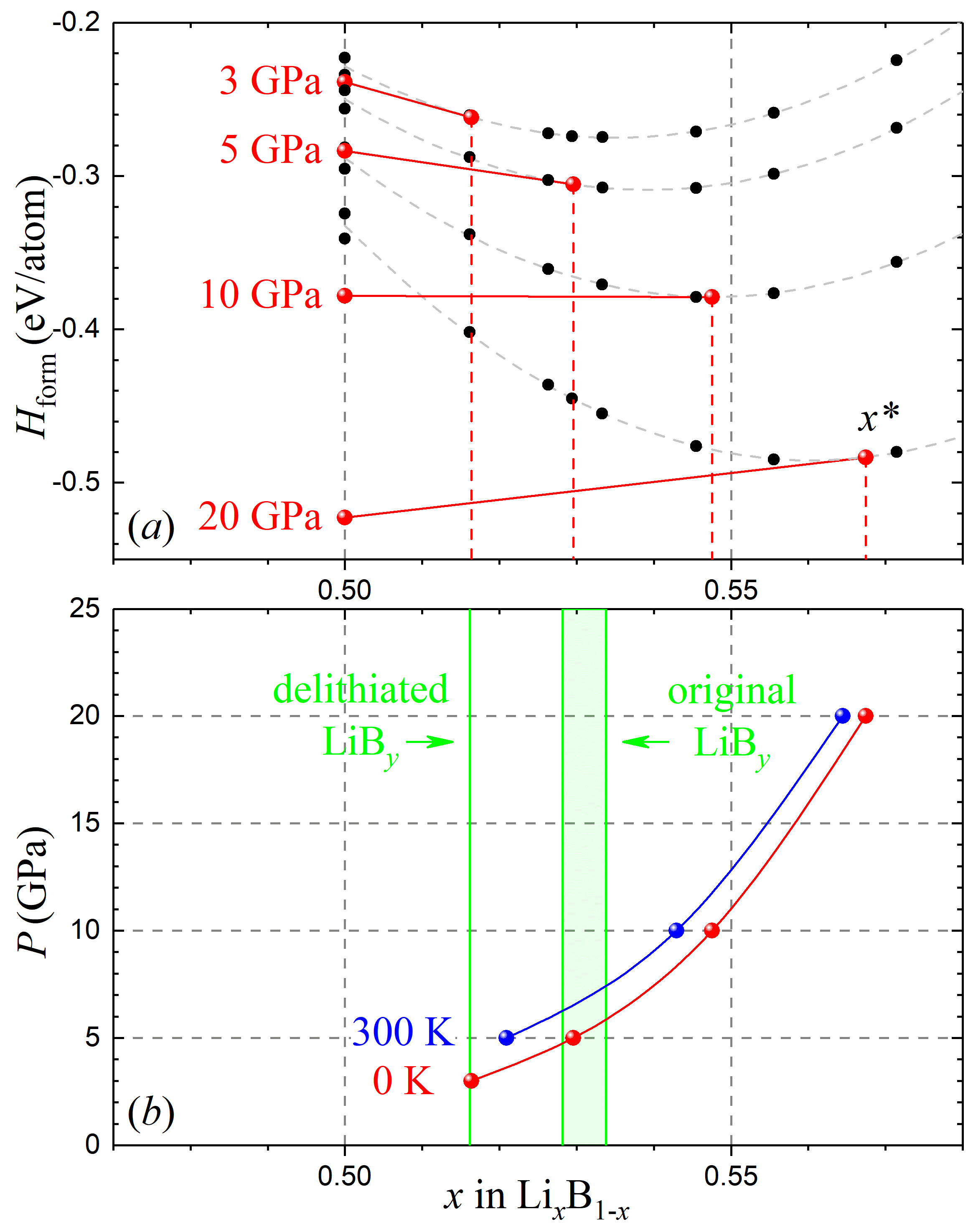}
	\caption{\label{fig-S1-LiBy} Analysis of the LiB$_y$ stability with respect to decomposition into LiB and LiB$_{y^*}$ (y*$\textless$ y) at $zero$ temperature and different pressures. (a) Fitted parabolas (dashed grey lines) of LiB$_y$ formation enthalpies (black circles) with tangents (solid red lines) connecting the formation enthalpies of LiB (large red circles). (b) The pressure-dependent concentrations $x^*$ = (1+$y^*$)-1 that define the products of the LiB$_y$ $\Rightarrow$ LiB + LiB$_{y^*}$ decomposition at 0 K (red points) and 300 K (blue points). The 300 K results were obtained via shifting the LiB formation enthalpies in (a) by the difference in the vibrational entropy corrections between the layered LiB and the most stable 1:1 member of the LiB$_y$ family. The shaded green range marks stable compositions for LiB$_y$, while the single green line corresponds to the proposed metastable delithiated LiB$_y$ with $y$=15/16.}
\end{figure*}

\begin{figure*}[h]
	\centering
	\includegraphics[width=0.95\linewidth]{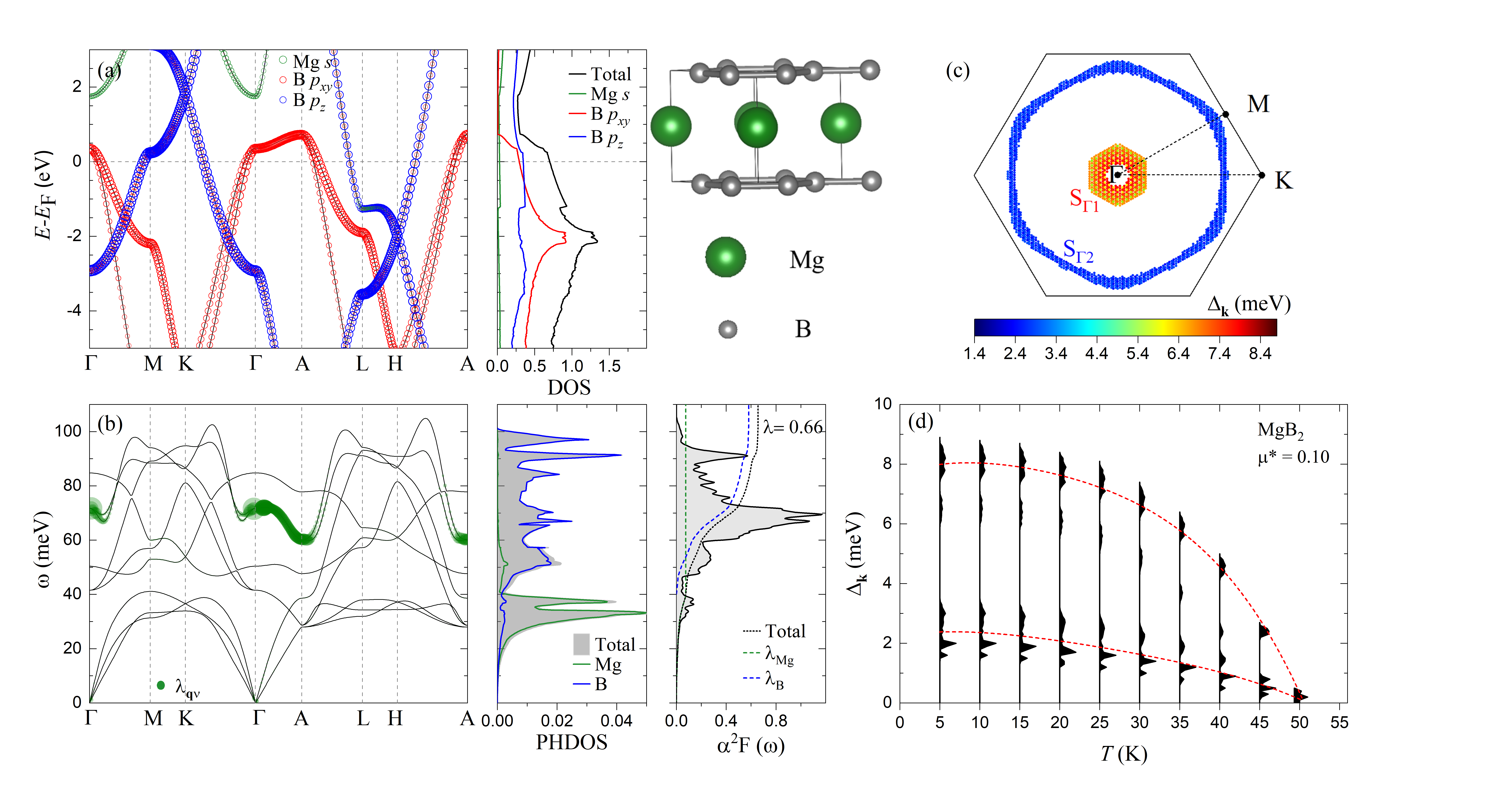}
	\caption{\label{fig-S2-MgB2} Summary of MgB$_2$ properties. (a) Calculated band structure and DOS [states/(eV f.u.)]. The size of the symbols is proportional to the contribution of each orbital character. (b) Calculated phonon dispersion, phonon DOS, and Eliashberg spectral function $\alpha^2F(\omega)$. The phonon branches are broadened by the $e$-ph coupling strength $\lambda_{\bq\nu}$, showcasing the modes and directions that have the strongest $e$-ph coupling. (c) The distribution of superconducting gap at 5~K on the cross-section Fermi surface plane through $z$=0. (d) Calculated anisotropic superconducting gaps $\Delta_\bk$ as a function of temperature (red dashed curves are a guide to the eye).}
\end{figure*}

\begin{figure*}[h]
	\centering
	\includegraphics[width=0.8\linewidth]{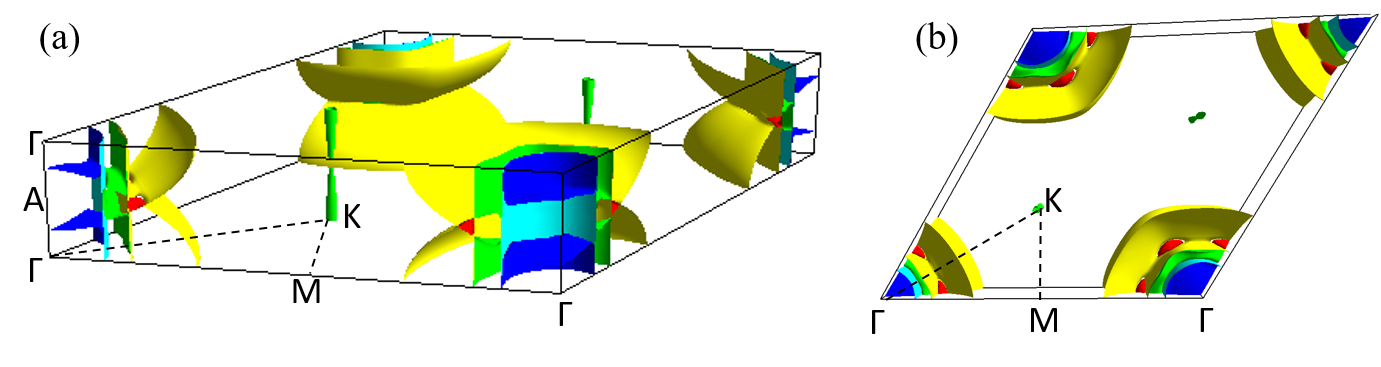}
	\caption{\label{fig-S3-fs-LiB} (a) Side and (b) top view of the Fermi surface of LiB at 0 GPa.}
\end{figure*}

\begin{figure*}[tb]
	\centering
	\includegraphics[width=0.99\linewidth]{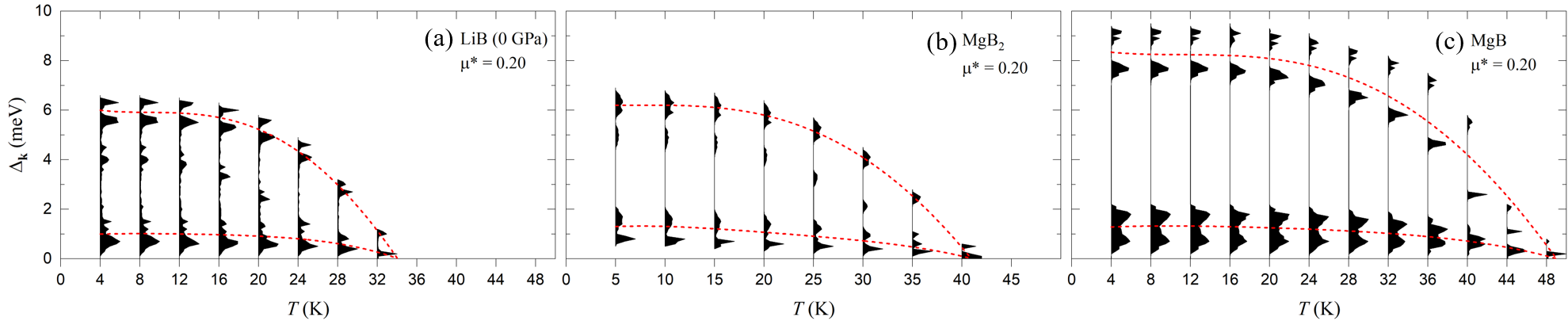}
	\caption{\label{fig-S4-mu02} Calculated anisotropic superconducting gaps $\Delta_\textbf{k}$ of (a) LiB, (b) MgB$_2$, and (c) MgB as a function of temperature with $\mu^*$ = 0.20. The red dashed curves are a guide to the eye.}
\end{figure*}




\begin{figure*}[h]
	\centering
	\includegraphics[width=0.8\linewidth]{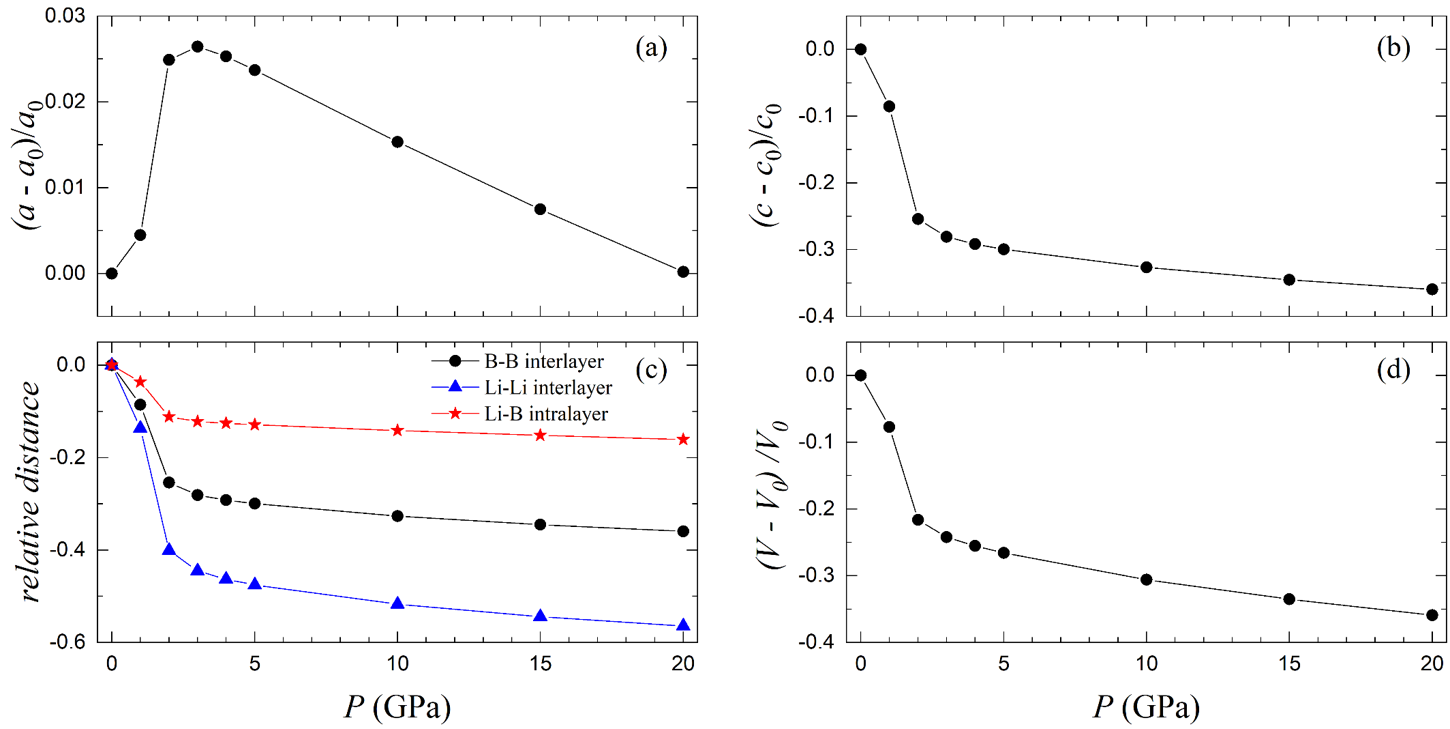}
	\caption{\label{fig-S5-lat-param} Pressure dependence of the relative change of (a)-(b) lattice parameters $a$ and $c$, (c) B-B interlayer, Li-Li interlayer, and Li-B intralayer distances, and (d) volume per formula unit of LiB.} 
\end{figure*}

\begin{figure*}[t]
	\centering
	\includegraphics[width=0.9\linewidth]{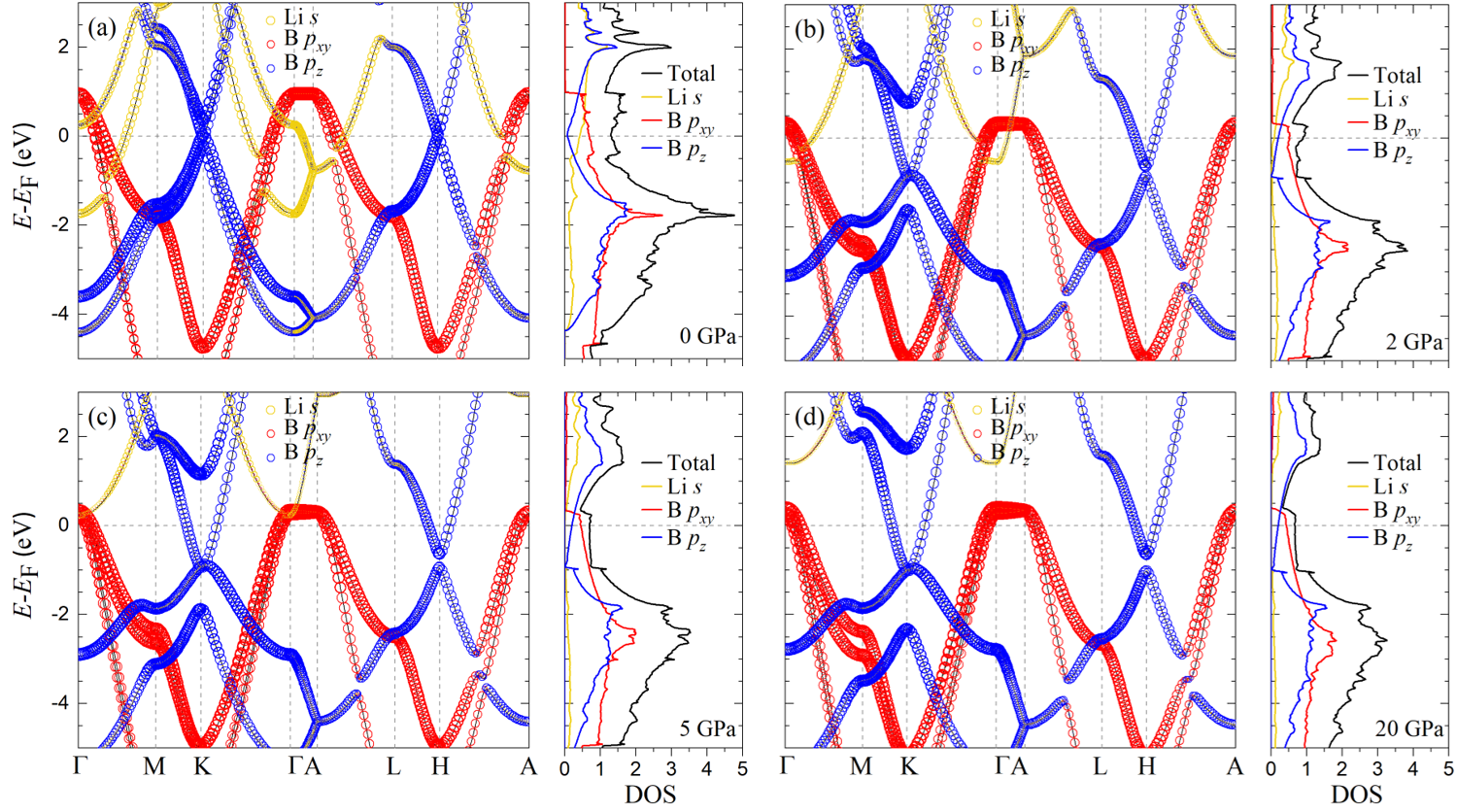}
	\caption{\label{fig-S6-band} Calculated band structure and DOS [states/(eV f.u.)] of LiB at (a) 0, (b) 2, (c) 5, and (d) 20 GPa. The size of the symbols is proportional to the contribution of each orbital character.}
\end{figure*}

\begin{figure*}[t]
	\centering
	\includegraphics[width=0.99\linewidth]{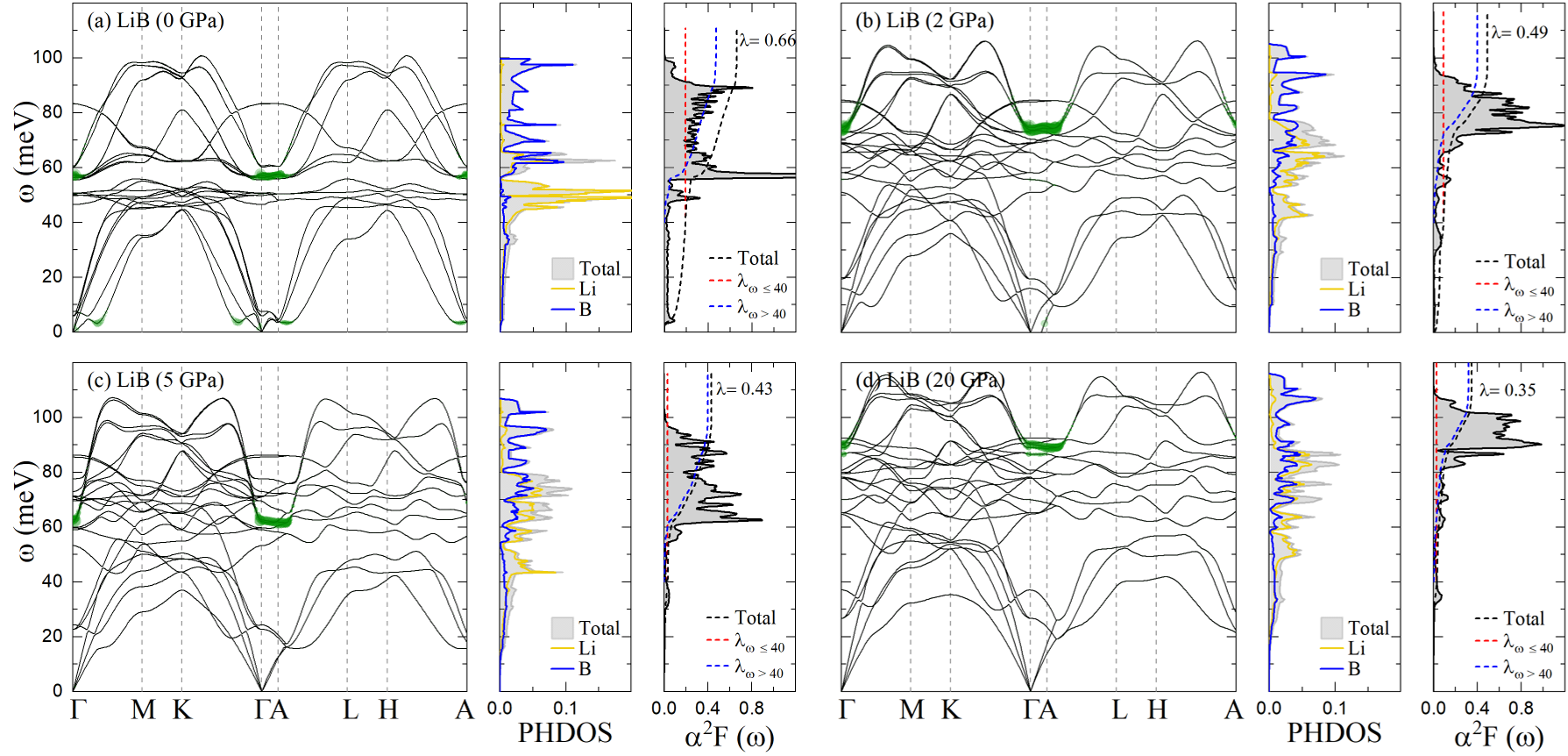}
	\caption{\label{fig-S7-phonon} Calculated phonon dispersion, PHDOS, and Eliashberg spectral function $\alpha^2F(\omega)$ of LiB at (a) 0, (b) 2, (c) 5, and (d) 20 GPa. The phonon branches are broadened by the $e$-ph coupling strength $\lambda_{\bq\nu}$, showcasing the modes and high-symmetry directions that have the strongest $e$-ph coupling.}
\end{figure*}

\begin{table}[h!]
	\caption{\label{table1} Band-resolved symmetrized $e$-ph coupling $\Lambda_{ij}$ and DOS for the two-band FS divisions of LiB under pressure. DOS is given in states/(eV spin).}
	\begin{tabular*}{0.4\textwidth}{c @{\extracolsep{\fill}} cccc}
		& \multicolumn{3}{l}{}                    \\ 
		\hline\hline
		Pressure (LiB) & FS regions  & $\sigma$+$\zeta$ & $\pi$ &             \\ 
		\hline
		0 GPa &$\sigma$+$\zeta$   & 0.643  & 0.008  &              \\
		$\lambda$=0.66 &$\pi$   & 0.008  & 0.000  &            \\
		&   &   &   &              \\
		&DOS & 0.680  & 0.020  &  \\  
		\hline\hline
		&   & $\sigma$+$\zeta$ & $\pi$  &             \\ 
		\hline
		2 GPa       &$\sigma$+$\zeta$   & 0.374  & 0.046   &          \\
		$\lambda$=0.49            &$\pi$   & 0.046  & 0.023    &         \\
		&    &        &      &             \\
		&DOS & 0.355  & 0.111   &          \\
		\hline\hline
		&   & $\sigma$+$\zeta$ & $\pi$  & \\
		\hline
		5 GPa       &$\sigma$+$\zeta$   & 0.324  & 0.038    &         \\
		$\lambda$=0.43            &$\pi$   & 0.038  & 0.029    &         \\
		&    &        &       &            \\
		&DOS & 0.228  & 0.132      &       \\
		\hline\hline
		&   & $\sigma$ & $\pi$  &    \\
		\hline
		20 GPa      &$\sigma$   & 0.240  & 0.041      &       \\
		$\lambda$=0.35            &$\pi$   & 0.041  & 0.024  &           \\
		&    &        &        &           \\
		&DOS & 0.221  & 0.120    &         \\
		\hline
		\hline
	\end{tabular*}
\end{table}

\begin{figure*}[t]
	\centering
	\includegraphics[width=0.9\linewidth]{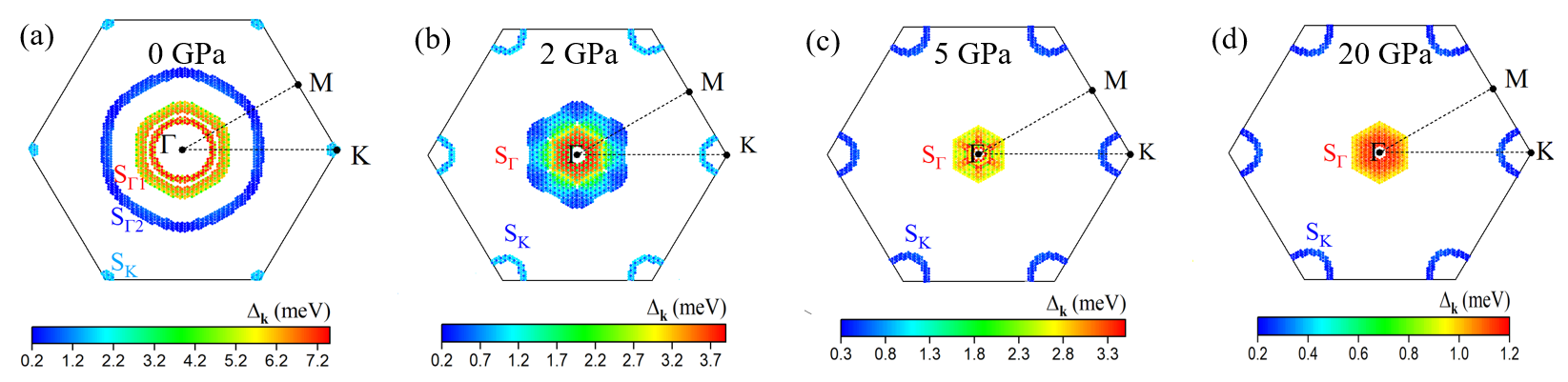}
	\caption{\label{fig-S8-gap-FS} The distribution of superconducting gap at 4~K on the cross-section Fermi surface plane through $z$=0 of LiB for (a) 0, (b) 2, (c) 5, and (d) 20 GPa.}
\end{figure*}

\begin{figure*}[tb]
	\centering
	\includegraphics[width=0.95\linewidth]{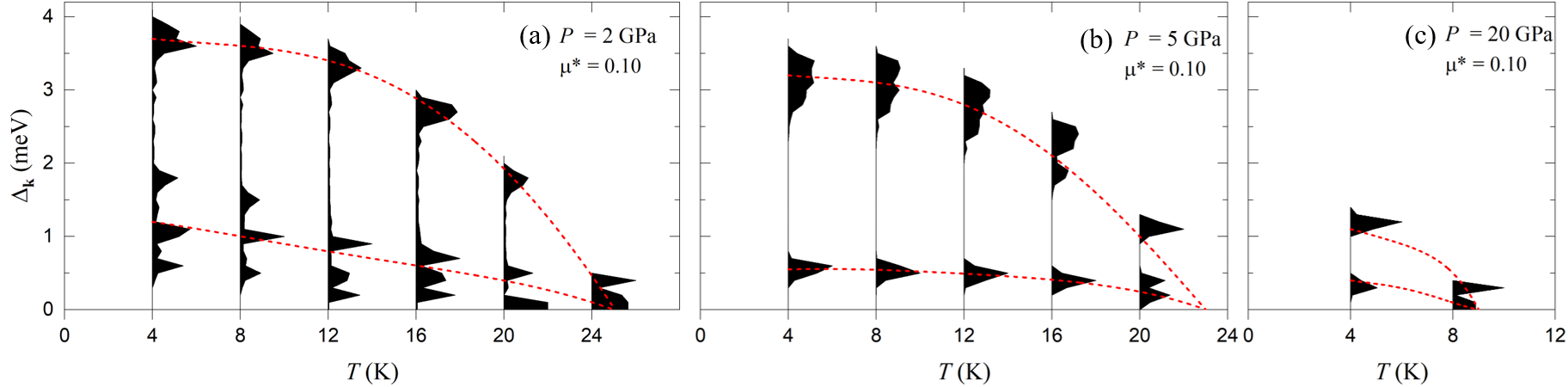}
	\caption{\label{fig-S9-aniso-pressure} Calculated anisotropic superconducting gaps $\Delta_\bk$ of LiB as a function of temperature at (a) 2, (b) 5, and (c) 20 GPa  with $\mu^*$ = 0.10. The red dashed curves are a guide to the eye.}
\end{figure*}

\begin{figure*}[h!]
	\centering
	\includegraphics[width=0.98\linewidth]{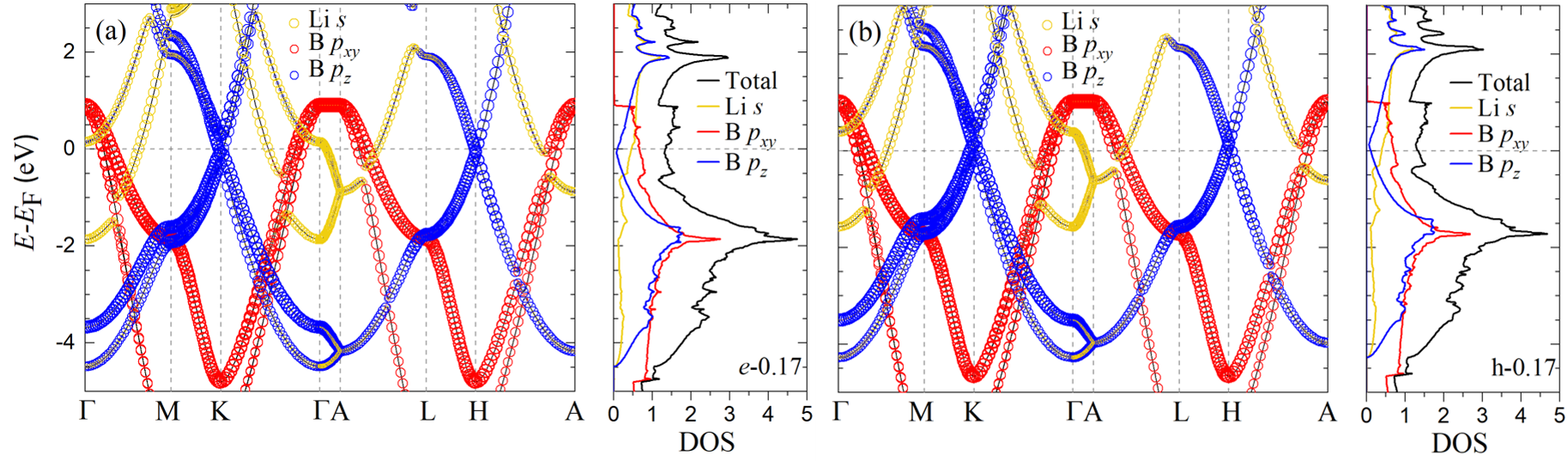}
	\caption{\label{fig-S10-band-doping} Calculated band structure and DOS [states/(eV f.u.)] of LiB for the doping level of (a) $e$-0.17,  and (b) h-0.17. The size of the symbols is proportional to the contribution of each orbital character.}
\end{figure*}

\begin{figure*}[h!]
	\centering
	\includegraphics[width=0.98\linewidth]{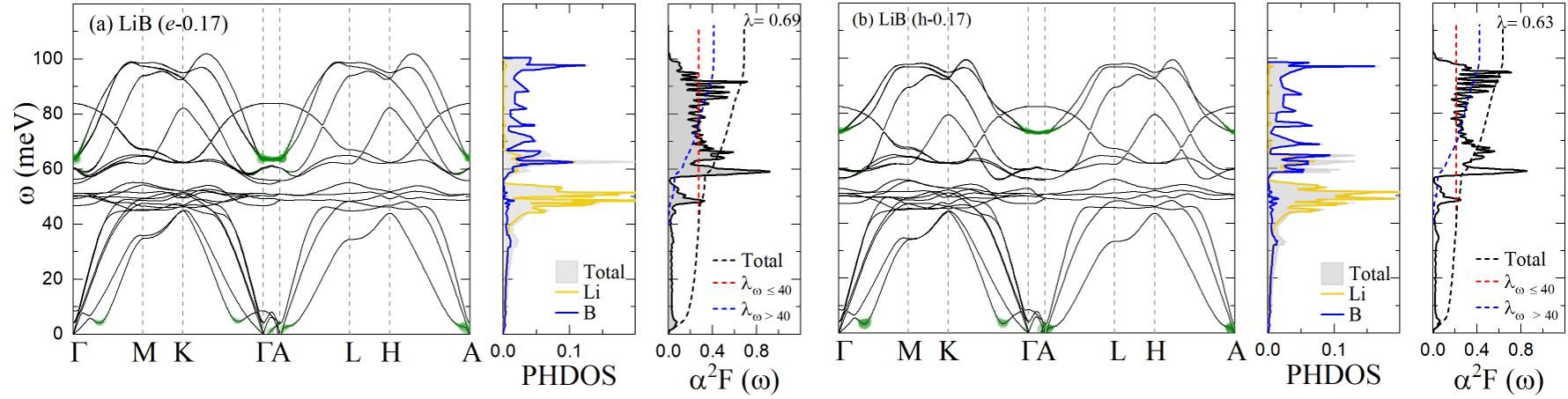}
	\caption{\label{fig-S11-ph-doping} Calculated phonon dispersion, PHDOS, and Eliashberg spectral function $\alpha^2F(\omega)$ of LiB for the doping level of (a) $e$-0.17,  and (b) h-0.17. The phonon branches are broadened by the $e$-ph coupling strength $\lambda_{\textbf{q}\nu}$, showcasing the modes and directions that have the strongest $e$-ph coupling.}
\end{figure*}

\begin{figure*}[h!]
	\centering
	\includegraphics[width=0.9\linewidth]{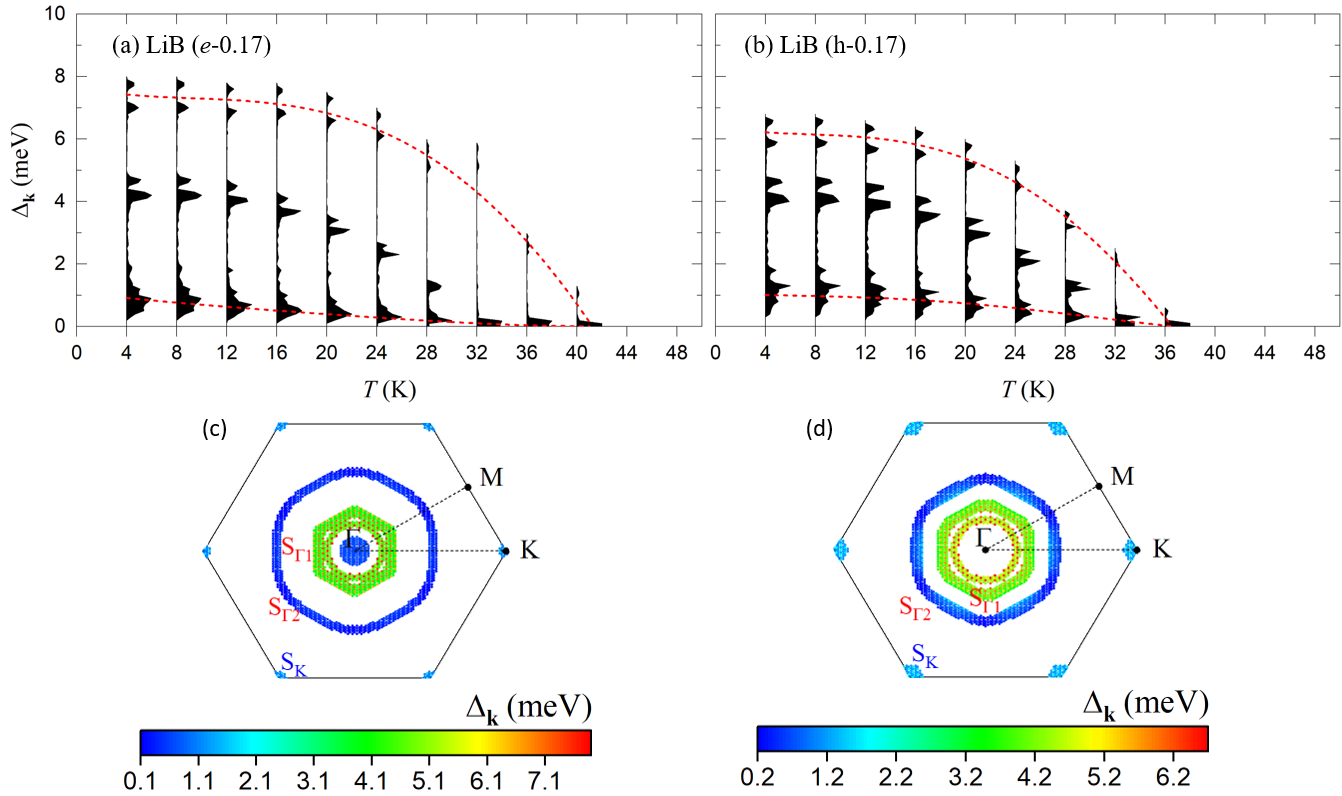}
	\caption{\label{fig-S12-aniso-doping} Calculated anisotropic superconducting gaps $\Delta_\textbf{k}$ of LiB as a function of temperature under (a) $e$-0.17 and (b) h-0.17 doping with $\mu^*$ = 0.10. The red dashed curves are a guide to the eye. The distribution of superconducting gap at 4~K on the cross-section Fermi surface plane through $z$=0 under (c) $e$-0.17 and (d) h-0.17 doping.}
\end{figure*}


\begin{table}[h!]
	\caption{\label{table1} Band-resolved symmetrized $e$-ph coupling $\Lambda_{ij}$ and DOS for the three-band FS divisions of LiB at 0 GPa and under doping. DOS is given in states/(eV spin).}
	\begin{tabular*}{0.4\textwidth}{c @{\extracolsep{\fill}} cccc}
		& \multicolumn{4}{l}{}  \\ 
		\hline\hline
		LiB & FS regions  & $\sigma$+$\zeta$ & $\zeta$ & $\pi$              \\ 
		\hline
		undoped &$\sigma$+$\zeta$   & 0.506  & 0.059  &  0.006             \\
		$\lambda$=0.66&$\zeta$    & 0.059  & 0.019  &  0.002             \\
		&$\pi$   & 0.006  & 0.002  & 0.000           \\
		&   &   &   &              \\
		&DOS & 0.507  & 0.173  & 0.019 \\  
		\hline
		$e$-0.17 &$\sigma$+$\zeta$   & 0.523  & 0.065  &  0.005             \\
		$\lambda$=0.69&$\zeta$   & 0.065  & 0.019  &  0.002             \\
		&$\pi$    & 0.005  & 0.002  & 0.000           \\
		&   &   &   &              \\
		&DOS & 0.481  & 0.168  & 0.020   \\
		\hline
		$h$-0.17 &$\sigma$+$\zeta$    & 0.503  & 0.043  &  0.014             \\
		$\lambda$=0.63&$\zeta$   & 0.043  & 0.009  &  0.004             \\
		&$\pi$    & 0.014  & 0.004  & 0.002           \\
		&   &   &   &              \\
		&DOS & 0.531  & 0.133  & 0.042  \\
		\hline
		\hline
	\end{tabular*}
\end{table}

\begin{figure*}[h]
	\centering
	\includegraphics[width=0.8\linewidth]{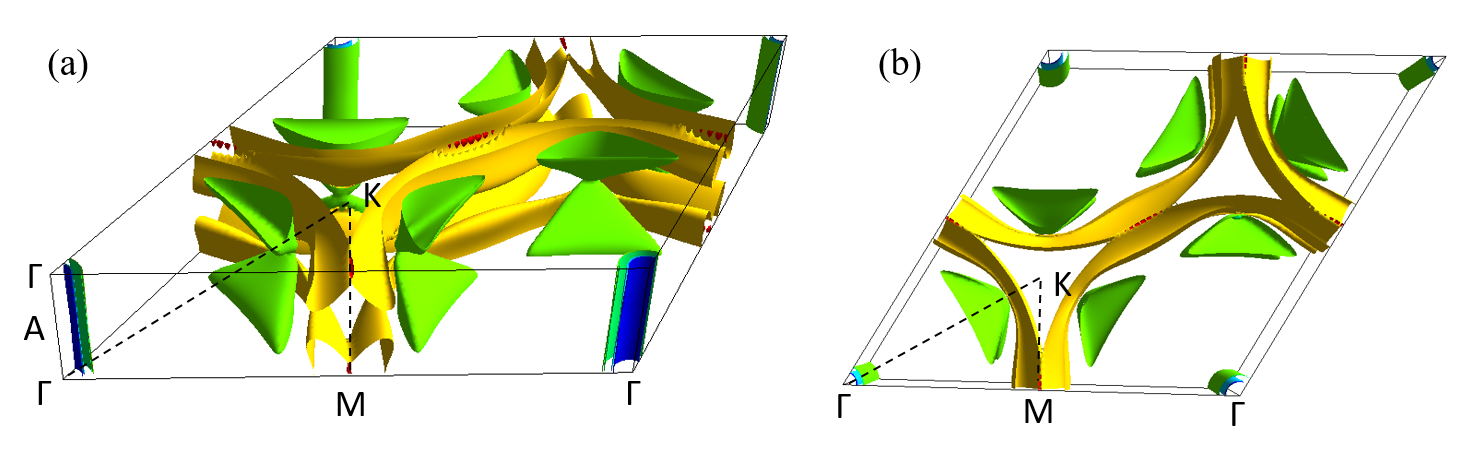}
	\caption{\label{fig-S13-fs-MgB} (a) Side and (b) top view of the Fermi surface of MgB. }
\end{figure*}

\begin{figure*}[h]
	\centering
	\includegraphics[width=0.8\linewidth]{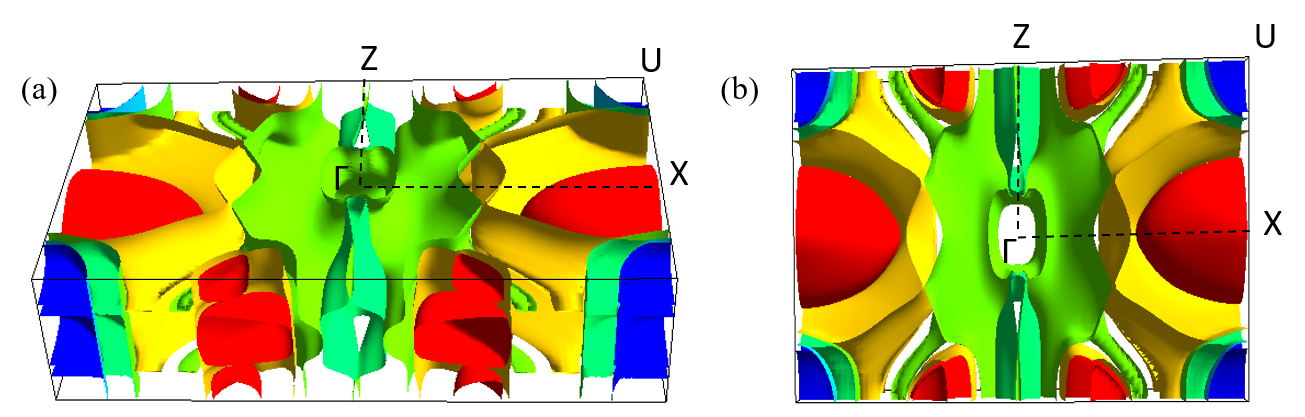}
	\caption{\label{fig-S14-fs-LiMgB2} (a) Side and (b) top view of the Fermi surface of LiMgB$_2$.}
\end{figure*}


\end{document}